\documentclass[twocolumn,showpacs,preprintnumbers,amsmath,amssymb,superscriptaddress]{revtex4}
\usepackage{graphicx}
\usepackage{dcolumn}
\usepackage{bm}
\usepackage[figuresright]{rotating}
\usepackage{xspace}
\usepackage{multirow}
 
\newcommand{\numu}{\mbox{$\nu_{\mu}$}}                   
\newcommand{\pizero}[1]{\ensuremath{\pi^0}}              
\newcommand{\dms}       {\Delta m^2}
\newcommand{\sstt}      {\sin^2 2 \theta}
\newcommand{\Rnqe}      {R_{\mathrm{nQE}}}
\newcommand{\pot}{POT\xspace}
\newcommand{\fn}{$F/N$\xspace}

\newcommand{\nsk} {\ensuremath{N^\mathrm{SK}_\mathrm{exp}}} 
\newcommand{\enu} {\ensuremath{E_\nu}}   
\newcommand{\enurec} {\ensuremath{E_\nu^\mathrm{rec}}} 
\newcommand{\rfn} {\ensuremath{R^\mathrm{F/N}}}

\newcommand{\ktint} {\ensuremath{N^\mathrm{1KT}_\mathrm{int}}}
\newcommand{\effkt} {\ensuremath{\epsilon^\mathrm{1KT}}}

\newcommand{\cND}       {\chi^{2}_{\mathrm{ND}}}
\newcommand{\cKT}       {\chi^{2}_{\mathrm{1KT}}}
\newcommand{\cSF}       {\chi^{2}_{\mathrm{SF}}}
\newcommand{\cSB}       {\chi^{2}_{\mathrm{SB}}}
\newcommand{\MC}        {\mathrm{MC}}
\newcommand{\QE}        {\mathrm{QE}}
\newcommand{\nQE}       {\mathrm{nQE}}
\newcommand{\fphi}      {f^\phi}

 \newcommand{\Ptrack}       {P^{\mathrm{SB}}_{\mathrm{2trk/1trk}}}
 \newcommand{\Pdtheta}      {P^{\mathrm{SB}}_{\mathrm{nonQE/QE}}}
 \newcommand{\Ppscale}      {P^{\mathrm{SB}}_{\mathrm{p-scale}}}

 \newcommand{\exv}[2]    {#1 \! \times \! 10^{#2}}

 \newcommand{\Lnorm}     {\mathcal{L}_{\mathrm{norm}}}
 \newcommand{\Lshape}     {\mathcal{L}_{\mathrm{shape}}}
 \newcommand{\Lsyst}     {\mathcal{L}_{\mathrm{syst}}} 

 \newcommand{\Nobs}     {N_{\mathrm{obs}}} 
 \newcommand{\Nexp}     {N_{\mathrm{exp}}} 
 \newcommand{\Sysf}     {\mbox{\boldmath $f$}}

 \newcommand{\LSK}     {\Phi^{\mathrm{SK}}}
 
  \newcommand{\LND}    {\Phi^{\mathrm{ND}}}

 \newcommand{\POTSK}    {\mathrm{POT}^{\mathrm{SK}}} 
 \newcommand{\POTKT}    {\mathrm{POT}^{\mathrm{1KT}}} 
 \newcommand{\Effsrm}    {\epsilon_{1\mathrm{R}\mu}^{\mathrm{SK}}}

 \newcommand{\MSK}    {\mathrm{M}^{\mathrm{SK}}} 
 \newcommand{\MKT}    {\mathrm{M}^{\mathrm{1KT}}} 

 \newcommand{\lnL}       {{\ln\Lcal}}
 \newcommand{\Lcal}      {\mathcal{L}}

\begin{document}


\title{Measurement of Neutrino Oscillation by the K2K Experiment}

\newcommand{\BCN}{\affiliation{Institut de Fisica d'Altes Energies, Universitat Autonoma de Barcelona, E-08193 Bellaterra (Barcelona), Spain}}
\newcommand{\BU}{\affiliation{Department of Physics, Boston University, Boston, Massachusetts 02215, USA}}
\newcommand{\UBC}{\affiliation{Department of Physics \& Astronomy, University of British Columbia, Vancouver, British Columbia V6T 1Z1, Canada}}
\newcommand{\UCI}{\affiliation{Department of Physics and Astronomy, University of California, Irvine, Irvine, California 92697-4575, USA}}
\newcommand{\SACLAY}{\affiliation{DAPNIA, CEA Saclay, 91191 Gif-sur-Yvette Cedex, France}}
\newcommand{\CNU}{\affiliation{Department of Physics, Chonnam National University, Kwangju 500-757, Korea}}
\newcommand{\DU}{\affiliation{Department of Physics, Dongshin University, Naju 520-714, Korea}}
\newcommand{\DUKE}{\affiliation{Department of Physics, Duke University, Durham, North Carolina 27708, USA}}
\newcommand{\GENEVA}{\affiliation{DPNC, Section de Physique, University of Geneva, CH1211, Geneva 4, Switzerland}}
\newcommand{\UH}{\affiliation{Department of Physics and Astronomy, University of Hawaii, Honolulu, Hawaii 96822, USA}}
\newcommand{\KEK}{\affiliation{High Energy Accelerator Research Organization(KEK), Tsukuba, Ibaraki 305-0801, Japan}}
\newcommand{\HIR}{\affiliation{Graduate School of Advanced Sciences of Matter, Hiroshima University, Higashi-Hiroshima, Hiroshima 739-8530, Japan}}
\newcommand{\INR}{\affiliation{Institute for Nuclear Research, Moscow 117312, Russia}}
\newcommand{\KOBE}{\affiliation{Kobe University, Kobe, Hyogo 657-8501, Japan}}
\newcommand{\KOR}{\affiliation{Department of Physics, Korea University, Seoul 136-701, Korea}}
\newcommand{\KYO}{\affiliation{Department of Physics, Kyoto University, Kyoto 606-8502, Japan}}
\newcommand{\LSU}{\affiliation{Department of Physics and Astronomy, Louisiana State University, Baton Rouge, Louisiana 70803-4001, USA}}
\newcommand{\MIT}{\affiliation{Department of Physics, Massachusetts Institute of Technology, Cambridge, Massachusetts 02139, USA}}
\newcommand{\MIYAGI}{\affiliation{Department of Physics, Miyagi University of Education, Sendai 980-0845, Japan}}
\newcommand{\NAG}{\affiliation{Solar-Terrestrial Environment Laboratory, Nagoya University, Nagoya, Aichi 464-8601, Japan}}
\newcommand{\NIIGATA}{\affiliation{Department of Physics, Niigata University, Niigata, Niigata 950-2181, Japan}}
\newcommand{\OKAYAMA}{\affiliation{Department of Physics, Okayama University, Okayama, Okayama 700-8530, Japan}}
\newcommand{\OSAKA}{\affiliation{Department of Physics, Osaka University, Toyonaka, Osaka 560-0043, Japan}}
\newcommand{\ROME}{\affiliation{University of Rome La Sapienza and INFN, I-000185 Rome, Italy}}
\newcommand{\SNU}{\affiliation{Department of Physics, Seoul National University, Seoul 151-747, Korea}}
\newcommand{\SOLTAN}{\affiliation{A.~Soltan Institute for Nuclear Studies, 00-681 Warsaw, Poland}}
\newcommand{\TOHOKU}{\affiliation{Research Center for Neutrino Science, Tohoku University, Sendai, Miyagi 980-8578, Japan}}
\newcommand{\SB}{\affiliation{Department of Physics and Astronomy, State University of New York, Stony Brook, New York 11794-3800, USA}}
\newcommand{\TUS}{\affiliation{Department of Physics, Tokyo University of Science, Noda, Chiba 278-0022, Japan}}
\newcommand{\KAM}{\affiliation{Kamioka Observatory, Institute for Cosmic Ray Research, University of Tokyo, Kamioka, Gifu 506-1205, Japan}}
\newcommand{\RCCN}{\affiliation{Research Center for Cosmic Neutrinos, Institute for Cosmic Ray Research, University of Tokyo, Kashiwa, Chiba 277-8582, Japan}}
\newcommand{\TRIUMF}{\affiliation{TRIUMF, Vancouver, British Columbia V6T 2A3, Canada}}
\newcommand{\VAL}{\affiliation{Instituto de F\'{i}sica Corpuscular, E-46071 Valencia, Spain}}
\newcommand{\UW}{\affiliation{Department of Physics, University of Washington, Seattle, Washington 98195-1560, USA}}
\newcommand{\WARSAW}{\affiliation{Institute of Experimental Physics, Warsaw University, 00-681 Warsaw, Poland}}

\BCN
\BU
\UBC
\UCI
\SACLAY
\CNU
\DU
\DUKE
\GENEVA
\UH
\KEK
\HIR
\INR
\KOBE
\KOR
\KYO
\LSU
\MIT
\MIYAGI
\NAG
\NIIGATA
\OKAYAMA
\OSAKA
\ROME
\SNU
\SOLTAN
\TOHOKU
\SB
\TUS
\KAM
\RCCN
\TRIUMF
\VAL
\UW
\WARSAW

\author{M.~H.~Ahn}\SNU
\author{E.~Aliu}\BCN                
\author{S.~Andringa}\BCN 
\author{S.~Aoki}\KOBE 
\author{Y.~Aoyama}\KOBE
\author{J.~Argyriades}\SACLAY 
\author{K.~Asakura}\KOBE 
\author{R.~Ashie}\KAM 
\author{F.~Berghaus}\UBC
\author{H.~G.~Berns}\UW 
% ...
\collaboration{The K2K Collaboration}\noaffiliation

\date{\today}

\begin{abstract}
  We present measurements of $\nu_\mu$ disappearance 
  in K2K, the KEK to Kamioka long-baseline neutrino
  oscillation experiment.  One hundred and twelve beam-originated
  neutrino events are observed in the fiducial volume of
  Super-Kamiokande with an expectation of $158.1^{+9.2}_{-8.6}$ events
  without oscillation.  A distortion of the energy spectrum is also
  seen in 58 single-ring muon-like events with reconstructed
  energies.  The probability that the observations are explained by 
  the expectation for no neutrino oscillation is
  $0.0015\%$~($4.3\sigma$).  In a two flavor oscillation scenario, the
  allowed $\dms$ region at $\sstt=1$ is between $1.9$ and $3.5 \times
  10^{-3}$~$\rm eV^2$ at the 90~\% C.L.  with a best-fit value of $2.8
  \times 10^{-3}$~$\rm eV^2$.
\end{abstract}


\maketitle
\newpage


\section{Introduction}
The oscillation of $\nu_\mu$ neutrinos into other neutrino flavors is
now well established.  By using the angle and energy distribution of
atmospheric neutrinos, the Super-Kamiokande collaboration has measured
the parameters of oscillation and observed the sinusoidal
disappearance signature predicted by
oscillations~\cite{Ashie:2005ik,Ashie:2004mr}.  The K2K collaboration
has previously reported evidence of neutrino oscillations in a
man-made neutrino beam which was directed 250~km across
Japan~\cite{Ahn:2002up,Aliu:2004sq}.

For neutrinos of a few GeV, the dominant oscillation 
is between $\nu_\mu$ and $\nu_\tau$ flavor states and two-flavor oscillations
suffice to describe and analyze the data.  In the two-flavor neutrino
oscillation framework the probability that a neutrino of energy
$E_\nu$ with a flavor state $\nu_\mu$ will later be observed in the
$\nu_\tau$ flavor eigenstate after traveling a distance $L$ in vacuum
is:

\begin{equation}
  \label{eqn:oscillation}
  P(\nu_\mu \rightarrow \nu_\tau) = \sin^2 2 \theta \sin^2{\Big(}\frac{1.27
    \Delta m^2(\textrm{eV}^2) L (\textrm{km})}{E_\nu(\textrm{GeV})}\Big{)},
\end{equation}

\noindent
where $\theta$ is the mixing angle between the mass eigenstates and
the flavor eigenstates and $\dms$ is the difference of the
squares of the masses of the mass eigenstates.

The KEK to Kamioka long-baseline neutrino oscillation experiment
(K2K)~\cite{Ahn:2001cq} uses an accelerator-produced beam of nearly
pure $\numu$ with a neutrino flight distance of 250~km to probe the
same $\dms$ region as that explored with atmospheric neutrinos.  The
neutrinos are measured first by a suite of detectors located
approximately 300 meters from the proton target and then by the
Super-Kamiokande (SK) detector 250~km away. The near detector complex
consists of a 1 kiloton water Cherenkov detector (1KT) and a fine
grained detector system.  SK is a 50 kiloton water Cherenkov detector,
located 1000~m underground~\cite{Fukuda:2002uc}.

The K2K experiment is designed to measure neutrino oscillations using
a man-made beam with well controlled systematics, complementing and
confirming the measurement made with atmospheric neutrinos.  In this
paper we report a complete description of the observation of neutrino
oscillations in the K2K long-baseline experiment, and present a
measurement of the $\dms$ and mixing angle parameters.

Neutrino oscillation causes both a suppression in the total number of
$\numu$ events observed at SK and a distortion of the measured energy spectrum
compared to that measured at the production point.  Therefore, all of
the beam-induced neutrino events observed within the fiducial volume
of SK are used to measure the overall suppression of flux.  In
addition, in order to study the spectral distortion, the subset of these
events for which the incoming neutrino energy can be reconstructed are
separately studied.

If the neutrino interaction which takes place at SK is a charged-current (CC)
quasi-elastic(QE)($\nu_\mu + n \rightarrow \mu + p$)
the incoming neutrino energy can be reconstructed using two-body
kinematics, and the spectral distortion studied.  
At the energy of the K2K experiment typically only the muon in this
reaction is energetic enough to produce Cherenkov light and be
detected at SK but kinematics of the muon alone are enough to
reconstruct the energy for these events.  

In order to select the charged-current (CC) quasi-elastic ~(QE) events
in the data sample, one-ring events identified as a muon~($1R \mu$) are
chosen which have a high fraction of CC-QE at the K2K energy.  For
these events, the energy of the parent neutrino can be calculated by
using the observed momentum of the muon, assuming QE interactions, and
neglecting Fermi momentum:
\begin {eqnarray}
E_\nu ^{\rm rec}=\frac {m_NE_\mu-m^2_\mu/2} {m_N-E_\mu+P_\mu\cos\theta_\mu},
\label{eq:Enurec}
\end {eqnarray}

\noindent
where $m_N$, $E_\mu$, $m_\mu$, $P_\mu$ and $\theta_\mu$ are the
nucleon mass, muon energy, the muon mass, the muon momentum and the
scattering angle relative to the neutrino beam direction,
respectively.

In this paper, all data taken in K2K between June 1999 and November 2004 are
used to measure the suppression of events and energy distortion and to measure
the parameters of oscillation.


\section{Neutrino beam}
\subsection{K2K neutrino beam and beam monitor}
\begin{figure}
  \begin{center}
    \includegraphics[width=\columnwidth]{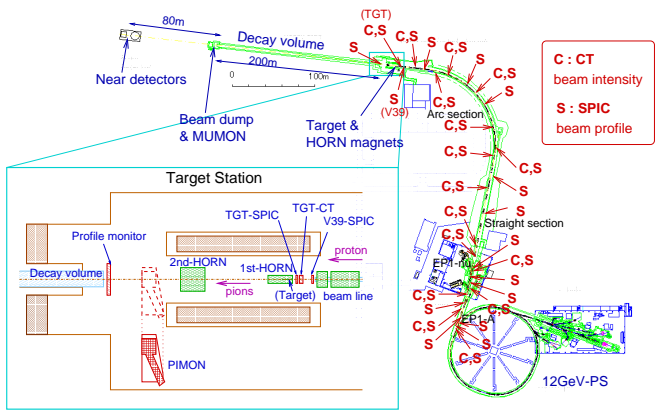}
    \caption[Schematic view of KEK-PS and neutrino beam line and location of
      beam line components.]
            {A schematic view of the KEK-PS and neutrino beam line and the location
              of beam line components. The EP1 neutrino beam line leads protons
              through a distance of 400~m from the EP1-A extraction point to
              the target station via the straight and arc sections.
              The characters ``C'' and ``S'' in the figure show the locations
              of the CT and SPIC installations, respectively.
              The lower-left inset is a magnified view of the target station.
              The production target and a set of horn magnets are located in
              the target station.
              A pion monitor was installed on two occasions downstream the horn
              magnets.}
	    \label{fig:beam:KEK-PS}
  \end{center}
\end{figure}

The accelerator and the neutrino beam line for K2K consist of
a 12~GeV proton synchrotron~(KEK-PS), a primary proton transportation line,
a hadron production target, a set of focusing horn magnets for secondary
particles, a decay volume, and a beam dump.
A schematic view of the KEK-PS and neutrino beam line is shown in
Fig.~\ref{fig:beam:KEK-PS}.
In this section, we describe each beam line component in order, from
upstream to downstream.

\subsubsection{Primary proton beam}
Protons are accelerated by the KEK-PS to a kinetic energy of 12~GeV.
After acceleration, all protons are extracted in a single turn to
the neutrino beam line.
The duration of an extraction, or a ``spill'', is 1.1~$\mu$sec, which contains
9 bunches of protons with a 125~ns time interval between them.
As shown in Fig.~\ref{fig:beam:KEK-PS},
the beam is extracted toward the north, bent by 90$^{\circ}$
toward the direction of SK, and transported to the target station.
There is a final steering magnet just before the target which directs the 
beam to SK at an angle of about 1$^\circ$ downward from horizontal.

The beam intensity is monitored by 13 current transformers (CTs)
installed along the neutrino beam line as shown in Fig.~\ref{fig:beam:KEK-PS}.
The CTs are used to monitor the beam transportation efficiency.
The overall transportation efficiency along the beam line is about 85\%.
A CT placed just in front of the production target
is used to estimate the total number of protons delivered to the target.
A typical beam intensity just before the target is
about $5\times10^{12}$ protons in a spill.

In order to measure the profile and the position of the beam,
28 segmented plate ionization chambers (SPICs) are also installed
(Fig.~\ref{fig:beam:KEK-PS}).
They are used to steer and monitor the beam, while the last two SPICs
in front of the target are used to estimate the beam size
and divergence, which is used as an input to our beam Monte Carlo (MC) simulation.

\subsubsection{Hadron production target and horn magnets}
A hadron production target and a set of horn magnets are placed in
the target station.
Protons hit the target and a number of secondary particles are generated
at the production target.
Two toroidal magnetic horns are employed to focus positively charged particles,
mainly $\pi^+$'s, in the forward direction by the magnetic field.
A typical focusing of transverse momentum by the horn magnets is
about 100 MeV/$c$ per meter.
The momenta of focused pions are around 2$-$3~GeV/$c$,
which corresponds to about 1.0$-$1.5~GeV of  energy for those neutrinos
decaying in the forward direction.
According to our Monte Carlo simulation, the
flux of neutrinos above 0.5~GeV is 22 times greater with 
horn magnets with 250~kA current 
than without the horn current.

\begin{figure}
 \begin{center}
  \includegraphics[width=\columnwidth]{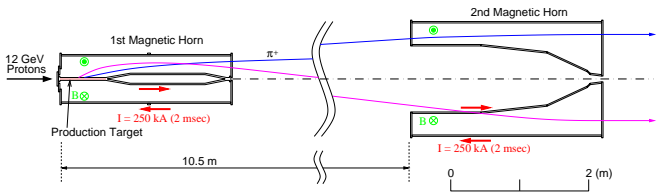}
  \caption[Schematic view of the two horn magnets]
          {Schematic view of the two horn magnets.
           An electrical current of 250~kA is supplied to both horns,
           creating a toroidal magnetic field inside the horns.
           The production target, an aluminum rod of 66~cm in length
           and 3~cm in diameter, is embedded inside the first horn magnet,
           which also plays the role of inner conductor of the horn.
           The second horn is located 10.5~m downstream of the first horn.
           }
  \label{fig:beam:horn-magnets}
 \end{center}
\end{figure}

A schematic view of the horn magnets is shown in
Fig.~\ref{fig:beam:horn-magnets}.
The dimensions of the first horn are 0.70~m in diameter and 2.37~m in length,
while those of the second horn are 1.65~m in diameter and 2.76~m in length.
Both horns are cylindrically symmetric in shape.
The production target, a rod of a length of 66~cm and diameter of 3~cm,
made of aluminum alloy 6061-T, is embedded inside
the first horn.
The target diameter was 2~cm in June 1999 and was changed to 3~cm
in November 1999 for improved mechanical strength.
The target also plays the role of inner conductor of the first horn, making
a strong magnetic field inside the horn to achieve high focusing efficiency.
The second horn is located 10.5~m downstream from the first horn,
playing the role of a reflector, which re-focuses
over-bent low energy pions, and in addition further focuses under-bent
high energy pions.

Pulsed current with a duration of 2~msec and an amplitude of 250~kA
(200~kA in June 1999) is supplied by four current feeders to each horn.
The peaking time of the current is adjusted to match the beam timing.
The maximum magnetic field in the horn is 33~kG at the surface of the
target rod with 3~cm diameter target and 250~kA horn current.

The values of the current supplied to the horn magnets are
read out by CTs put in between current feeders
and recorded by a flash analog-to-digital converter (FADC) on a spill-by-spill basis.
Overall current and current balance between feeders are monitored
to select good beam spills.
The magnetic field inside the prototype of the first horn
was measured using pickup coils; results showed that the radial distribution
of the field was in agreement with the design distribution and
the azimuthal symmetry was confirmed to within a measurement error of 15\%.
Detailed descriptions of the horn magnets are found
in~\cite{Yamanoi:1997pi,Yamanoi:2000hg,Kohama:master}.

A pion monitor~(PIMON) was installed on two occasions just downstream of the
horn magnets, as shown in Fig.~\ref{fig:beam:KEK-PS}, in order to measure the
momentum and angular distributions of pions coming through the horn magnets.
The PIMON will be described in detail later in Sec.~\ref{sec:farnear}.

\subsubsection{Decay volume, beam dump, and muon monitors}
The positive pions focused by the horn magnets go into a 200~m long decay
volume which starts 19~m downstream of the production target,
where the $\pi^+$ decay:  
$\pi^+\to\mu^+\,\nu_\mu$.
The decay volume is cylindrical in shape and is separated into
three sections with different dimensions.
The diameters of the pipe are 1.5~m, 2~m, and 3~m in the first 10~m,
the following 90~m, and the remaining 100~m sections, respectively.
The decay volume is filled with helium gas of 1~atm (rather than air) 
to reduce the loss
of pions by absorption and to avoid uncontrollable pion production in the gas.
The beam dump is located at the end of the decay volume
to absorb all the particles except for neutrinos.
It consists of 3.5~m thick iron, 2~m thick concrete, and
a region of soil about 60~m long.

\begin{figure}
 \begin{center}
  \includegraphics[width=\columnwidth]{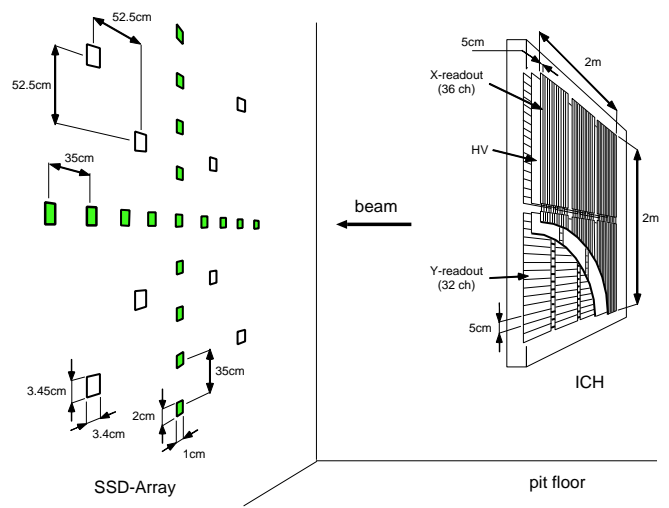}
  \caption[A schematic view inside the muon-pit]
          {A schematic view inside the muon-pit.
           An ionization chamber (ICH) and an array of silicon pad detectors
           (SPD) are located inside the muon-pit.
          }
  \label{fig:beam:inside-muon-pit}
 \end{center}
\end{figure}

There is a pit called the ``muon-pit'' just downstream of the iron
and concrete shields.
Muons with momentum greater than 5.5~GeV/$c$ can reach the muon-pit.
The flux at the pit is roughly $10^4 ~\mathrm{muons/cm^2/spill}$.
The parent particles of both muons and neutrinos are pions, so
the profile center of muons corresponds to that of neutrinos.
A change in the beam direction by 3~mrad corresponds to a change in
the neutrino flux and spectrum at SK of about 1\%, and hence
it must be controlled and monitored to be within 3~mrad.
Fig.~\ref{fig:beam:inside-muon-pit} shows a schematic view inside the pit.
Two detectors (MUMONs) are installed in it:
one is an ionization chamber (ICH) and the other is an array of silicon pad
detectors (SPD).
The purpose of these detectors is to measure the profile and intensity of muons
penetrating the shields on spill-by-spill basis.

An ICH is a segmented plate chamber
with a size of 190~cm (horizontal) $\times$ 175~cm (vertical).
It consists of six modules of size
60~cm $\times$ 90~cm,
3 modules in the horizontal direction and 2 modules in the vertical direction.
The gap between modules is 25~cm in horizontal and 15~cm in vertical
(Fig.~\ref{fig:beam:inside-muon-pit}).
The corresponding strip lines of adjoining modules
are electrically connected over the gaps
to make long strip lines of length of $\sim$180~cm.
There are 36 horizontal readout channels and 32 vertical channels.
The channel-to-channel uniformity is calibrated by moving ICH horizontally
and vertically~\cite{Maruyama:PhD} assuming stability of the muon beam.
The relative gain of the channels has been stable within an accuracy of
several percent.

Two types of SPDs are used:
one is a small SPD which has a sensitive area of 1~cm $\times$ 2~cm
with a depletion layer thickness of 300~$\mu$m,
and the other is a large SPD which has a sensitive area of
3.4~cm $\times$ 3.05~cm with a depletion layer thickness of 375~$\mu$m.
Seventeen small SPDs are arranged along the horizontal and
the vertical axes at 35~cm intervals
while nine large SPDs are in diagonal arrays at 74.2~cm intervals.
The sensitivity of each small SPD was measured using an
LED light source at a test bench and it was found that
all the small SPDs agree within 6\%~\cite{Maruyama:PhD}.
The sensitivity difference between the large SPDs was measured using the
muon beam at the muon-pit.
All the large SPDs were aligned along the beam axis simultaneously
and the output charge from each SPD was compared to obtain
the relative gain factor.
The gain factors have an uncertainty of 10\% 
due to the $z$-dependence of the muon beam intensity~\cite{Maruyama:PhD}.

\subsection{Summary of beam operation}
\label{sec:beam:summary}
The construction of neutrino beam line was completed
early in 1999 and beam commissioning started in March 1999.
The beam line and all the components were constructed and aligned
within an accuracy of 0.1 mrad with respect to a nominal beam axis
which was determined based on the results of a global positioning system (GPS) survey
accurate to 0.01~mrad
between KEK and Kamioka sites~\cite{Noumi:1997iq}.
In June 1999, the neutrino beam and detectors were ready
to start data-taking for physics.
We took data on and off over the period from June 1999 to November 2004,
which is divided into five subperiods according to
different experimental configurations:
June 1999~(Ia),
November 1999 to July 2001~(Ib),
December 2002 to June 2003~(IIa), 
October 2003 to February 2004~(IIb), and
October 2004 to November 2004~(IIc).
The horn current was 200~kA~(250~kA) and the diameter of the production
target was 2~cm~(3~cm) in the Ia~(other) period.
The SK PMTs were full density for Ia and Ib, but were half density
for IIa, IIb and IIc.
There was a lead-glass calorimeter~(LG) installed in between
a scintillating-fiber/water-target tracker~(SciFi) and
a muon range detector~(MRD) during the Ia and Ib periods;
it was replaced by a totally active fine-segmented scintillator
tracker~(SciBar) for IIa, IIb and IIc.
Only the first four layers of the SciBar detector were installed for
IIa while it was in its full configuration for IIb and IIc.
Furthermore, the water target in the SciFi was replaced by aluminum rods
during IIc.
The different experimental configurations for the different periods
are briefly summarized in Table~\ref{table:beam:beamsummary}.

\begin{figure}
 \begin{center}
  \includegraphics[width=\columnwidth]{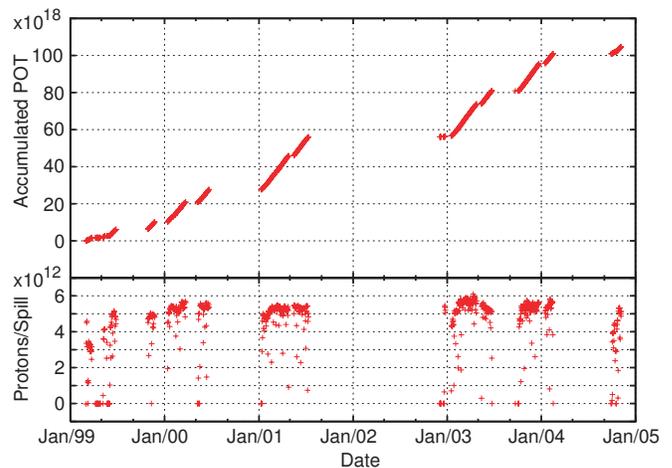}
  \caption{The number of protons delivered to the production target
            in the period from March 1999 to November 2004.
            The horizontal axis corresponds to the date.
            The upper figure shows the total number of protons on target (\pot)
            accumulated since March 1999, and the lower figure shows the
            \pot per spill averaged in a day. In total,
            $104.90\times10^{18}$ protons were delivered
            during the entire period including beam commissioning and
            tuning periods.}
  \label{fig:beam:intpot}
 \end{center}
\end{figure}

The number of protons delivered to the target is summarized in
Table~\ref{table:beam:beamsummary}, and shown as a function of time
in Fig.~\ref{fig:beam:intpot}.
Among the delivered spills, spills which satisfy the following
criteria are used for the physics analysis:
(1) beam spills with normal machine status. Spills during machine studies,
beam tuning, and several beam studies are discarded.
(2) Beam spills with no trouble in the beam components and data acquisition
systems.
(3) Beam spills with the proton intensity greater than
$1\times10^{12}$ protons.
(4) Beam spills with the horn current greater than 240~kA (190~kA) for
the period other than Ia (for the Ia period).
The number of protons on target (\pot) for the physics analysis
is summarized in Table~\ref{table:beam:beamsummary} as well as the total
number of protons delivered.
In total, $1.049\times10^{20}$~protons were delivered to the production
target while $0.922\times10^{20}$~\pot are used in our physics analysis.
\begin{table*}
  \begin{center}
    \caption{Summary of the number of protons on target and the experimental
             configuration for each running period. The row labeled
             ``LG/SciBar configuration'' indicates the detector installed
             between the SciFi and MRD detectors.
             For the row ``SK configuration'',
             ``SK-I'' refers to the configuration with full PMT density
             while ``SK-II'' refers to that with half density.
             See the text for more detailed description of
             the experimental configurations.
             The delivered \pot shown in the table includes the beam
             delivered during commissioning and beam tuning work before
             the physics runs.}
    \label{table:beam:beamsummary}
    \begin{tabular}{r|ccccc|r}
      \hline\hline
      \multirow{2}{2cm}{\rightline{Periods}}
                    &    Ia &    Ib &   IIa &   IIb &   IIc
                    & \multirow{2}{1.1cm}{\rightline{total}} \\
                    & \multirow{1}{2.4cm}{\centerline{Jun.'99}}
                    & \multirow{1}{2.4cm}{Nov.'99$-$Jul.'01}
                    & \multirow{1}{2.4cm}{Dec.'02$-$Jun.'03}
                    & \multirow{1}{2.4cm}{Oct.'03$-$Feb.'04}
                    & \multirow{1}{2.4cm}{Oct.'04$-$Nov.'04} & \\
      \hline
      Delivered POT $(\times10^{18})$
                    &  6.21 & 49.85 & 24.91 & 20.15 &  3.78 & 104.90 \\
      POT for analysis $(\times10^{18})$
                    &  3.10 & 44.83 & 22.57 & 18.61 &  3.12 &  92.23 \\
      \hline
      Horn current & 200~kA & 250~kA & 250~kA & 250~kA & 250~kA & \\
      Target diameter & 2~cm &  3~cm &   3~cm &   3~cm &   3~cm & \\
      SK configuration    & SK-I  & SK-I & SK-II & SK-II & SK-II & \\
      LG/SciBar configuration & LG & LG & SciBar (4 layers) & SciBar & SciBar & \\
      Target material in SciFi & water & water & water & water & aluminum & \\
      \hline\hline
    \end{tabular}
  \end{center}
\end{table*}

\begin{figure}
 \begin{center}
  \includegraphics[width=\columnwidth]{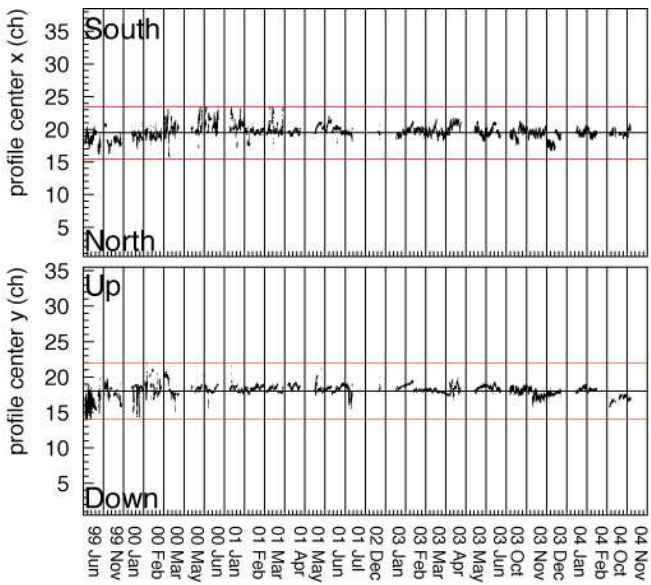}
  \caption{Stability of the center of muon profile measured by the ionization
            chamber (ICH) in MUMON.
            The upper figure shows the profile center of the horizontal
            direction and the lower figure shows that of the vertical
            direction. In each figure, the beam direction to SK measured by 
            GPS and $\pm1~{\rm mrad}$ off the center are indicated by horizontal lines.
            The data shown here are after good beam selection.}
  \label{fig:beam:mucen}
 \end{center}
\end{figure}

During these periods, the direction of the neutrino beam was monitored
by MUMON in the muon-pit.
Fig.~\ref{fig:beam:mucen} shows the stability of the center of the
muon profile measured by the ionization chamber (ICH) in MUMON.
The beam was pointed to the direction of SK within $\pm1~{\rm mrad}$
during the whole run period, so that the neutrino flux and spectrum at SK
was stable within much better than 1\%.

\subsection{K2K neutrino beam simulation}
\label{sec:beam:simulation}
We use a neutrino beam Monte Carlo (beam MC) simulation program
to study our neutrino beam properties.
The beam line geometry is implemented in GEANT~\cite{brun:1987ma}
and particles are tracked in materials
until they decay into neutrinos or are absorbed in the material.
The tracks of neutrinos are extrapolated along a straight line to
the near detector (ND) and Super-Kamiokande (SK) and
the fluxes and the energy spectrum at these locations are determined.

In the simulation program, protons with a kinetic energy of 12~GeV are
injected into the aluminum production target.
The profile and divergence are assumed to be Gaussian-like
and the values measured by two SPICs in front of the target
are used as inputs.
An empirical formula for the differential cross-section by J. R. Sanford and
C. L. Wang~\cite{Sanford-Wang:1967sw,Wang:1970bn} is used to simulate
the primary hadron production in the target.
The Sanford-Wang formula is expressed as following:
\begin{eqnarray}
\displaystyle{\frac{d^2\sigma}{d\Omega dp}}& = &
C_1 \, p^{C_2} \left( 1 - \displaystyle{\frac{p}{p_{B}}} \right)\,\times \\
\nonumber
&& \exp \! \left( - \displaystyle{\frac{C_3 \,p^{C_4}}{p_{B}^{C_5}}}
- C_6 \, \theta \, (p - C_7 \, p_{B} \, \cos^{C_8}\! \theta )
\right)\, ,
\label{eqn:Sanford-Wang}
\end{eqnarray}
where ${d^2\sigma}/{d\Omega dp}$ is the double differential cross section
of particle production per interacting proton in the unit of
${\rm mb}~{\rm sr}^{-1}~({\rm GeV}/c)^{-1}$,
$\theta$ is the angle between the secondary particle and the beam axis
in the laboratory frame, $p$ and $p_{B}$ are the momenta of
the secondary particle and the incident proton, respectively.
The $C_i$'s are parameters fitted to existing hadron production data.
For the production of positively charged pions, we use as a reference model
the $C_i$'s obtained from a fit designated the ``Cho-CERN compilation'',
in which the data used in the compilation mainly come from the measurement
of proton-beryllium interactions performed by Cho et al.~\cite{Cho:1972jq}.
The values for $C_i$'s are shown in Table~\ref{table:beam:cho-cern}.
\begin{table}
  \begin{center}
    \caption{The fitted parameters, $C_i$'s, in the Sanford-Wang formula
      for the production of positively charged pions in the
      Cho-CERN compilation
      and for the HARP results~\cite{Catanesi:2005rc}.
      The target nucleus is beryllium in Cho-CERN compilation while
      it is aluminum in the HARP results. The values in the table are
      before the nuclear scaling is applied.
      }
    \label{table:beam:cho-cern}
    \begin{tabular}{rccccccccc}\hline\hline
      & & $C_1$ & $C_2$ & $C_3$ & $C_4$ & $C_5$ & $C_6$ & $C_7$ & $C_8$ \\
      \hline
      HARP     & & 440 & 0.85 & 5.1  & 1.78 & 1.78 & 4.43 & 0.14 & 35.7 \\
      Cho-CERN & & 238 & 1.01 & 2.26 & 2.45 & 2.12 & 5.66 & 0.14 & 27.3 \\
      \hline\hline
    \end{tabular}
  \end{center}
\end{table}
A nuclear rescaling is then applied to convert the pion production
cross section on beryllium to that on aluminum.
The scaling factor, $w$, is defined as
\begin{eqnarray}
w\equiv \left(\frac{A_{\rm Al}}{A_{\rm Be}} \right)^{\alpha(x_F)}\, ,
\end{eqnarray}
where $A_{\rm Al}$ and $A_{\rm Be}$ are atomic masses for aluminum and
beryllium, respectively, and an index $\alpha(x_F)$ is expressed as
\begin{eqnarray}
\alpha(x_F) = 0.74 + x_F(-0.55 + 0.26 x_F)
\end{eqnarray}
as a function of the Feynman $x$ variable, $x_F$.

Negatively charged pions and charged and neutral kaons are generated
as well as positively charged pions using the same Sanford-Wang formula
with different sets of $C_i$'s.
For negative pion production, the parameters in \cite{Cho:1972jq} are
used, while those described in \cite{Yamamoto:1981zr} are used
for the kaon production.

Generated secondary particles are tracked by GEANT with the
GCALOR/FLUKA~\cite{Gabriel:1977hf,Zeitnitz:1994bs,Fasso:1993kr} hadron model
through the two horn magnets and the decay volume
until they decay into neutrinos or are absorbed in materials.

Since GEANT treats different types of neutrinos identically,
we use a custom-made simulation program to
treat properly the type of neutrinos emitted by particle decays.
Charged pions are treated so that they decay into muon and neutrino
($\pi^+\to\mu^+\,\nu_\mu, \pi^-\to\mu^-\,\overline{\nu}_\mu$,
called $\pi^\pm_{\mu2}$)
with branching fraction of 100\%.
The kaon decays considered in our simulation are so-called
$K^{\pm}_{\mu2}$, $K^{\pm,0}_{e3}$ and $K^{\pm,0}_{\mu3}$ decays.
Their branching ratios are taken from the Particle Data
Group~\cite{Eidelman:2004wy}.  Other decays are ignored.  Neutrinos
from $K^0_S$ are ignored since the branching ratio for $K^0_S$
decaying to neutrinos is quite small.  The Dalitz plot density of
$V$$-$$A$ theory~\cite{Eidelman:2004wy,Commins:WIQL} is employed
properly in $K_{\ell3}$ decays.  Muons are considered to decay via
$\mu^{\pm}\to e^{\pm}\,\,\,\nu_{e}(\overline{\nu}_{e})\,\,\,
\overline{\nu}_{\mu}(\nu_{\mu})$, called $\mu^{\pm}_{e3}$, with 100\%
branching fraction.  The energy and angular distributions of the muon
antineutrino (neutrino) and the electron neutrino (antineutrino)
emitted from a positive (negative) muon are calculated according to
Michel spectra of $V$$-$$A$ theory~\cite{Commins:WIQL}, where the
polarization of the muon is taken into account.

The produced neutrinos are extrapolated to the ND and SK according to a
straight line
and the energy and position of the neutrinos entering the ND and SK
are recorded and used in our later simulations for neutrino interaction
and detector simulators.

\begin{figure}
 \begin{center}
   \includegraphics[width=0.4944\columnwidth]{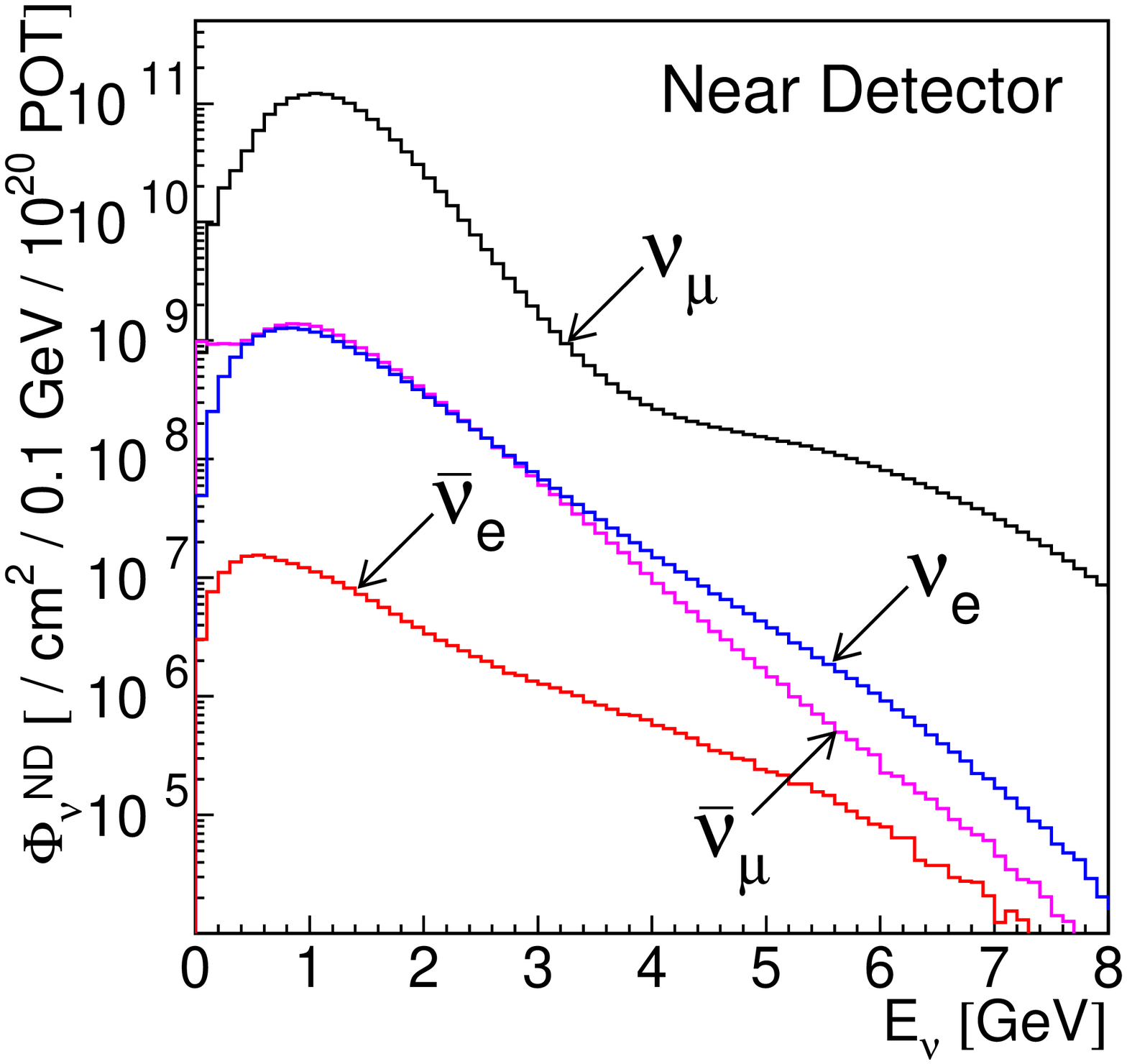}
   \includegraphics[width=0.4944\columnwidth]{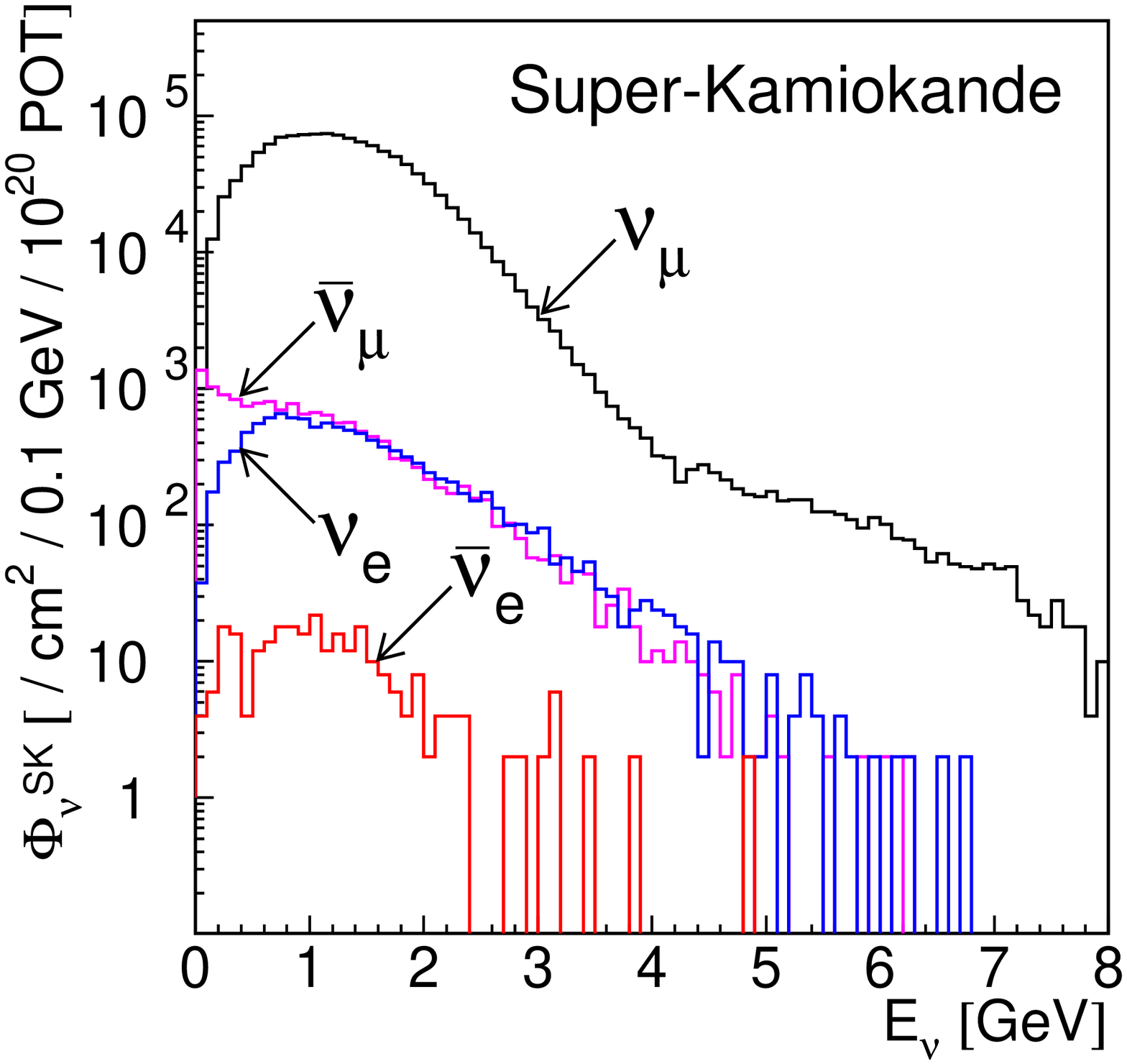}
  \caption{The energy spectrum for each type of neutrino at ND (left)
           and SK (right) estimated by the beam MC simulation.
           The neutrino beam is 97.3\% (97.9\%) pure muon neutrino
           with contaminations of $\nu_e/\nu_\mu\sim0.013~(0.009)$,
           $\overline{\nu}_\mu/\nu_\mu\sim0.015~(0.012)$, and
           $\overline{\nu}_e/\nu_\mu\sim1.8\times10^{-4}~(2.2\times10^{-4})$
           at ND (SK).
          }
  \label{fig:beam:BeamMC-EnuNeutrinoType}
 \end{center}
\end{figure}

The composition of the neutrino beam is dominated by muon neutrinos
since the horn magnets mainly focus the positive pions.
Figure~\ref{fig:beam:BeamMC-EnuNeutrinoType} shows the energy spectra of 
each type of neutrino at ND and SK estimated by the beam MC simulation.
About 97.3\% (97.9\%) of neutrinos at ND (SK)
are muon neutrinos decayed from positive pions,
and the beam is contaminated with a small fraction of neutrinos
other than muon neutrinos;
$\nu_e/\nu_\mu\sim0.013~(0.009)$,
$\overline{\nu}_\mu/\nu_\mu\sim0.015~(0.012)$, and
$\overline{\nu}_e/\nu_\mu\sim1.8\times10^{-4}~(2.2\times10^{-4})$ at ND (SK).
The validity of our beam MC simulation has been confirmed by
both the HARP experiment and PIMON measurements, which will be described
in detail in Sec.~\ref{sec:farnear}.

\section{Neutrino detectors}
\begin{figure}
\begin{center}
\includegraphics[width=\columnwidth]{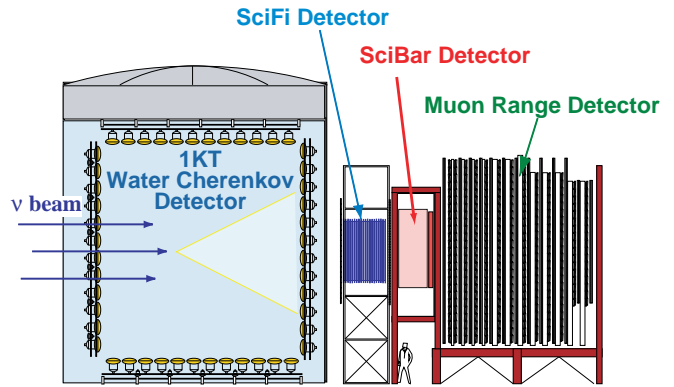} 
\end{center}
\caption{The schematic view of the near neutrino detectors for K2K-IIb period.
         In K2K-I, the Lead-Glass calorimeter was located at the position
         of the SciBar detector.}
\label{fig:ND}
\end{figure}

A near neutrino detector system (ND) is located 300~m downstream 
from the proton target.
The primary purpose of the ND is to measure the direction, flux, and the 
energy spectrum of neutrinos at KEK before they oscillate. 
The schematic view of the ND during the K2K-IIb period is shown in Fig.~\ref{fig:ND}.
The ND is comprised of two detector systems; a one
kiloton water Cherenkov detector~(1KT) and
a fine-grained detector~(FGD) system. 
The FGD consists of a scintillating-fiber/water-target tracker~(SciFi),
a Lead-Glass calorimeter~(LG) in K2K-I period,
a totally active fine-segmented scintillator tracker~(SciBar) in K2K-IIb
and K2K-IIc periods, and a muon range detector~(MRD). 
The far detector is the 50 kiloton water Cherenkov detector, Super-Kamiokande~(SK),
which is located 250~km away from KEK and 1000~m (2700~m water equivalent)
below the peak of Mt. Ikeno-yama in Gifu prefecture.

\subsection{1 kiloton water Cherenkov detector}
A one kiloton water Cherenkov detector (1KT) is located in the
experimental hall at KEK as the upstream detector.
The 1KT detector is a miniature version of SK, and uses
the same neutrino interaction target material and instrumentation.
The primary role of the 1KT detector is to measure
the $\nu_\mu$ interaction rate and the $\nu_\mu$ energy spectrum.
The 1KT detector also provides a high statistics measurement of
neutrino-water interactions.

The cylindrical tank, 10.8~m in diameter and 10.8~m in height,
holds approximately 1000\,tons of pure water.
The center of the water tank
is 294~m downstream of the pion production target.
The water tank is optically separated into
the inner detector (ID) and the outer detector (OD)
by opaque black sheets and reflective Tyvek$^{\textregistered}$
(a material manufactured by DuPont) sheets.
The ID of the 1KT detector is
a cylinder of 8.6~m in diameter and 8.6~m in height.
This volume is viewed by 680 photomultiplier tubes (PMTs)
of 50~cm diameter facing inward to detect
Cherenkov light from neutrino events.
The PMTs and their arrangement are identical to those of
SK;
70~cm spacing between PMTs gives a 40\% photocathode coverage.
The fiducial volume used for selecting neutrino events in the 1KT
is defined as a 25~ton cylindrical region with a diameter of 4~m
and a length of 2~m oriented along the beam axis.
The OD covers the upstream third of the barrel wall
and the whole of the bottom wall.
The OD volume is viewed by 68 PMTs of 20~cm diameter,
facing outward to veto the incoming particles.
The OD is also used to trigger through-going/stopping
cosmic ray muon events for detector calibrations.

To compensate for the geomagnetic field which affects the PMT response,
nine horizontal Helmholtz coils and seven vertical Helmholtz coils are
arranged surrounding the water tank.
The water purification system for the 1KT detector
circulates about 20~tons/hour of water.
The electrical resistance ($\sim$10~M$\Omega$/cm)
and water temperature ($\sim$11~$^oC$) are kept constant by the system.

The 1KT detector data acquisition (DAQ) system
is similar to that of SK.
The signal from each PMT is processed using
custom electronics modules called ATMs,
which were developed for the SK experiment
and are used to record
digitized charge and timing information for each PMT hit
over a threshold of about $1/4$ photoelectrons.
The DAQ trigger threshold is about 40 PMT hits
within a 200~nsec time window in a 1.2~$\mu$sec beam spill gate,
where the beam spill gate is issued to all near detectors,
synchronized with the beam timing, by the accelerator.
The 40 hit threshold is roughly equivalent to the signal of a 6~MeV electron.
The pulse shape of the analog sum of all 680 PMTs' signals (PMTSUM)
is also recorded for every beam spill by a 500~MHz flash analog to
digital converter (FADC) which enables us
to identify multiple interactions in a spill gate.
We determine the number of interactions in each spill
by counting the peaks in PMTSUM greater than
a threshold equivalent to a 100~MeV electron signal.

The physical parameters of an event in the 1KT detector
such as the vertex position, the number of Cherenkov rings,
particle types and momenta are determined using the same algorithms as
in SK~\cite{Ashie:2005ik}.
First, the vertex position of an event is determined from the PMT
timing information.
With knowledge of the vertex position,
the number of Cherenkov rings and their directions
are determined by a maximum-likelihood procedure.
Each ring is then classified as
$e$-like, representing a showering particle ($e^{\pm}$,\,$\gamma$), or
$\mu$-like, representing a non-showering particle ($\mu^{\pm}$,\,$\pi^{\pm}$),
using its ring pattern and Cherenkov opening angle.
On the basis of this particle type information,
the vertex position of a single-ring event is further refined.
The momentum corresponding to each ring is determined from the
Cherenkov light intensity.
Fully contained (FC) neutrino events,
which deposit all of their Cherenkov light inside the inner detector,
are selected by requiring the maximum number of photoelectrons on a single PMT
at the exit direction of the most energetic particle to be less than 200.
The events with the maximum number of photoelectrons greater than 200
are identified as a partially contained (PC) event.  This criterion is
used because a muon passing through the wall produces a lot of light in
the nearest PMTs.

The reconstruction quality, especially the vertex position 
and angular resolution, are estimated with a MC simulation.
The vertex resolution is estimated to be 14.7~cm for FC single-ring events
and 12.5~cm for PC single-ring events, 
while those for multi-ring FC and PC events are 39.2~cm and 34.2~cm,
respectively.
The angular resolution for single-ring CC-QE events is estimated to be
1.05$^\circ$ for FC events and 0.84$^\circ$ for PC events.
As for the capability of the particle identification,
0.3\% of muon neutrino CC quasi-elastic events with a single ring are
misidentified as $e$-like while 3.3\% of electron neutrino CC quasi-elastic
events with a single ring are misidentified as $\mu$-like.
The momentum resolution for muons is estimated to be 2.0-2.5\% in the
whole momentum range of the 1KT.

The gain and timing of each PMT are calibrated using
a Xe lamp and a N$_2$ laser as light sources, respectively.
The absorption and scattering coefficients of water
are measured using laser calibration,
and the coefficients in the detector simulation are further tuned
to reproduce the observed charge patterns of cosmic ray muon events.
The energy scale is calibrated and checked 
by cosmic ray muons with their decay electrons
and neutral current $\pi^0$s produced by the K2K neutrino beam.
The absolute energy scale uncertainty is $^{+3}_{-4}$\% while the
vertical/horizontal detector asymmetry of the energy scale is 1.7\%. 
The energy scale is stable within about 1\% from 2000 to 2004.

The performance of vertex reconstruction is experimentally studied
by special cosmic ray muon data
utilizing a PVC pipe with scintillating strips at each end inserted
vertically into the tank.
Cosmic ray muons going through the pipe
emulate the neutrino-induced muons whose vertex position
is defined at the bottom end of the pipe.
This study demonstrates that the vertex reconstruction works
as well as we expected from the Monte Carlo simulation.
We find a vertex bias difference between data and MC simulation of less than 
4~cm for both FC and PC events.
\label{sec:1kt}

\subsection{Scintillating Fiber detector}
The scintillating fiber (SciFi) detector is a 6 ton tracking detector with
integral water target layers.
Details of the design and performance of the detector are
described in Refs.\cite{Suzuki,BKJ}.
The SciFi detector is used to measure the neutrino spectrum,
and to reconstruct with high resolution the charged particle tracks
produced in neutrino interactions. 
It can estimate the rates
for quasi-elastic and inelastic interactions
and is sensitive to higher energy events, and hence has
complementary capabilities to the 1KT detector.
The SciFi detector has been in stable operation since 1999 when
the first K2K neutrino beam was delivered.

The SciFi detector consists of 20 layers of 2.6~m $\times$ 2.6~m
tracking modules, placed 9 cm apart.
Each layer contains a double layer of sheets of scintillating fibers
arranged, one each, in the horizontal and vertical directions;
each sheet is itself two fibers thick.
The diameter of each fiber is 0.692~mm.
In between the fiber modules, there are 19 layers of water target
contained in extruded aluminum tanks.
The water level was monitored;
it has stayed constant within 1\% throughout the experiment,
except for a few tanks which drained following an earthquake.
This monitoring, as well as measurements when the tanks were filled and 
later drained, give a fiducial mass of 
5590~kg with 1\% accuracy.  The fiducial mass fractions
are 0.700 H$_2$0, 0.218 Al, and 0.082 HC ($\pm$ 0.004).

The fiber sheets are coupled to an image intensifier tube (IIT) with a
CCD readout system.
The relative position between the fibers and the CCD coordinate system
is monitored periodically by illuminating every 10th or 20th fiber
with an electro-luminescent plate placed at the edge of each fiber sheet.
In addition, cosmic-rays were used to monitor the gain of the system
on a weekly basis.

Hit fibers are extracted using the CCD images.
The raw data consists of hit CCD pixels and their digitized brightness.
Neighboring hit pixels are grouped to make a pixel cluster.
Those clusters are then combined and matched to the location
of specific scintillating fibers.
The efficiency to identify a fiber through which a charged particle passed
is estimated using cosmic ray muons
to be about 95\%, but closer to 90\% at angles within 30 degrees of 
the beam.
After hit fibers are reconstructed, tracks with three or more 
hit layers are reconstructed using conventional fitting techniques.
The efficiency to find a track is also estimated using
cosmic ray muons, and is $\sim70\%$ for tracks with length of three 
layers,
$\sim87\%$ for four layers, and approaches 100\% for longer tracks.

Surrounding the SciFi are two plastic scintillator hodoscope systems.
One is placed downstream of SciFi and gives track timing and
position information.
It also serves as a pre-shower detector for the Lead Glass calorimeter.
The other is upstream of SciFi and is used to veto muons and
other particles from the beam, primarily from
neutrino interactions in the upstream 1KT detector, 
but also from cosmic rays.

The downstream system consists of 40 scintillator units placed
one upon another having a total height of 4~m.
Each unit is made of a plastic scintillator 466~cm long,
10.4~cm high, and 4~cm thick.
A PMT is attached to each end of the scintillator.
The horizontal position of the charged particle
can be calculated with 5~cm resolution from the timing information read 
out by the both end PMT's.
The upstream veto wall is similar,
but pairs of scintillators are joined together by optical cement and
share a single light guide for each PMT.
Thus there are fewer readout channels and the vertical resolution
is twice as coarse, but the hodoscope covers
the same total area as the one downstream.
The charge and timing information from each of the 120 total PMT's 
are recorded.
The energy deposit measured in the downstream hodoscope is used to
select electron neutrino events as described later.
The energy resolution of these hodoscopes is 
estimated using cosmic ray muons to be 7.4\%
for minimum ionizing particles.

A more detailed description of the hodoscope system 
can be found
in~\cite{Ahn2000:Veto}.


\subsection{Scintillating Bar detector}
The SciBar detector~\cite{Nitta:2004nt} was constructed as an upgrade of
the near detector system.
The purposes of the SciBar detector are to measure the neutrino energy
spectrum and to study the neutrino interaction with high detection efficiency
for low momentum particles.
The main part of the SciBar detector consists of an array of plastic
scintillator strips.
Its totally active and finely segmented design allows us to detect all
the charged particles produced in a neutrino interaction.

We use extruded scintillator strips produced by FNAL~\cite{Pla-Dalmau:2001en}.
The dimensions of a strip are 1.3~cm thick, 2.5~cm wide, and 300 cm~long.
In total, 14,848 scintillator strips are arranged in 64 layers of
alternating vertical and horizontal planes. 
The dimension of the detector is $3 \times 3 \times 1.7 ~{\rm m^3}$
providing the total weight of about 15~tons.

The scintillation light is guided to multi-anode PMTs 
by wavelength shifting fibers inserted into the holes of
scintillator strips.
Sixty-four wavelength shifting fibers are bundled together and glued to an attachment
to be precisely coupled between fibers and the photo cathode of the multi-anode PMT.
Both charge and timing of the PMT outputs are recorded using
custom-made electronics~\cite{Yoshida:2004mh}.
The noise level and the timing resolution for minimum-ionizing particle signal are about
0.3 photoelectrons  and 1.3~nsec, respectively.

The gain of all multi-anode PMT channels was measured at a test bench
prior to the installation.
In order to monitor and correct gain drift during operation,
the SciBar is equipped with a gain calibration system
using LED~\cite{Hasegawa:PhD}.
The gain stability is monitored with precision better than 1\%.
Cosmic-ray data are collected between beam spills to calibrate
the multi-anode PMT gain and scintillator light yield \textit{in-situ}.
The light yield has been stable within 1~\% during operation.
The light attenuation length of the wavelength shifting fiber is also measured with cosmic ray muons.
It is confirmed to be consistent with the test bench measurement
done prior to the installation.

An electromagnetic calorimeter (EC) is installed downstream the
tracker part of SciBar to study the amount of the electron neutrino
contamination in the beam and $\pi^0$ production in neutrino
interactions. The calorimeter is made of bars of dimensions
$262\times8\times4~{\rm cm}^3$. The bars, a sandwich of lead and
scintillating fibers, were originally built for the ``spaghetti''
calorimeter of the CHORUS neutrino experiment at
CERN~\cite{Buontempo:1994yp}. Each bar is read out by two
PMTs per side. In the SciBar-EC, 32 bars are assembled
to form a plane of vertical elements, followed by a plane of 30
horizontal bars. The two planes, each 4 cm thick, cover an area of
$270\times262~{\rm cm}^2$ and $262\times250~{\rm cm}^2$, respectively. The EC
adds eleven radiation lengths to the tracker part which has about
four radiation lengths. The response linearity of the EC is
understood to be better than 10\%. The energy resolution is about
$14\%/\sqrt{E\textrm{[GeV]}}$ as measured with a test
beam~\cite{Buontempo:1994yp}.

To reconstruct neutrino events, hit scintillator strips in SciBar 
with more than or equal
to two photoelectrons (corresponding to about 0.2~MeV) are selected.
Charged particles are reconstructed by looking for track projections
in each of two dimensional view ($x$-$z$ and $y$-$z$).
using a cellular automaton algorithm~\cite{Glazov:1993ur}.
Then, track candidates in two views are combined based on
matching of the track edges in $z$ direction and timing information.
Reconstructed tracks are required to have hits in more than or equal to
three consecutive layers.
The minimum length of reconstructible track is, therefore, 8~cm,
which is corresponding to 450~MeV/c for protons.
The reconstruction efficiency for an isolated track longer than 10~cm is 99\%.


\subsection{Muon range detector}
The muon range detector~(MRD)~\cite{Ishii:2002nj} has two purposes.
One is to monitor the stability of the neutrino beam direction, 
profile and spectrum by measuring the energy, angle and production point
of muons produced by charged-current neutrino interaction by utilizing
its huge mass of the iron as the target.
The other is to identify the muons produced in the upstream detectors
and to measure their energy and angle with combination of other
fine grain detectors.
This enables us to measure the energy of the incident neutrino.

MRD consists of 12 layers of iron absorber
sandwiched in between 13 sets of vertical and horizontal
drift-tube layers.
The size of a layer is approximately $7.6~{\rm m} \times 7.6~{\rm m}$.  
In order to have a good energy resolution for the whole energy region, 
the upstream four iron plates are 10~cm thick while
the downstream eight plates are 20~cm thick.  
The total iron thickness is 2.00~m covering the muon energy up to 2.8~GeV.
MRD has 6,632 drift tubes, each of which is made of aluminum
with a cross section of $5~{\rm cm} \times 7~{\rm cm}$.
P10 gas (${\rm Ar}:{\rm CH}_{4}=90\%:10\%$) is supplied to all the tubes.
The maximum drift time in a tube is about $1~\mu{\rm sec}$.
The drift time is digitized by 20~MHz 6-bit TDCs.
The total weight of iron is 864 tons and 
the total mass of MRD including the aluminum drift tubes is 915 tons.

A conventional track finding algorithm is employed to reconstruct tracks
from hits.
The track finding efficiency is 66\%, 95\% and 97.5\%
for tracks with one, two and three traversed iron plate(s), respectively,
and it goes up to 99\% for longer tracks.
The range of track is estimated using the path length of 
the reconstructed track in iron.

Accurate knowledge of the iron-plate weight is necessary for
the measurements of both neutrino interaction rate and track range.
Relative thickness of each plate was studied by comparing the event rate
using the neutrino beam data.
Also, the density was measured directly using a sample of the same iron.
Combining these studies, we quote the weight of the iron plates
with an accuracy of $1\%$.
The relation between the muon energy and the muon range in iron
was calculated using a GEANT based Monte Carlo code.
There is at maximum $1.7\%$ difference in the muon range among
various calculations.
We quote the error on energy scale in the range measurement
to be $2.7\%$ by linearly adding these two errors.

The energy acceptance and resolutions of the MRD were studied 
by a Monte Carlo simulation.
The acceptance is ranging from 0.3~GeV to 2.8~GeV while
the resolution is 0.12~GeV for forward-going muons.
The track angular resolution is about 5~degrees and the resolution
of the vertex point perpendicular to the beam direction is about 2~cm.

\subsection{Lead glass calorimeter}
\label{sec:LGsetup}
The Lead Glass (LG) calorimeter was located between SciFi and MRD
in K2K-I period.
The purpose of LG is to distinguish electrons from muons by measuring
the energy deposit. 
The LG calorimeter is made up of 600 cells. 
A LG cell of approximately
$12~{\rm cm} \times 12~{\rm cm} \times 34~{\rm cm}$ is viewed
by 3~inch-in-diameter PMT(Hamamatsu, R1652) through a light guide cylinder
made also by lead glass.
This LG calorimeter was once used in the TOPAZ
experiment~\cite{Kawabata:1987fj} and reused for the K2K experiment.

The LG detector system reads out only the charge information for each cell.
The absolute energy scale of 9 standard LG cells out of 600 were
calibrated prior to installation by using an electron beam
from the electron synchrotron with the energy range from 50 MeV to 1.1 GeV.
The resolution was estimated by this pre-calibration to be 10\% at 1~GeV.
Position dependence for the energy resolution were also measured to be 4\%.
The other LG cells were relatively calibrated to the standard cells
by cosmic-ray muons.

Responses for muons were also calibrated by using cosmic-ray muons at KEK
prior to installation.
The relative peak pulse height for PMTs was adjusted to each other
within 2\%.
The responses for charged pions were checked at different momenta
(0.3$-$2.0~GeV/$c$) by using the KEK test beam, confirmed to be
in good agreement with the expectation by an MC simulation.

\subsection{Super-Kamiokande}
The far detector of the K2K experiment is Super-Kamiokande, which is located
in the Kamioka Observatory, operated by the 
Institute for Cosmic Ray Research, University of Tokyo.
The SK detector is a cylindrically shaped water Cherenkov detector 
which is 41~m
in height, 39~m in diameter and has a total mass of 50~kilotons of water.
The water tank is optically separated into a cylindrically-shaped
inner detector~(ID)
and outer detector~(OD) by opaque black sheets and Tyvek$^{\textregistered}$ sheets attached
to a supporting structure.
The ID is viewed by 11,146 20-inch PMTs facing inward covering
40\% of the ID surface from June 1999 to 2001 (called SK-I and K2K-I),
while it is viewed by 5,182 PMTs enclosed in a fiber reinforced plastic and
sealed with acrylic covers on their front surface, covering 19\%
of the ID surface from December 2002 (SK-II and K2K-II).
The transparency and the reflection of these covers in water
are 97\% and 1\%, respectively.
In the OD region, outward-facing 1,885 8-inch PMTs are attached
to the outer side of the supporting structure.
The performance of OD PMTs is improved in SK-II.
The fiducial volume is defined to be a cylinder whose surface is 2~m away
from the ID wall providing a fiducial mass of 22.5~kilotons.
Details of the detector performance and systematic uncertainties in SK-I
are written in~\cite{Fukuda:2002uc,Ashie:2005ik}.
For SK-II, these quantities are estimated using similar methods 
as used in SK-I.
Momentum resolution for SK-II is slightly worse than SK-I;
2.4\% and 3.6\% for 1~GeV/$c$ muons in SK-I and SK-II, respectively.
This is because the number of ID PMTs in SK-II is about a half of SK-I.
However, the performance of the vertex reconstruction, the ring counting,
and the particle identification in SK-II are almost the same as in SK-I.
The purity of the QE interaction in 1-ring $\mu$-like events is 58\%.
The uncertainty in the energy scale is estimated to be 2.0\% for SK-I
and 2.1\% for SK-II.

In this long baseline experiment, timing information is used to distinguish
between beam neutrino events and cosmic ray induced background events
in the SK detector.  The GPS is used to 
synchronize the timing of the beam spill between KEK and SK.  At both sites
are a free running 50 MHz (32-bit) local time counter connected to a GPS
receiver and an event trigger (at Super-K) or the beam spill trigger (at KEK).
At first, a quartz oscillator was used with good results,
and later oscillator drift was improved further with a rubidium clock.
This counter is synchronized using the one pulse-per-second signal from 
the GPS.  In this way, events can be synchronized within approximately
50 ns, after compensating for oscillator drift.  This is confirmed by 
comparing a second, 
independent timing system at each site which gives the same
result as the primary system within 35 ns 99\% of the time.  As described
later in this paper, this accuracy is sufficient to observe the neutrino 
beam's bunch structure in the SK neutrino data.  The system is described
more completely in \cite{Berns:2000}.

\section{Neutrino interaction simulation}
The neutrino interaction simulation plays an important role both in
estimating the expected number of neutrino interactions and in
deriving the energy spectrum of neutrinos from the data.  The Monte
Carlo program simulates neutrino interactions with protons, oxygen,
carbon and iron, which are the target materials of the neutrino
detectors.

In the simulation program, we include the following charged and
neutral current neutrino interactions:
quasi-elastic scattering ($\nu\ N \rightarrow \ell\ N'$),
single meson production ($\nu\ N \rightarrow \ell\ N'\ m$),
coherent $\pi$ production
($\nu\ ^{16}{\rm O}(^{12}{\rm C},^{56}{\rm Fe}) \ 
\rightarrow\ \ell\ \pi\ ^{16}{\rm O}(^{12}{\rm C},^{56}{\rm Fe})$),
and deep inelastic scattering ($\nu N\ \rightarrow \ell\ N'\ hadrons$).
In these reactions, $N$ and $N'$ are
the nucleons (proton or neutron), $\ell$ is the lepton, and $m$ is the
meson. For the single meson production processes, the $K$ and $\eta$
are simulated as well as the dominant $\pi$ production processes.  If
the neutrino interaction occurs in oxygen or other nuclei,
the re-interactions of the resulting particles with the remaining
nucleons in the nucleus are also simulated.

\subsection{Quasi-elastic scattering}
The formalism of quasi-elastic scattering off a free neutron used in
the simulation programs is described by
Llewellyn-Smith~\cite{LlewellynSmith:1972zm}.
For scattering off nucleons in the nucleus, we use the relativistic
Fermi gas model of Smith and Moniz~\cite{Smith:1972xh}.
The nucleons
are treated as quasi-free particles and the Fermi motion of nucleons
along with the Pauli exclusion principle is taken into account.  The
momentum distribution of the target nucleon is assumed to be flat up
to a fixed Fermi surface momentum of 225~MeV/$c$ for carbon and
oxygen and 250MeV/$c$ for iron.
  The same Fermi momentum
distribution is also used for all of the other nuclear interactions.
The nuclear potential is set to 27~MeV for carbon and oxygen and
32~MeV for iron.

\subsection{Single meson production}

Rein and Sehgal's model is used to simulate the resonance production
of single $\pi$, $K$ and $\eta$~\cite{Rein:1981wg,Rein:1983pf,Rein:1987cb}.
This model divides the interaction into two parts.  First there is the
interaction
\begin{eqnarray*}
  \nu + N \rightarrow \ell + N^*,
\end{eqnarray*}
which is then followed by
\begin{eqnarray*}
  N^* \rightarrow \pi (\rm{ \ or\ } K \rm{ \ or\ } \eta) + N',
\end{eqnarray*}

\noindent
where $N$ and $N'$ are the nucleons, and $N^*$ is the baryon
resonance like $\Delta(1232)$.  The mass of the intermediate 
resonance is restricted to be
less than 2~GeV/$c^2$.  To determine the direction of the pion in the
final state, we also use Rein and Sehgal's method for the dominant
resonance $P_{33}$(1232). For the other resonances, the directional
distribution of the generated pion is set to be isotropic in the
resonance rest frame.  The angular distribution of $\pi^+$ has been
measured for the $\nu p \rightarrow \mu^- p \pi^+$
mode~\cite{Kitagaki:1986ct} and the results agree well with the Monte
Carlo prediction.  The Pauli blocking effect in the decay of the
baryon resonance is taken into account by requiring that the momentum
of the nucleon should be larger than the Fermi surface momentum.  
In addition, the delta may be absorbed by the nucleus.  
For these events there is no pion in the final state, and only a lepton 
and nucleon are emitted~\cite{Singh:1998ha}. 
We explicitly make this happen for 20\% of the deltas produced.
Single $K$ and $\eta$ productions are
simulated using the same framework as for single $\pi$ production
processes.

Both the quasi-elastic and single-meson production models contain a
phenomenological parameter (the axial vector mass, $M_{A}$), that must
be determined by experiment.
As the value of $M_{A}$ increases, interactions with higher $Q^2$
values (and therefore larger scattering angles) are enhanced.
The $M_{A}$ parameters in our Monte Carlo simulation program are set
to be 1.1~GeV for both the quasi-elastic and single-meson production
channels based on the analysis of the near detector
data ~\cite{Ahn:2002up}.

Coherent single $\pi$ production, the interaction between a neutrino
and the entire  nucleus, is simulated using the formalism developed
by Rein and Sehgal \cite{Rein:1983pf}. Here, only the neutral current
interactions are considered because the cross-section of the charged
current coherent pion production was found to be very small at the K2K
beam energy~\cite{Hasegawa:2005td}.

\subsection{Deep inelastic scattering}
In order to calculate the cross-section for deep inelastic scattering,
we use the GRV94 parton distribution functions\cite{Gluck:1994uf}.
Additionally, we have included the corrections in the small $q^2$
region developed by Bodek and Yang~\cite{Bodek:2002vp}. In the
calculation, the hadronic invariant mass, $W$, is required to be
larger than 1.3~GeV/$c^2$.
Also, the multiplicity of pions is restricted to be larger than or
equal to two for $1.3<W<2.0~{\rm GeV}/c^2$, because single pion
production is already taken into account as previously described.  In
order to generate events with multi-hadron final states, two models
are used. For $W$ between 1.3 and 2.0~GeV/$c^2$, a custom-made
program~\cite{Nakahata:1986zp} is employed
while PYTHIA/JETSET~\cite{Sjostrand:1994yb} is used for the events whose
$W$ is larger than 2~GeV/$c^2$.

\begin{figure}
\begin{center}
\includegraphics[width=7.2cm]{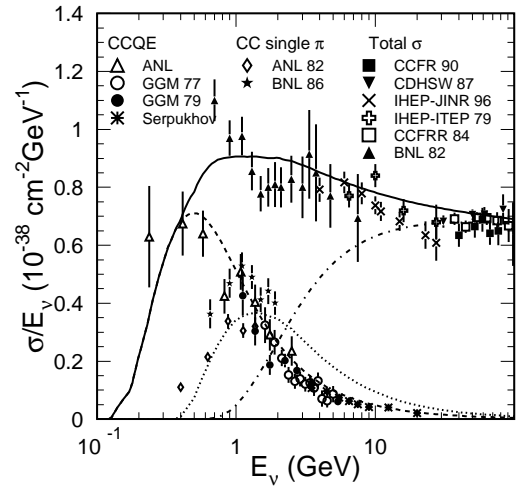}
\caption{ Charged current total cross section divided by $E_\nu$ for
  neutrino nucleon charged current interactions. The solid line shows
  the calculated total cross section. The dashed, dot and dash-dotted
  lines show the calculated quasi-elastic, single-meson and
  deep-inelastic scatterings, respectively.  The data points are
  taken from the following experiments:
  \mbox{({$\triangle$})ANL\protect\cite{Barish:1977qk}},
  \mbox{({$\bigcirc$})GGM77\protect\cite{Bonetti:1977cs}},
  \mbox{({$\bullet$})GGM79(a)\protect\cite{Ciampolillo:1979wp},(b)\protect\cite{Armenise:1979zg}},
  \mbox{({$\ast$})Serpukhov\protect\cite{Belikov:1985kg}},
  \mbox{({$\Diamond$})ANL82\protect\cite{Radecky:1982fn}},
  \mbox{({$\star$})BNL86\protect\cite{Kitagaki:1986ct}},
  \mbox{({$\blacksquare$})CCFR90\protect\cite{Auchincloss:1990tu}},
  \mbox{({$\blacktriangledown$})CDHSW87\protect\cite{Berge:1987zw}},
  \mbox{({$\times $})IHEP-JINR96\protect\cite{Anikeev:1996dj}},
  \mbox{({$+$})IHEP-ITEP79\protect\cite{Mukhin:1979bd}},
  \mbox{({$\Box$})CCFRR84\protect\cite{MacFarlane:1984ax}}, and
  \mbox{({$\blacktriangle$})BNL82\protect\cite{Baker:1982ty}}.  }
\label{plot_tot}
\end{center}
\end{figure}

The total charged current cross sections including quasi-elastic
scattering, single meson production and deep inelastic scattering are
shown in Fig.~\ref{plot_tot} overlaid with data from several
experiments.

\subsection{Nuclear effects}
The intra-nuclear interactions of the mesons and nucleons produced in
neutrino interactions in the carbon, oxygen or iron nuclei are also
important to consider for this analysis.  Any absorption or change of
kinematics of these particles will affect the event type
classification.  Therefore, the interactions of $\pi$, $K$, $\eta$ and
nucleons are also simulated in our program.  These interactions are
treated using a cascade model, and each of the particles is traced
until it escapes from the nucleus.

Among all the interactions of mesons and nucleons, the interactions of
pions are most important, since both the cross sections for pion
production for neutrino energies above 1 GeV and also the interaction
cross sections of pions in the nucleus are large.  In our simulation
program, the following pion interactions in nucleus are considered:
inelastic scattering, charge exchange and absorption. The actual
procedure to simulate these interactions is as follows: first the
generated position of the pion in nucleus is set according to the
Woods-Saxon nucleon density distribution~\cite{Woods:1954}. Then, the
interaction mode is determined by using the calculated mean free path
of each interaction.  To calculate these mean free paths, we adopt the
model described by Salcedo et al.~\cite{Salcedo:1988md}. The
calculated mean free paths depend not only on the momentum of the pion
but also on the position of pion in the nucleus.

If inelastic scattering or charge exchange occurs, the direction and
momentum of pion are determined by using the results of a phase shift
analysis obtained from $\pi-N$ scattering
experiments~\cite{Rowe:1978fb}.  When calculating the pion scattering
amplitude, the Pauli blocking effect is also taken into account by
requiring the nucleon momentum after the interaction to be larger than
the Fermi surface momentum at the interaction point.

\begin{figure}
\includegraphics[width=7cm]{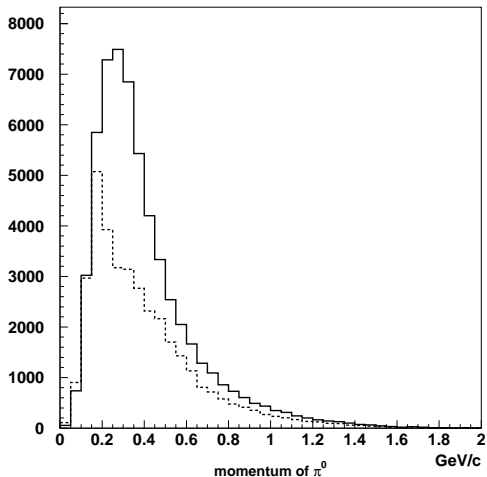}
\caption{ The $\pi^0$ momentum distributions
for neutral current single $\pi$ production processes off the 
water target and for the K2K neutrino beam at the near detector.
The solid and dashed lines show the spectrum without and with the
pion nuclear effects.
}
\label{fig:pi0-momentum}
\end{figure}

This pion interaction simulation is tested by comparison with data
using the following three interactions: $\pi ^{12}$C scattering, $\pi
^{16}$O scattering and pion photo-production ($\gamma + ^{12}$C
$\rightarrow \pi^- + X$).  The importance of including the proper
treatment of nuclear effects is illustrated in
Fig.~\ref{fig:pi0-momentum} which shows the momentum distribution for
neutral current single $\pi^0$ production in the water target both
with and without having them applied.

The re-interactions of the recoil protons and neutrons
produced in the neutrino interactions are also important,
because the proton tracks are used to select quasi-elastic
like events. This is done with the SciFi and SciBar
near detectors, and allows us to estimate the neutrino energy.
Nucleon-nucleon interactions modify the outgoing nucleon's
momentum and direction, which also affects whether the
nucleon will be above detection threshold~\cite{Walter:2002sa}. 
Both elastic scattering and pion
production are considered. In order to simulate these interactions, a
cascade model is again used and the generated particles in the nucleus
are tracked using the same code as for the mesons.

\section{The far/near flux ratio}\label{sec:farnear}
\subsection{Definition of the far-to-near ratio}
The effects of neutrino oscillation appear as a reduction
in the number of neutrino events and a distortion of the neutrino
energy spectrum in SK.
The observations for these quantities are compared to their expectations
in SK to study neutrino oscillation.
The ND measures the neutrino flux and spectrum before neutrinos oscillate.
Those measurements are then extrapolated by the expected ratio of
muon neutrino fluxes at the far and near detector locations,
the far-to-near (\fn) flux ratio, to predict the number of neutrino
events and energy spectrum in SK.

The neutrino flux at any distance from its source can be predicted
when the geometry of the decay volume and the momenta and directions
of the pion parents of neutrinos are provided.
Due to the finite size of the decay volume and the detectors,
the neutrino flux does not simply obey an $L^{-2}$ rule
(where $L$ is distance from the neutrino source);
rather the flux ratio between far and near detectors has some dependence
on neutrino energy.
Therefore, we define the \fn flux ratio, $\rfn$, as
\begin{eqnarray}
\rfn = \frac{\Phi^{\rm SK}(E_\nu)}{\Phi^{\rm ND}(E_\nu)}\, ,
\end{eqnarray}
where $\Phi^{\rm SK(ND)}(E_\nu)$ is the neutrino energy spectrum at SK (ND).

The \fn flux ratio is estimated by our beam MC simulation.
In this simulation, while we use the Cho-CERN compilation as a reference model,
we employ the HARP experiment~\cite{Catanesi:2005rc} result
as an input for simulation of pion production.
The pion production measurement done by HARP is of direct relevance
for K2K, since it uses the same beam
proton momentum and the same production target, and it covers a large
fraction of the 
phase space contributing to the K2K neutrino flux.
The details of the HARP measurements are described
in Sec.~\ref{sec:farnear:harp}.
The pion monitor~(PIMON) measurement is performed for a confirmation
of the validity of the beam MC simulation.
It gives us \textit{in-situ} information on the momentum and the direction
of pions entering the decay volume after they are focused by the horn
magnetic fields although the PIMON is not sensitive to pions below 2~GeV/$c$
(corresponding to neutrinos below 1~GeV) due to its threshold.
A description of the PIMON measurement is given
in Sec.~\ref{sec:farnear:pimon}.

\subsection{Prediction of far-to-near ratio from the HARP result}
\label{sec:farnear:harp}
The dominant uncertainty in neutrino flux predictions
for conventional neutrino beams is due to the
pion production uncertainty in the hadronic interactions of primary
beam protons with the nuclear target material.
In this analysis, we use the results provided by the HARP experiment
at CERN as input to the pion production simulation. The HARP experiment
precisely measured the positively-charged pion production in the interactions
of 12.9~GeV/$c$ protons in a thin aluminum target~\cite{Catanesi:2005rc}. 

The HARP experiment took data in 2001 and 2002 in the CERN PS T9 beamline,
in order to study in a systematic and accurate way hadron production for
a variety of produced hadrons (pions and kaons in particular) with
large phase space coverage.  Data were 
taken as a function of incident beam particle type
(protons, pions), beam momentum (from 1.5 to 15~GeV/$c$),
nuclear target material (from hydrogen to lead), and nuclear target
thickness (from 2\% to more than 100\% hadronic interaction length fraction).
Secondary tracks are efficiently reconstructed in the HARP forward
spectrometer via a set of drift chambers located upstream and downstream with
respect to a dipole magnet. Particle identification for forward tracks is
obtained with a time-of-flight system, a Cherenkov threshold detector,
and an electromagnetic calorimeter.

In particular, the recent HARP pion
production measurement~\cite{Catanesi:2005rc} is directly relevant 
for the K2K \fn flux ratio because it is obtained for the same proton
beam momentum (12.9~GeV/$c$) and nuclear target
material (aluminum) as those used to produce the K2K neutrino beam.
Moreover, beam MC simulations show that the forward pion
production region measured in HARP,
$30<\theta_{\pi}<210$~mrad, $0.75<p_{\pi}<6.5$~GeV/$c$,
matches well the pion production phase space responsible
for the dominant fraction of the K2K muon neutrino fluxes at
both the near and far detector locations.

\begin{figure}
\begin{center}
\includegraphics[width=\columnwidth]{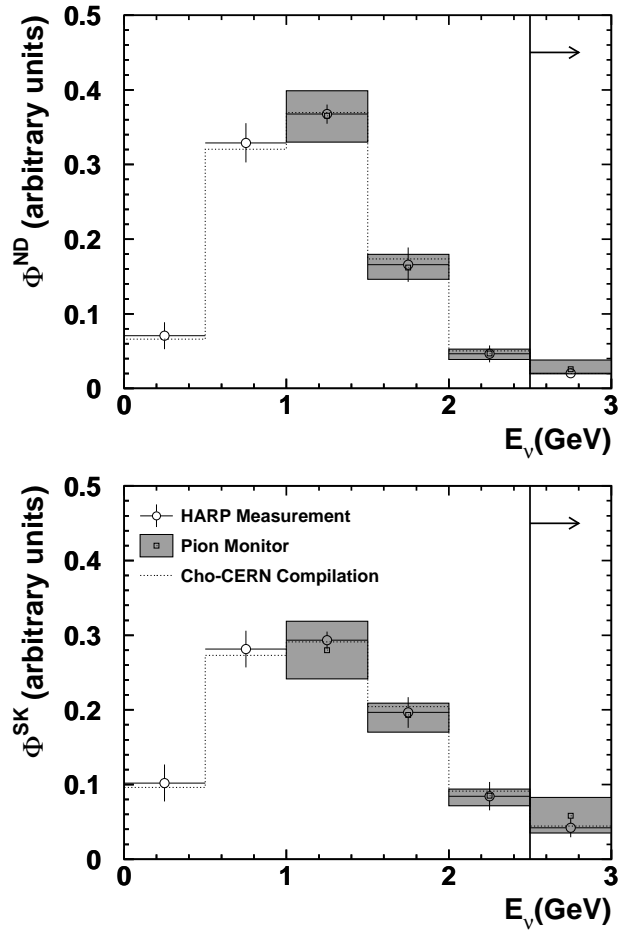}
\caption{\label{fig:harprelfluxes}
{Relatively-normalized muon neutrino flux predictions at the
 near (top) and far (bottom) detectors. The empty circles with error bars show
 the central values and shape-only errors based on the HARP $\pi^+$ production
 measurement, the empty squares
 with shaded error boxes show the central values and errors from the pion
 monitor (PIMON) measurement, and the dotted histograms show the central values
 from the Cho-CERN compilation of older (non-HARP) $\pi^+$ production data.
 The PIMON predictions
 are normalized such that the integrated fluxes above 1~GeV neutrino energy
 match the HARP ones, at both the near and far detectors.}
}
\end{center}
\end{figure}

The result of the pion production measurements described in
\cite{Catanesi:2005rc} is incorporated into our beam MC simulation
to estimate the neutrino spectra at ND and SK and the energy dependence
of the \fn flux ratio in the absence of neutrino oscillations.
The relatively-normalized fluxes at ND and SK,
$\Phi^{\hbox{\small ND}}$ and $\Phi^{\hbox{\small SK}}$, respectively,
predicted by HARP measurement, are shown in Fig.~\ref{fig:harprelfluxes},
together with the associated total systematic uncertainties,
by the empty circles with error bars.
Uncertainties in the primary and secondary hadronic interactions,
in the pion focusing performance in the horn magnetic fields, and in
the primary beam optics, are considered.
Here, primary hadronic interactions are defined as hadronic interactions
of protons with more than 10~GeV total energy in aluminum,
while secondary hadronic interactions are defined to be 
hadronic interactions that
are not primary ones.
In the following,  the assumptions on systematic uncertainties
affecting neutrino flux predictions are summarized.

The uncertainty in the multiplicity and kinematics of $\pi^+$ production
in primary hadronic interactions is estimated based on the accurate HARP
results. In this case, the HARP $\pi^+$ Sanford-Wang parameters' uncertainties
and
correlations given in \cite{Catanesi:2005rc} are propagated into flux
uncertainties using standard error matrix propagation methods:
the flux variation in each energy bin is estimated by varying a given
Sanford-Wang parameter by a unit standard deviation in the beam MC simulation.
An uncertainty of about 30\% is assumed for the uncertainty in
the proton-aluminum hadronic interaction length.
The uncertainty in the overall charged and neutral kaon production
normalization is assumed to be 50\%.

The systematic uncertainty due to our imperfect knowledge of
secondary hadronic interactions, such as $\pi^+$ absorption in the
target and horns, is also considered.
We take the relatively large differences between the
GCALOR/GFLUKA~\cite{Gabriel:1977hf,Zeitnitz:1994bs,Fasso:1993kr} and
GHEISHA~\cite{Fesefeldt:1985yw} descriptions of secondary interactions,
also in comparison to available experimental data, to estimate this
uncertainty.

We account for the uncertainties in our knowledge
of the magnetic field in the horn system.
We assume a 10\% uncertainty in the absolute field strength,
which is within the experimental uncertainty on the magnetic
field strength and the horn current measured using inductive coils
during horn testing phase~\cite{Kohama:master}.
Furthermore, a periodic perturbation in azimuth of up to $\pm15\%$ amplitude
with respect to the nominal field strength is assumed as the uncertainty in
the field homogeneity, which is also based on the experimental accuracy
achieved in the measurement of the magnetic field mapping in azimuth during
horn testing~\cite{Maruyama:PhD}.

Finally, beam optics uncertainties are estimated based on measurements
taken with
two segmented plate ionization chambers (SPICs) located upstream of the
target.
An uncertainty of 1.2~mm and 2.0~mrad in the mean transverse impact point
on target and in the mean injection angle, respectively, are assumed
based on long-term beam stability studies~\cite{Inagaki:PhD}.
The uncertainty on the beam profile width at the target and angular divergence
is also estimated, based on the $\sim$20\% accuracy with which the beam profile
widths are measured at the SPIC detector locations~\cite{Inagaki:PhD}.

\begin{figure}
\begin{center}
\includegraphics[width=\columnwidth]{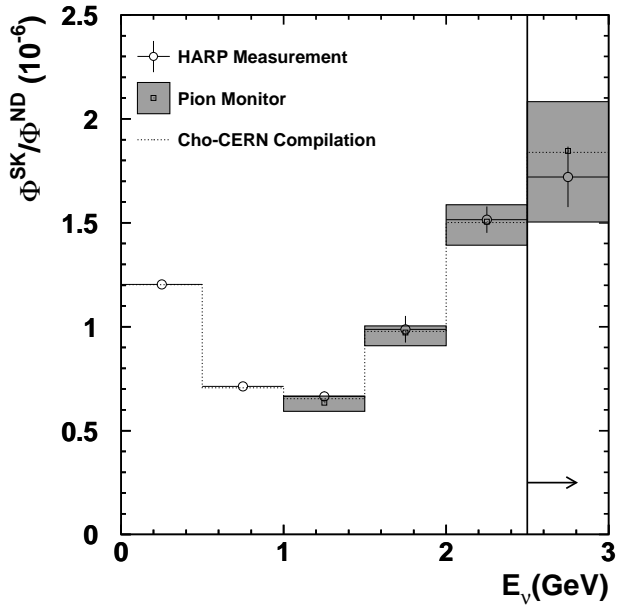}
\caption{\label{fig:harpfluxratio}
{Prediction for the K2K muon neutrino \fn flux ratio in absence
 of oscillations. The empty circles with error bars show
 the central values and systematic errors on the muon neutrino flux
 predictions from the
 HARP $\pi^+$ production measurement discussed in the text, the empty squares
 with shaded error boxes show the central values and errors from the pion
 monitor measurement, and the dotted histograms show the central values
 from the Cho-CERN compilation of older (non-HARP) $\pi^+$ production data.}
}
\end{center}
\end{figure}

The \fn flux ratio, $\Phi^{\hbox{\small SK}}/\Phi^{\hbox{\small ND}}$,
predicted by the HARP $\pi^+$ production measurement for primary hadronic
interactions with the systematic error evaluation discussed above,
in the absence of neutrino oscillations, is shown in
Fig.~\ref{fig:harpfluxratio} as a function of neutrino energy.
We estimate that the flux ratio uncertainty as a function of the neutrino
energy binning used in this analysis is at the 2-3\% level below 1~GeV
neutrino energy, while it is of the order of 4-9\% above 1~GeV.
We find that the dominant contribution to the uncertainty
in \fn comes from the HARP $\pi^+$ measurement itself.
In particular, the uncertainty in the flux ratio prediction
integrated over all neutrino energies is 2.0\%,
where the contribution of the HARP $\pi^+$ production uncertainty is 1.4\%.
Table \ref{tab:fn_errorcontributions} shows the contributions of all
systematic uncertainty sources discussed above on the far-to-near flux ratio
prediction for each neutrino energy bin.

\begin{table*}
\begin{center}
\caption{\label{tab:fn_errorcontributions}
Contributions to the uncertainty in the far-to-near flux ratio
prediction. The uncertainties are quoted in \%. The six columns refer
to different bins in neutrino energy, as shown in the table in units of GeV.}
\begin{tabular}{llrrrrrr} \hline\hline
\multicolumn{2}{l}{Source}
 & 0.0$-$0.5 & 0.5$-$1.0 & 1.0$-$1.5 & 1.5$-$2.0 & 2.0$-$2.5 & 2.5$-$ \\
\hline
\multicolumn{2}{l}{Hadron interactions} & & & & & & \\
$\hspace{5mm}$ & Primary interaction rate & 0.3 & 0.9 & 0.9 & 2.1 & 0.2 & 0.3 \\
$\hspace{5mm}$ & $\pi^+$ mult. and kinematics & 0.7 & 2.0 & 1.8 & 2.1 & 2.9 & 4.7 \\
$\hspace{5mm}$ & Kaon multiplicity & 0.1 & $<$0.1 & 0.1 & $<$0.1 & 0.1 & 4.9  \\
$\hspace{5mm}$ & Secondary interactions & 0.3 & 1.2 & 2.0 & 2.1 & 0.4 & 0.7 \\
\hline
\multicolumn{2}{l}{Horn magnetic field} & & & & & & \\
$\hspace{5mm}$ & Field strength & 1.1 & 0.8 & 1.4 & 4.2 & 2.8 & 3.9 \\
$\hspace{5mm}$ & Field homogeneity & 0.3 & 0.2 & 0.5 & 0.3 & 0.6 & 0.3 \\
\hline
\multicolumn{2}{l}{Primary beam optics} & & & & & & \\
$\hspace{5mm}$ & Beam centering & 0.1 & $<$0.1 & $<$0.1 & $<$0.1 & 0.1 & 0.1 \\
$\hspace{5mm}$ & Beam aiming & 0.1 & $<$0.1 & $<$0.1 & 0.1 & 0.4 & 0.2 \\
$\hspace{5mm}$ & Beam spread & 0.1 & 0.7 & 1.7 & 3.4 & 1.0 & 3.2 \\
\hline
\multicolumn{2}{l}{Total} & 1.4 & 2.7 & 3.6 & 6.5 & 4.2 & 8.5 \\
\hline\hline
\end{tabular}
\end{center}
\end{table*}

The dotted histograms in Figures~\ref{fig:harprelfluxes} and
\ref{fig:harpfluxratio} show the central value predicted by using
the ``Cho-CERN'' compilation for primary hadronic interactions,
which was used in K2K prior to the availability of HARP data.
In this case, the same Sanford-Wang functional form of $\pi^+$
production
is employed to describe a CERN compilation of $\pi^+$ production
measurements in proton-beryllium interactions, which is mostly
based on Cho {\it et al.} data~\cite{Cho:1972jq}.
A nuclear correction to account for the different pion production kinematics
in different nuclear target materials is applied.
The details of the Cho-CERN compilation are described in
Sec.~\ref{sec:beam:simulation}.  We find that the predictions of \fn
flux ratio by HARP and Cho-CERN are consistent with each other for
all neutrino energies.
Note that the difference between Cho-CERN and HARP central
values represents a difference in hadron production treatment only.

\subsection{Confirmation of far-to-near ratio by pion monitor measurement}
\label{sec:farnear:pimon}
A confirmation for the validity of the \fn ratio has been
performed by \textit{in-situ} pion monitor (PIMON) measurements.
The PIMON was installed on two occasions just downstream
the horn magnets to measure the momentum~($p_{\pi}$) versus
angle~($\theta_{\pi}$) 2-dimensional distribution of pions entering
the decay volume.
The PIMON measurements were done twice:
once measurement was done in June 1999 for the configuration of Ia period
(200~kA horn current with 2~cm target diameter)
and the other was done 
in November 1999 for the configuration of the other periods
(250~kA horn current with 3~cm target diameter).

\begin{figure}
  \begin{center}
    \includegraphics[width=\columnwidth]{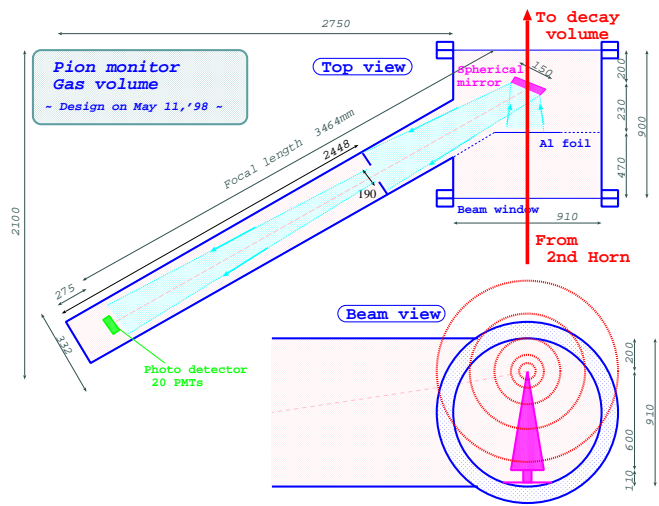}
    \caption{A schematic view of the pion monitor (PIMON).  The PIMON
      consists of a gas vessel, a spherical mirror, and an array of 20
      photomultiplier tubes.  The gas vessel is filled with freon gas
      R-318 (${\rm C}_4{\rm F}_8$).  A wedge-shaped spherical mirror
      is set inside the gas vessel and Cherenkov light produced by the
      pions in the beam, represented by the concentric circles in the
      figure, is reflected by the mirror and directed to the array of
      photo-multiplier tubes which is set in the focal plane.  }
    \label{fig:farnear:PIMON-schematicView}
  \end{center}
\end{figure}
A schematic view of PIMON is shown in
Fig.~\ref{fig:farnear:PIMON-schematicView}.
PIMON is a gas Cherenkov imaging detector which consists of a gas vessel,
a spherical mirror, and an array of 20 photomultiplier tubes.
The Cherenkov photons emitted by pions passing through the gas
vessel are reflected toward and focused onto the PMT array
by the spherical mirror.
Then, the PMT array on the focal plane detects the Cherenkov image.
Due to the characteristics of the spherical mirror,
photons propagating in the same direction are focused to the same position
on the focal plane, giving us information on the direction of the pions.
The pion momentum is also obtained from the size of the Cherenkov ring.
Furthermore, a momentum scan can be done by varying the refractive
index of the inner gas.
Therefore, the momentum and direction of pions can be measured
separately by looking at the Cherenkov light distribution on the focal plane.

As shown in Fig.~\ref{fig:farnear:PIMON-schematicView},
a wedge-shaped mirror is used as the spherical mirror to measure only 1/30 
of the beam assuming azimuthal symmetry of the distribution.
Its top is aligned to be on the beam center.
The reflection angle with respect to beam direction is 30$^{\circ}$.

An array of 20 PMTs (modified R5600-01Q made by Hamamatsu Corporation) is
set 3~m away from the beam center to avoid excess exposure
to radiation.
The size of the PMT outer socket is 15.5~mm in diameter and
the sensitive area of the photocathode is 8~mm in diameter.
They are arranged vertically at 35~mm intervals.
The array can be moved by a half pitch of the interval along the array,
and hence 40 data points (one point for every 1.75~cm)
are taken for a Cherenkov light distribution.
The relative gain among 20 PMTs was calibrated using Xe lamp before
the measurements.
The gain ratio between neighboring PMTs was also checked using Cherenkov
photons during the run.
The error on the relative gain calibration is estimated to be 10\% for
the June 1999 run and 5\% for the November 1999 run.
Saturation of the PMTs was observed in the June 1999 run,
which was corrected by a second order polynomial function.
The uncertainty due to
this correction was estimated to be 4\%~\cite{Maruyama:PhD}.

The gas vessel is filled with freon gas R-318 (${\rm C}_4{\rm F}_8$).
Its refractive index $n$ is varied by changing the gas pressure using
the external gas system.
The data are taken at several refractive indices ranging between
$n=1.00024\!-\!1.00242$ to make PIMON sensitive to different pion momenta.
The refractive index was not adjusted beyond $n=1.00242$ since the
primary protons also emit Cherenkov photons when $n$ exceeds this value,
and become a severe background to the pion measurement.
This corresponds to setting a momentum threshold of 2~GeV/$c$ for pions,
which corresponds to an energy threshold of 1~GeV for neutrinos.
The absolute refractive index is calibrated by the Cherenkov photon
distribution from 12~GeV primary protons with the refractive index set at
$n=1.00294$.

\begin{figure}
  \begin{center}
    \includegraphics[width=\columnwidth]{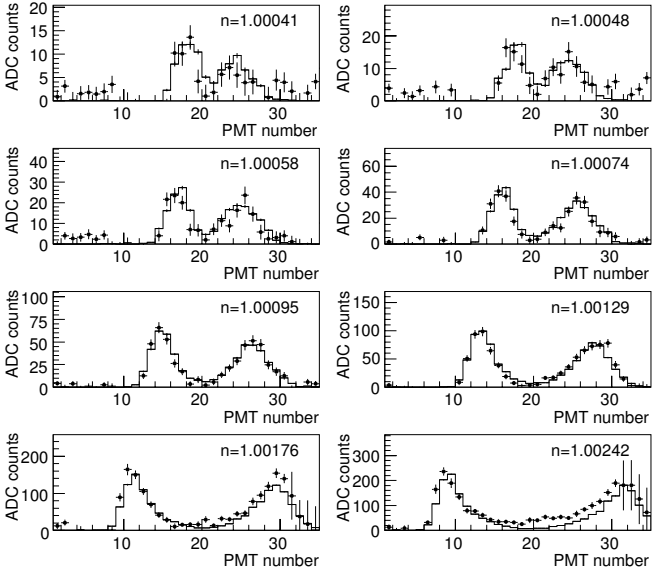}
    \caption{Cherenkov light distributions for various refractive indices
      measured in November 1999.
      Dots show data and the histograms show the MC simulation.
      The refractive indices for each plot are as written in the figure.
    }
    \label{fig:farnear:PIMON-dataChoCERN-Nov}
  \end{center}
\end{figure}

The Cherenkov light distribution for each refractive index
is taken by the PMT array.
For the background subtraction, a measurement with the mirror directed away
from the direction of PMT array was performed.
There is still non-negligible background from electromagnetic showers
which mainly come from the decay of neutral pions, $\pi^0\to2\gamma$.
The light distribution for this background is estimated using a MC simulation.
The normalization in the subtraction is done by using the distribution
measured at the lowest refractive index,
where the contribution from the electromagnetic components is dominant.
After all backgrounds are subtracted,
the distribution of the Cherenkov light emitted from pions is obtained as
shown in Fig.~\ref{fig:farnear:PIMON-dataChoCERN-Nov}.
The prediction of the MC simulation is superimposed as well.

\begin{figure}
  \begin{center}
    \includegraphics[width=0.485\columnwidth]{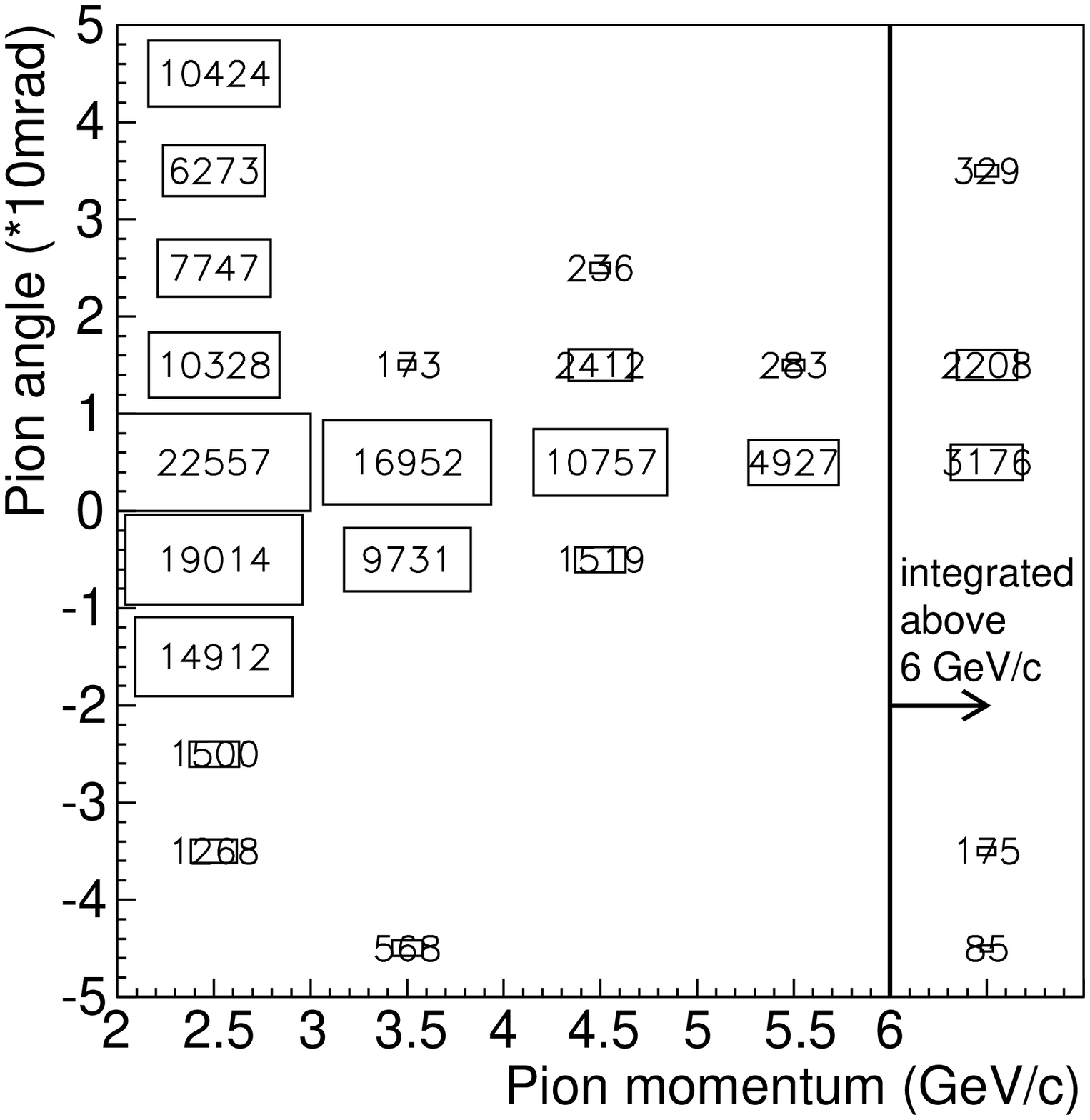}
    \hspace{\fill}
    \includegraphics[width=0.485\columnwidth]{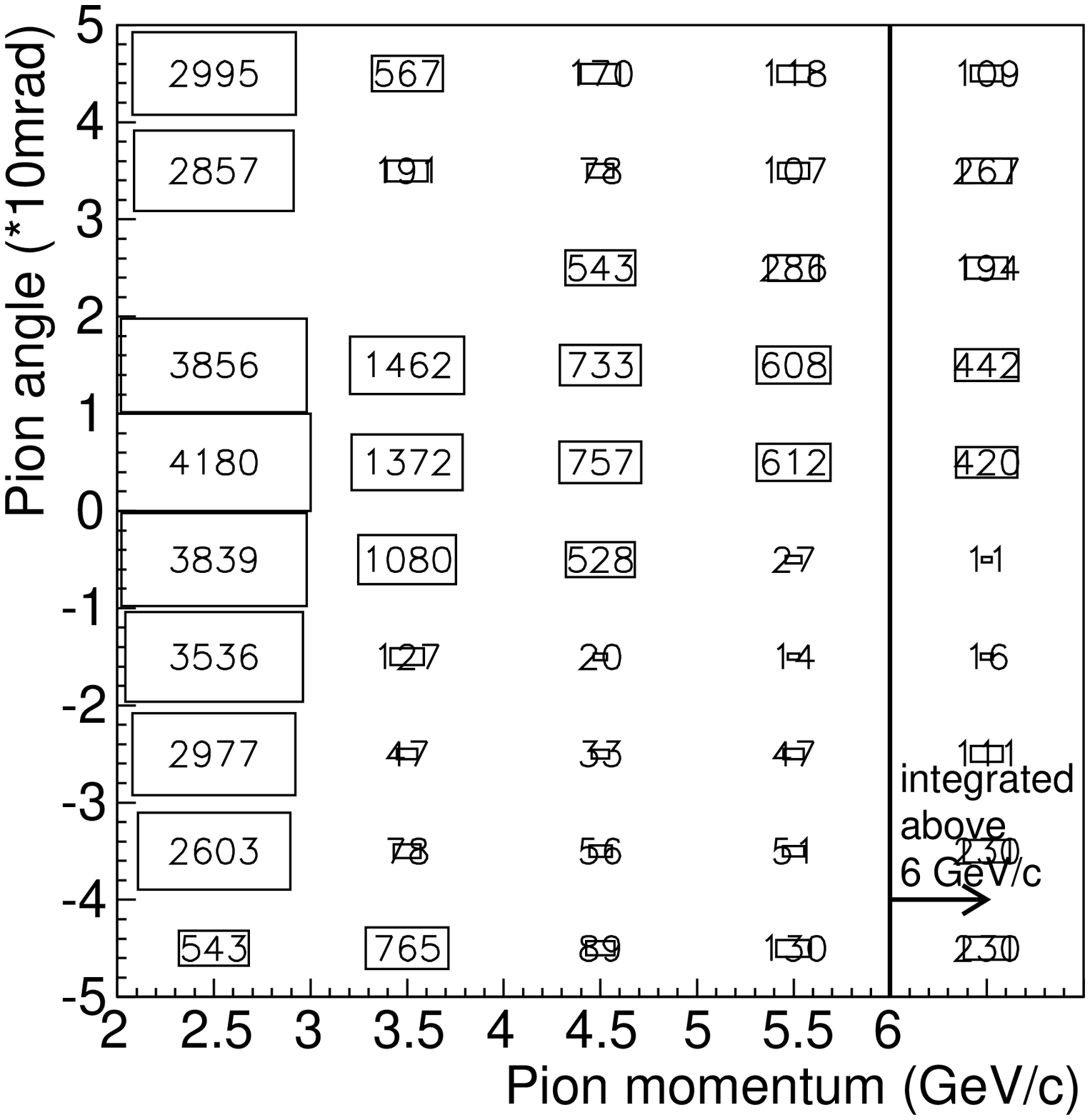}
    \caption{The fitting results of pion $(p_\pi,\theta_\pi)$ distribution
      in November 1999 run.
      The left figure shows the resulting central value of the
      weighting parameters, and the right figure shows the
      estimated fitting errors on them (no box means fitting errors are
      negligibly small.)
    }
    \label{fig:farnear:PIMON-KineResults}
  \end{center}
\end{figure}

A $\chi^2$-fitting is employed to extract the $(p_\pi,\theta_\pi)$ 2-dimensional
distribution from the Cherenkov light distributions with various refractive
indices.
The $(p_\pi,\theta_\pi)$-plane is binned into $5\times10$ bins;
5 bins in $p_\pi$ above 2~GeV/$c$ with 1~GeV/$c$ slice
(the last bin is integrated over $p_\pi>6$~GeV/$c$)
and 10 bins in $\theta_\pi$ from $-50$~mrad to $50$~mrad with 10~mrad slices.
Templates of the Cherenkov light distributions emitted by pions in
these bins are produced for each refractive index using a MC simulation.
Then, the weight of the contribution from each $(p_\pi,\theta_\pi)$ bin
being the fitting parameter,
the MC templates are fit to observed Cherenkov light distributions.
The fitting is done for the data in June 1999 and in November 1999, separately.
The resulting values of fitting parameters and errors on them in November 1999
run are shown in Fig.~\ref{fig:farnear:PIMON-KineResults}.

The neutrino energy spectra at ND and SK are derived by using the weighting
factors obtained above and a MC simulation.
The neutrino energy is binned into 6 bins: 0.5~GeV bins up to 2.5~GeV,
and integrated above 2.5~GeV.
The contribution of pions in each $(p_\pi,\theta_\pi)$ bin to neutrino energy
bins is estimated by a MC simulation, where
to a good approximation it depends only on the
pion kinematics and the geometry of the decay volume.
Then, the neutrino spectrum is obtained by summing up these contributions
weighted by fitted factors.
Finally, the ratio of the neutrino spectra at SK to that at ND yields
the \fn ratio.

The extracted neutrino spectra and the $F/N$ ratio from the PIMON data
taken in November 1999 are shown in Fig.~\ref{fig:harprelfluxes}
and \ref{fig:harpfluxratio} with empty squares and shaded error boxes.
All the systematic uncertainties in deriving them from the PIMON measurement
are included in the errors, where the most dominant contributions
to the error on the \fn flux ratio come from the fitting error,
the uncertainty in the analysis methodology,
and the uncertainty in the azimuthal symmetry of the horn magnetic field.
Further details on the systematic uncertainties in the PIMON measurement are
described in~\cite{Maruyama:PhD}.

\subsection{The far-to-near ratio in K2K }
The \fn flux ratio used to extrapolate the measurements in ND
to the expectation in SK is obtained in three independent ways: using
the HARP measurement, the Cho-CERN model, and the PIMON measurement,
as described in the previous sections.
We find that all three predictions of the \fn ratio are consistent
with each other within their measurement uncertainties.
Among these measurements, we use the one predicted by the HARP measurement
in our neutrino oscillation analysis described in this paper,
since the HARP pion production measurement was done for the same conditions
as K2K experiment: the proton beam momentum and the relevant
phase space of pions responsible for the neutrinos in K2K are the same.
In particular, the measured momentum region by the HARP experiment
reaches below 2~GeV/$c$ down to 0.75~GeV/$c$ where the PIMON is
insensitive.
The HARP measurement also gives us the most accurate measurements
on hadron production.

The central values for the \fn flux ratio as a function of neutrino energy
obtained from the HARP $\pi^+$ production results, $\overline{R}_{i}$,
are given in Tab.~\ref{tab:harpfluxratio_centralvalues}, where
the index $i$ denotes an energy bin number.
The total systematic uncertainties on the \fn flux ratio
as a function of neutrino energy are given in
Tab.~\ref{tab:harpfluxratio_errormatrix},
together with the uncertainty correlations among different energy bins,
expressed in terms of the fractional error matrix
$\langle \delta R_i \delta R_j \rangle/ (\overline{R}_i \overline{R}_j)$,
where $i,j$ label neutrino energy bins.
The \fn central values and its error matrix are used in the analysis for
neutrino oscillation described later.
\begin{table}
\caption{Predictions for the \fn muon neutrino flux ratio as a function
 of neutrino energy, for the HARP model for $\pi^+$ production in primary
 hadronic interactions. The neutrino energy binning is also indicated.}
\label{tab:harpfluxratio_centralvalues}
\begin{tabular}{crc} \hline \hline\\
Energy Bin Number $i$ & $E_\nu$~[GeV] & $\overline{R}_i$~$(\times10^{-6})$ \\
\hline
1 & 0.0$-$0.5 & 1.204 \\
2 & 0.5$-$1.0 & 0.713 \\
3 & 1.0$-$1.5 & 0.665 \\
4 & 1.5$-$2.0 & 0.988 \\
5 & 2.0$-$2.5 & 1.515 \\
6 & 2.5$-$    & 1.720 \\ \hline \hline
\end{tabular}
\end{table}
\begin{table}
\caption{Fractional error matrix
 $\langle \delta R_i \delta R_j \rangle / (\overline{R}_i \overline{R}_j)$
 obtained from the systematic uncertainties on the \fn flux predictions.
 The neutrino energy binning is the same as in
 Tab.~\ref{tab:harpfluxratio_centralvalues}.
 The values are given in units of $10^{-3}$.}
\label{tab:harpfluxratio_errormatrix}
\begin{tabular}{ c | r r r r r r } \hline \hline
Energy Bin & 1 & 2 & 3 & 4 & 5 & 6          \\ \hline
1 &  $0.187$ &  $0.002$ & $-0.036$ & $-0.372$ & $-0.281$ &  $0.240$ \\
2 &  $0.002$ &  $0.728$ &  $0.868$ &  $1.329$ &  $0.698$ & $-1.398$ \\
3 & $-0.036$ &  $0.868$ &  $1.304$ &  $2.122$ &  $1.041$ & $-2.040$ \\
4 & $-0.372$ &  $1.329$ &  $2.122$ &  $4.256$ &  $2.165$ & $-3.799$ \\
5 & $-0.281$ &  $0.698$ &  $1.041$ &  $2.165$ &  $1.779$ & $-2.678$ \\
6 &  $0.240$ & $-1.398$ & $-2.040$ & $-3.799$ & $-2.678$ &  $7.145$ \\
\hline \hline
\end{tabular}
\end{table}

While the neutrino flux predictions given in this section are appropriate
for most of the protons on target used in this analysis, a small fraction
of the data was taken with a different beam configuration. The
K2K-Ia period differed from the later configuration, as described in
Sec.~\ref{sec:beam:summary}.
As a result,
the far/near flux ratio for
June 1999 is separately estimated, in the same manner as
described above for later run periods. We find that the flux ratio
predictions for the two beam configurations, integrated over all
neutrino energies, differ by about 0.4\%. The flux ratio prediction
for the June 1999 beam configuration 
and the ND spectrum shape uncertainties are used to estimate the expected number
of neutrino events in SK and its error for the June 1999 period.

\section{Measurement of neutrino event rate at the near detector}
The integrated flux of the neutrino beam folded with the neutrino
interaction cross-section is determined by measuring the neutrino
event rate at the near site. The event rate at the 1KT detector is
used as an input to the neutrino oscillation study.
The stability of the neutrino beam is guaranteed by measuring the 
beam properties by the MRD detector.  In addition, the LG and SciBar detectors
measure the electron neutrino contamination in the beam to compare to our beam MC simulation.

\subsection{Neutrino event rate}
\begin{figure}
 \begin{center}
  \includegraphics[width=8cm,clip]{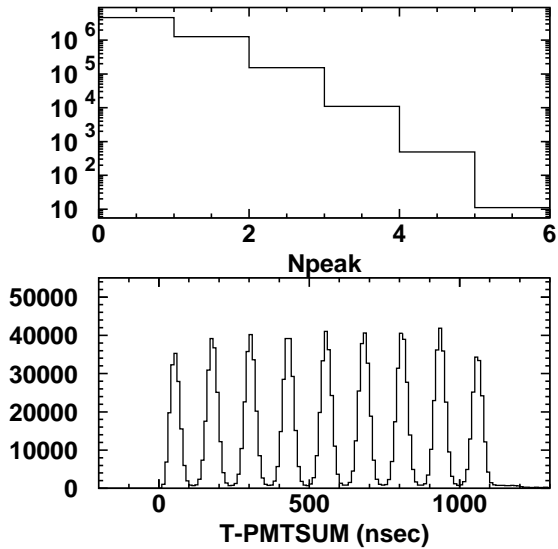}
 \end{center}
 \caption{The upper figure shows the number of neutrino interactions in a spill.
The lower figure shows the time distribution of
peaks of PMTSUM signal which are recorded by FADC. The beam's 9 micro bunch structure
 can be seen clearly.}
 \label{fig:npeak}
\end{figure}

As described in Section III.A, the 1KT water
cherenkov detector serves to measure the absolute number of neutrino
interactions in the near site and to predict the number of neutrino
interactions in the far site. Since the 1KT uses a water target and
almost the same hardware and software as SK, the systematic error in
the predicted number of interactions at the far site can be reduced.
The intensity of the neutrino beam is high enough that multiple neutrino
interactions per spill may occur in the 1KT. When this happens it is
difficult to reconstruct events. We employ Flash Analog-To-Digital
Converters (FADCs) to record the PMTSUM signal (see Section III.A) and
we can get the number of neutrino interactions by counting the number
of peaks above
a threshold. We set this threshold at 1000 photoelectrons (p.e.),
approximately equivalent to a 100 MeV electron signal, to reject low energy
background such as decay electrons from stopped muons. In Fig.
\ref{fig:npeak}, the upper and lower figures show the number of peaks
in a spill and the timing information of the peaks, respectively. We
can clearly see the 9 micro bunch structure of the beam in the lower
figure.  The fraction of multi-peak interactions in a spill is about
10\% of single-peak spills.\

 Sometimes the FADCs cannot identify multiple
 interactions if these events happen in the same bunch and the time gap between
 the interactions is too small. To correct for this possibility, 
 we employ a MC simulation
 with multiple interactions to estimate the misidentification probability.
 The PMTSUM signals recorded by the FADC are simulated, so the same
 method can be used for the MC simulation and data. 
We found that the number of interactions
in the fiducial volume is underestimated by 2.3\% for multiple
interactions.
The multiple interactions contribute 34\% of the total
number of interactions, and 
we have to correct the number of events by this multi-interaction
misidentification probability,
according to  $C_{multi} = 1+ 0.023\times0.34 = 1.008$.\

\begin{figure}
 \begin{center}
  \includegraphics[width=8cm,clip]{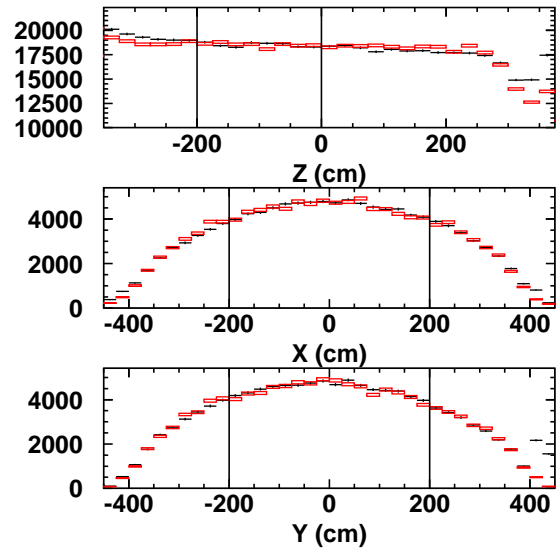}
 \end{center}
 \caption{The upper figure shows the vertex distribution in $z$
 with an $R_{xy} < 200~{\rm cm}$ cut. The beam direction is defined as the $z$-axis.
Crosses show data, boxes show the MC simulation normalized by area in the fiducial volume, and vertical black lines show the fiducial
 volume. The middle and lower figures show the $x$ and $y$ distributions, respectively.}
 \label{fig:vtx} 
\end{figure}

The fiducial volume in the 1KT is defined as a horizontal cylinder with
 axis along the beam direction (z-axis). The radius is 200~cm and 
the $z$ coordinate is limited to $-200~{\rm cm} < z < 0~{\rm cm}$,
where the center of the 1KT ID is defined as $z=0$~cm,
and the total fiducial mass is 25 tons. The fiducial volume
cut results in an almost pure neutrino sample, rejecting cosmic rays or
muons generated by the beam in the materials surrounding the 1KT
 (beam-induced muons).
Figure~\ref{fig:vtx} shows the vertex distributions of data 
and the MC simulation.  Because we simulate only neutrino interactions 
without beam-induced muons and without the cosmic-ray muons, we can
see excess events upstream of the $z$-distribution and the top part
of the detector ($y > 400~{\rm cm}$) in data.
The data and the MC simulation are in good agreement in the fiducial volume.

 Two major background sources are considered. Cosmic ray events
 usually have a
 vertex near the upper wall of the inner tank, but some events
 contaminate the
fiducial volume due to failure of the vertex reconstruction. To estimate the 
background rate, we run the detector without the beam, replacing the spill 
trigger by a periodical clock signal. The beam-off data are analyzed in the
 same way as the neutrino data; it is found that cosmic rays in the 
fiducial volume are 1.0\% of the neutrino data. 
The other important background source is beam-induced muons which can be
 tagged by PMTs located in the outer detector. 
After the vertex cut, the remaining events are scanned with a visual
event display and the fraction of beam induced muons found is 0.5\%.  
In addition, we had fake events which were produced by signal 
reflection due to an impedance mismatch of the cables in the 1999 runs only.
 The total background fraction is estimated to be 1.5\%  for runs
 starting in 2000, and 3.1\% before 2000.\

\begin{figure}
 \begin{center}
  \includegraphics[width=8cm,clip]{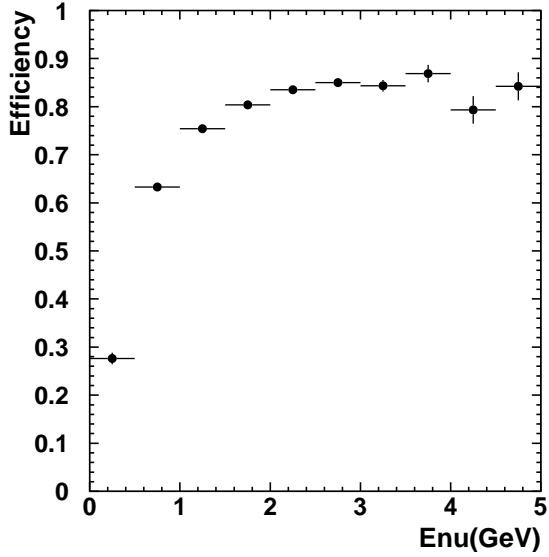}
 \end{center}
 \caption{Neutrino selection efficiency as a function of neutrino
 energy (GeV).}
 \label{fig:nu-eff}
\end{figure}

\begin{figure}
 \begin{center}
  \includegraphics[width=8cm,clip]{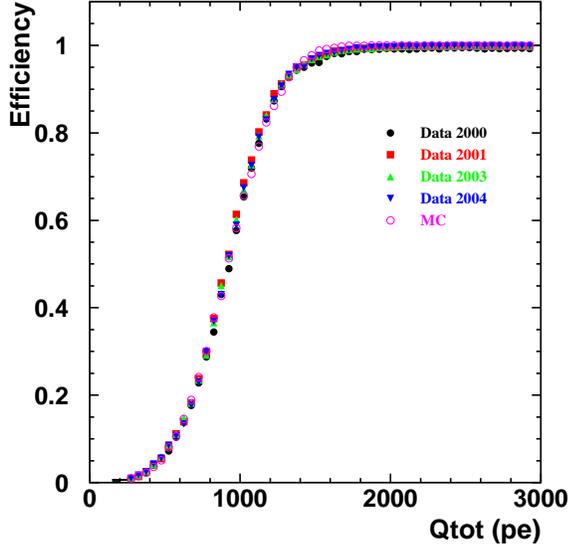}
 \end{center}
 \caption{FADC peak finding efficiency is shown as a function of total charge
 (p.e.). Data taken in 2000, 2001, 2003, and 2004 are shown with the MC
 simulation.}
 \label{fig:fadc-eff}
\end{figure}

 The neutrino event selection efficiency is calculated based on the MC simulation.
The efficiency is defined as:
\begin{equation}
   \effkt=\frac{\mbox{(\# of reconstructed events in 25t)}}
{\mbox{(\# of generated events in 25t)}}
\end{equation}
Figure~\ref{fig:nu-eff} shows the selection efficiency as a function of neutrino
energy. The overall efficiency including all energies and all interaction types
 is 75\% for the configuration with a 250kA horn current, and 71\% for
 the 200kA 
configuration. We had a problem with the FADC in November 1999 which corresponds to
 3\% of all data, and the efficiency in this period
was 5\% lower than the other 250kA configuration periods.
The dominant inefficiency comes from the single peak selection with a
1000 p.e. threshold by FADC. Figure~\ref{fig:fadc-eff} shows the peak-finding
 efficiency of the FADC as a function of total charge. The 2000-2004 data
 plotted in the figure shows that the efficiency curves are stable.
The MC event selection
efficiency is obtained by smearing the threshold assuming a gaussian
 distribution, in which the mean and width are obtained by fitting the
 data.\

While the FADC can count the event multiplicity, they do not
record information about each PMT channel. The ATM gives the timing and
charge information of each individual PMT channel which allows event reconstruction
if the spill has a single interaction. Both the FADC and the ATM are required
to derive the total number of neutrino interactions:\

\begin{equation}
N^{1KT}_{int} = N^{1KT}_{obs} \cdot \frac{N^{total}_{peak}}{N^{single}_{peak}} 
\cdot \frac{1}{\effkt} \cdot \frac{1}{1+R_{BKG}} \cdot C_{multi}
\label{eq:1ktint}
\end{equation}
\
where $N^{1KT}_{int}$ is the total number of neutrino interactions in
        the 25~t fiducial volume, 
        $N^{1KT}_{obs}$ is the number of events observed in the fiducial volume among single peak events,
        $N^{total}_{peak}$ is the total number of PMTSUM signal peaks,
        $N^{single}_{peak}$ is the number of single peak events,
        $\effkt$ is the selection efficiency of neutrino events in
        the 25t fiducial volume, $R_{BKG}$ is the background fraction,
        and $C_{multi}$ is the multiple interaction correction.

Table~\ref{table:Nkt} shows the total number of neutrino interactions
and the number of protons on target for each period.
\begin{table}[bp]
\caption{Number of neutrino interactions in the 1KT.}
\begin{center}
\begin{tabular}{|c|r|r|r|r|r|} \hline
Period & \multicolumn{1}{c|}{$POT_{1KT}(10^{18})$} & \multicolumn{1}{c|}{$N^{1KT}_{obs}$} & \multicolumn{1}{c|}{$N^{tot}_{peak}$} & \multicolumn{1}{c|}{$N^{single}_{peak}$} & \multicolumn{1}{c|}{$N^{1KT}_{int}$} \\\hline\hline
Ia & 2.6 & 4282 & 109119 & 89782 & 7206 \\\hline
Ib & 39.8 & 75973 & 1854781 & 1475799 & 130856 \\\hline
IIa & 21.6 & 43538 & 1061314 & 832112 & 73614 \\\hline
IIb & 17.1 & 34258 & 813599 & 644723 & 57308 \\\hline
IIc & 2.9 & 5733 & 137533 & 111834 & 9346 \\\hline
\end{tabular}
\end{center}
\label{table:Nkt}
\end{table}

Table~\ref{table:Sys-Nkt} shows the systematic uncertainty on the number
of neutrino
 interactions in the 1KT. The dominant error is the uncertainty
 of the fiducial volume. From the comparison
of neutrino interactions in data and the MC simulation, we quantitatively estimate the 
fiducial volume systematics. Varying the definition of the fiducial volume 
by about five times the vertex resolution, we observe a difference in the 
calculated event rate of 1.8\%. Most of the difference is due to the 
$z$-dependence of partially contained events.
To estimate this systematic effect, we used deposited energy from
neutrino interactions themselves for partially-contained events since
the deposited energy is roughly linear in the distance from the
vertex to the downstream wall of the ID.  We use the ``cosmic-ray
pipe'' muons, described in Section~\ref{sec:1kt}, to define the energy
scale for partially-contained events within 2.3\%.  We do not see
evidence for such a bias within the uncertainty of the energy
scale. The fiducial uncertainty arising from a vertex bias is
therefore 2.3\%.  We conservatively add those two numbers in
quadrature to obtain 3.0\% for the uncertainty of the fiducial
volume.

\begin{figure}
 \begin{center}
  \includegraphics[width=8cm,clip]{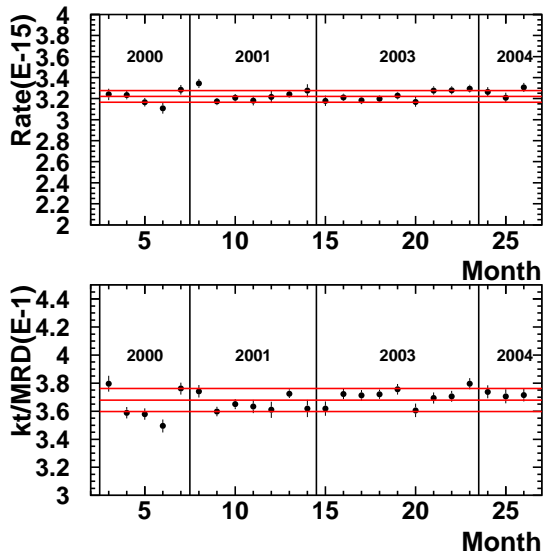}
 \end{center}
 \caption{The upper figure shows the event rate in the 1KT normalized by 
         the muon yield of the MUMON. One point corresponds to one month and 
         the 2000 - 2004 data are shown in the figure. The middle line shows 
         the average and the upper and lower lines show the RMS, 1.6\%.
         The lower figure shows the 1KT event rate divided by the MRD event rate,
         also showing the average and RMS (2.1\%).}
 \label{fig:ktrate}
\end{figure}

The energy scale uncertainty of the 1KT is estimated by using cosmic ray
muons which stopped inside of the detector and the reconstructed $\pi^0$ mass
which mostly comes from neutral current interactions. The absolute
energy uncertainty of the 1KT is estimated to be $^{+3\%}_{-4\%}$.
This energy uncertainty affects $N_{1KT}$ because of the FADC cut.
We changed the threshold of the FADC and the effect of energy scale
uncertainty is estimated to be 0.3\%.
The FADC charge is calibrated by the total charge recorded by ATM using single
interaction events with a lower threshold (200~p.e.).
The stability of the charge scale of FADC is $\pm5\%$ and its effect on
$N_{1KT}$ is $\pm0.8\%$.
 $N^{total}_{peak}$ , $N^{single}_{peak}$ and $N^{1KT}_{obs}$ depend on the 
FADC cut position, but $N^{1KT}_{int}$ should be independent of the cut
 if the efficiency correction is perfect. We calculate $N^{1KT}_{int}$  
changing FADC cut from 200 to 2000 p.e. and confirm the total number of
 neutrino interaction is stable within $\pm1.5\%$.  
In Fig.~\ref{fig:ktrate}, the upper figure shows the
1KT event rate normalized by muon yields at the SPD in the MUMON (see Section II.A.2).
We take 2.0\% for the uncertainty of the event
rate stability from the root-mean-square (RMS) of the distribution. 
A comparison between the event rates of the 1KT and the MRD is
shown in the lower
figure of Fig.~\ref{fig:ktrate} as a consistency check.
We assign statistical errors as systematic errors for the background 
and multiple interaction corrections due to the limited numbers of the
sample. In total, we quote a $\pm$4.1\% error on the number of 1KT
neutrino events over the entire K2K run.\

\begin{table}
\caption{Systematic errors on $N^{1KT}_{int}$.}
\begin{center}
 \begin{tabular}{lc}
  \ Source                    & Error (\%)   \\ \hline \hline
  \ Fiducial volume           & $\pm$ 3.0      \\ \hline
  \ Energy scale              & $\pm$ 0.3      \\ \hline
  \ FADC stability            & $\pm$ 0.8      \\ \hline
  \ FADC cut position         & $\pm$ 1.5      \\ \hline
  \ Event rate                & $\pm$ 2.0      \\ \hline
  \ Background                & $\pm$ 0.5      \\ \hline
  \ Multi-interaction         & $\pm$ 0.7      \\ \hline \hline
  \ Total                     & $\pm$ 4.1      \\ \hline
 \end{tabular}
\end{center}
 \label{table:Sys-Nkt}
\end{table}

\subsection{Neutrino beam stability}
The neutrino beam properties-- the profile, beam direction,  
energy spectrum and event rate-- are 
monitored by the MRD using $\nu_{\mu}$ interactions with iron in order 
to guarantee stability of the beam for the entire run period. 

A neutrino event is identified by a muon track in the MRD with the following 
selection criteria: only tracks within the time of the beam spill
are accepted; tracks with a common vertex and a common timing are taken 
as a single event with the longest track assumed to be a muon; muons
entering or exiting the detector are removed; and muons which traverse
one or more iron plates ($N_{layer} \geq 1$) are selected.  

\begin{figure}[]
 \begin{center}
   \includegraphics[width=\columnwidth]{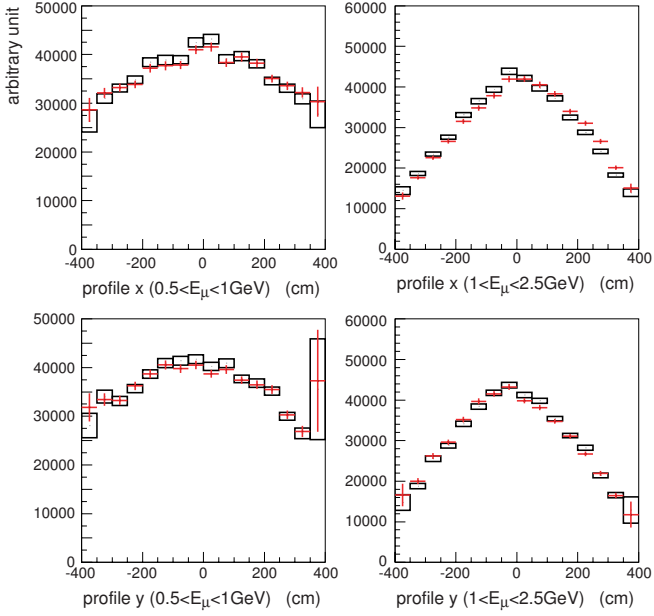}
 \caption{
        Neutrino beam profiles measured by the MRD. 
        Crosses show the measured profile and the boxes show the Monte 
        Carlo prediction. Normalization is by number of entries. 
        (Top left): horizontal profile for 0.5GeV $< E_{\mu} <$ 1.0GeV. 
        (Top right): horizontal profile for 1.0GeV $< E_{\mu} <$ 2.5GeV. 
        (Bottom left): vertical profile for 0.5GeV $< E_{\mu} <$ 1.0GeV. 
        (Bottom right): vertical profile for 1.0GeV $< E_{\mu} <$ 2.5GeV. 
  \label{fig:profile}}
 \end{center}
\end{figure}

\begin{figure}[]
 \begin{center}
   \includegraphics[width=\columnwidth]{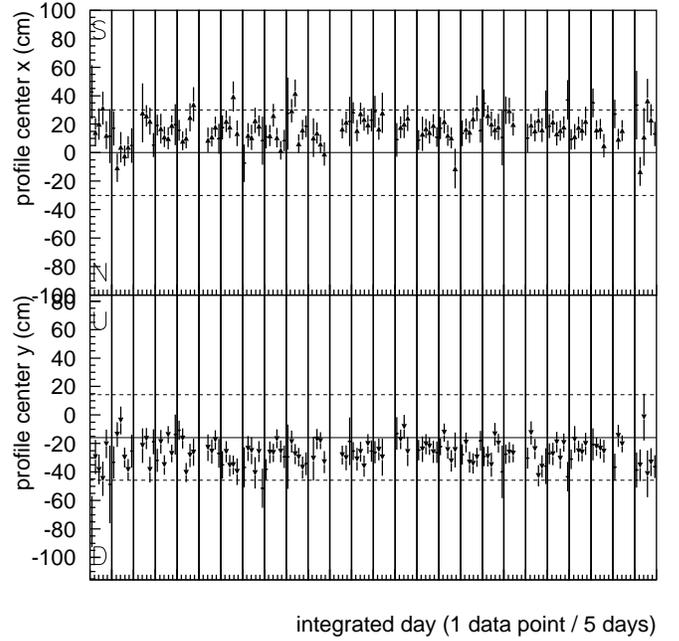}
 \caption{Stability of the neutrino beam direction measured by the MRD. 
          The direction is plotted every five days 
          for the entire experimental period. 
	  The solid line shows the SK direction and the 
          dashed lines show $\pm$1~mrad from the center.  
	  The direction is required to be stable within $\pm$3mr.  
         The top plot is for the horizontal direction; 
         the bottom for the vertical direction. 
  \label{fig:directionstab}}
 \end{center}
\end{figure}

The neutrino beam profile is obtained by measuring the vertex distribution 
in the MRD.  The location of the center of the profile gives the beam 
direction.  
For this purpose, any muons with reconstructed energy 
lower than 0.5 GeV or higher than 2.5 GeV are rejected 
and a cubic fiducial volume of ${\rm 6m} \times {\rm 6m}$ in 
the upstream 9 iron plates is used, which has a total mass of 419 tons. 
As shown in Fig.~\ref{fig:profile}, the profile is well reproduced 
by the MC simulation. 
The profile center is plotted as a function of time in 
Fig.~\ref{fig:directionstab}. 
The beam direction has been stable throughout the experiment 
within $\pm$1~mrad from the SK direction.  
This satisfies the requirement for the direction to be within 
$\pm$3~mrad, which is described in Section II.  
The stability of the profile width is also confirmed by the measurement.  

\begin{figure}[]
 \begin{center}
   \includegraphics[width=\columnwidth]{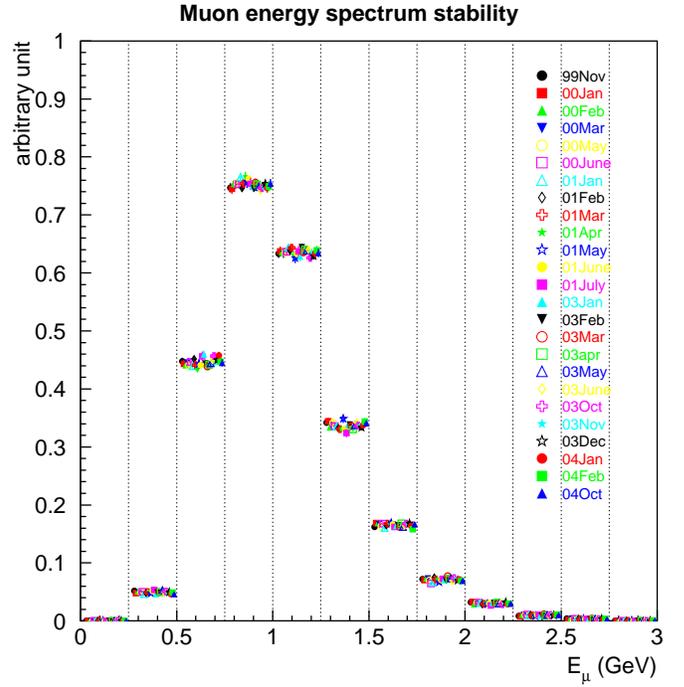}
 \caption{Stability of the muon energy distribution measured by the MRD.  
     Each bin is plotted every month for the entire experimental 
     period except for K2K-Ia.
     The distributions are normalized by number of entries. 
  \label{fig:emustab}}
 \end{center}
\end{figure}

\begin{figure}[]
 \begin{center}
   \includegraphics[width=\columnwidth]{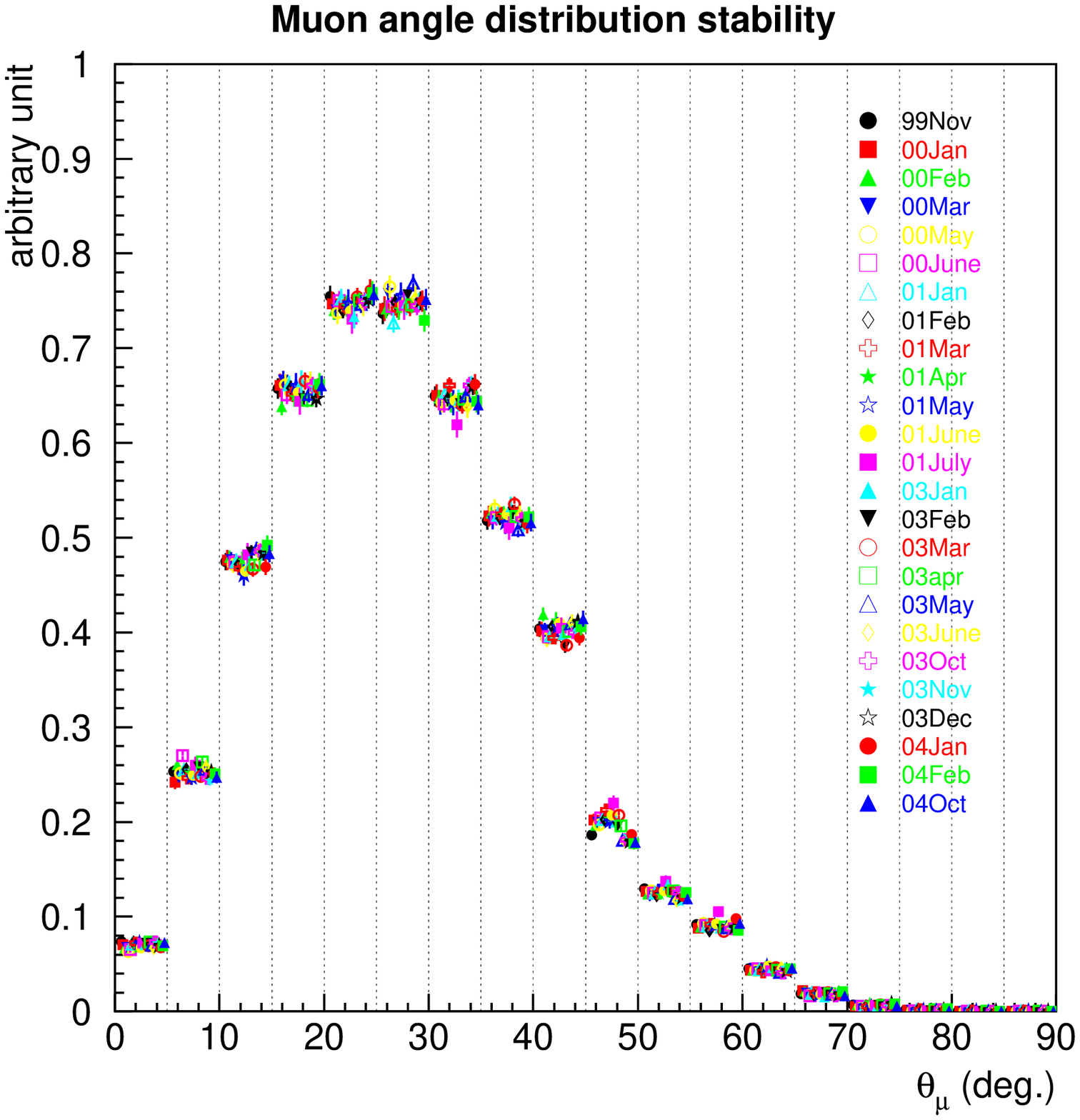}
 \caption{Stability of the muon angular distribution measured by MRD.  
     Each bin is plotted every month for the entire experimental 
     period except for K2K-Ia.  
     The distributions are normalized by their entries. 
  \label{fig:amustab}}
 \end{center}
\end{figure}

The muon energy and angular distributions are also continuously 
monitored.  
A cylindrical fiducial 
volume of radius 3m in the upstream 9 plates is used, where the mass 
is 329 tons.  
They show no change as a function of time beyond statistical uncertainty.  
The muon energy distribution and the angular distribution are plotted 
monthly in Fig.~\ref{fig:emustab} and Fig.~\ref{fig:amustab}, respectively. 
These results show that the $\nu_{\mu}$ energy spectrum
is stable within 2-4\% depending on the energy bin.  
This is well within the spectrum error quoted in the spectrum analysis 
described in Section VII.  

In order to check the 1KT event rate and 
to compare the neutrino cross sections in iron and water, 
the neutrino event rate in the MRD iron is derived.  
The fiducial mass used is 
72.8 tons, which is the 3m-radius cylinder of the upstream three 
iron plates only.  For this purpose, $N_{layer} \geq 2$ is required 
in order to reduce the hadronic background.  
The stability of the 1KT event rate divided by the MRD rate is plotted 
in the bottom figure of Fig.~\ref{fig:ktrate}.  

Since the absolute normalization is less certain, 
a double ratio of the MRD and 1KT event rates is calculated 
for the cross-section comparison between iron and water. 
The result for this double ratio is 
$({\rm data/MC})_{\rm MRD}/({\rm data/MC})_{\rm 1KT} 
= 1.04 \pm 0.003({\rm stat.}) ^{+0.08}_{-0.11}({\rm sys.})$.  
The average event rates for MRD and 1KT data for
the entire K2K run (except for K2K-Ia) are used.
This double ratio should be unity if we correctly 
understand the iron/water neutrino cross-section ratio 
in the K2K energy range regardless of the absolute cross section.  
The sources of the systematic error are summarized in 
Table~\ref{table:sys-MRD}.  
Here, the uncertainties due to the neutrino spectrum, NC/CC ratio, 
and non-QE/QE cross-section ratio have the cancellation between 1KT and MRD
taken into account.

\begin{table}
\caption{Systematic errors of the event-rate double ratio 
 $({\rm data/MC})_{\rm MRD}/({\rm data/MC})_{\rm 1KT}$ }. 
\begin{center}
 \begin{tabular}{l|rl|rl}
  \ Source                    & & Error +(\%) & & -(\%)  \\ \hline
  \hline
  \ Fiducial volume             & +1.6   & &  -5.7 &   \\
  \ Selection efficiency        & +1.2   & &  -5.7 &   \\
  \ Tracking efficiency         & +1.0   & &  -1.0 &   \\
  \ Beam direction              & +1.9   & &  -0.0 &   \\ \hline
  \ MRD detector oriented total & & +2.9   & &  -8.1   \\ \hline
  \ 1KT detector oriented total & & +4.1   & &  -4.1   \\ \hline
  \ Neutrino spectrum           & +0.9   & &  -0.9 &   \\
  \ NC/CC ratio                 & +4.0   & &  -3.7 &   \\
  \ Non-QE/QE ratio             & +5.3   & &  -3.7 &   \\ \hline
  \ Spectrum and neutrino int.  & & +6.7   & &  -5.3   \\ \hline
  \hline
  \ Grand total                 & & +8.4   & & -10.5   \\ \hline
 \end{tabular}
\end{center}
 \label{table:sys-MRD}
\end{table}

The stability of the detector itself is confirmed as a whole by 
analyzing the off-spill data which is essentially cosmic-ray data 
taken between each beam spill.  
The event rate, angular distributions and 
$N_{layer}$ distribution are found to be stable.

\subsection{Electron neutrino component}

The $\nu_e$ component in the beam is measured in the FGD system
independently by the LG and by the SciBar detectors. In each
detector we perform a search for $\nu_e$ interactions by looking
for events with an electron in the final state. ``Electron" events
come essentially only from the $\nu_e$ component of the beam, since
the ${\nu_e}/{\nu_{\mu}}$ flux ratio is about $1.3\times
10^{-2}$, while the ${\overline{\nu}_e}/{\nu_{\mu}}$ flux ratio
is about $1.8\times 10^{-4}$, and the cross section for $\nu_{\mu}$e
scattering is about a factor $1.5\times 10^{-4}$ smaller than that
for $\nu_{\mu}$ CC scattering on a nucleon. The measurement of the
$\nu_e$ events validates the prediction of the
${\nu_e}/{\nu_{\mu}}$ flux ratio at the near location obtained
from the beam MC simulation. The data-MC comparison also involves
the cross-sections of $\nu_e$ and $\nu_{\mu}$ interactions, and the
measurement is therefore an important check of the MC simulation
used to predict the number of $\nu_e$ interactions in SK.

The LG measurement~\cite{Yoshida} is performed by looking
for $\nu_e$ interactions taking place in the SciFi detector, with
the electron detected in the downstream scintillator hodoscope, and
its energy measured in the LG calorimeter. Electron events are
selected by requiring the following: an interaction vertex inside the SciFi
fiducial volume, an energy deposit in the downstream scintillator
hodoscope system greater than 20~MeV (2.5 times larger than that
expected from a muon), an energy deposit in the LG greater than
1~GeV; and finally, no hit in the MRD matching the electron direction.
For an exposure of $2.9\times 10^{19}$~POT, 51 electron
candidates are found with an estimated background of 24 $\nu_{\mu}$
induced events. The ${\nu_e}/{\nu_\mu}$ interaction ratio is
estimated to be $1.6 \pm 0.4({\rm stat.})^{+0.8}_{-0.6}({\rm
syst.})\%$, which is in agreement with the beam MC prediction of
1.3\%.  The dominant source of the $\nu_e$-induced component of CC interactions
is muon decay (87\%) in the beam, and the remainder comes from kaon decay.

The measurement of the $\nu_{e}$ contamination in the beam has also
been performed using the SciBar plus EC detector, with a statistics
corresponding to $2.1\times 10^{19}$~POT. The search for electrons
is mainly based on the signals from the EC, the electromagnetic
calorimeter which follows the tracking section of the SciBar. The two
planes of the EC correspond to approximately 5.5 radiation
lengths (X$_{0}$) each, so
that electrons in the 1 GeV energy range are almost fully contained.
The average energy lost by a muon or pion traversing one plane is
small, of the order of 50 MeV (with $\sim$ 60 MeV full width).
Therefore, to look for electrons we select events with a large
signal in a restricted region of the first EC plane.  We require
$E_1>350$~MeV, where $E_1$ is the energy in a cluster of 20 cm width
centered in the module with maximum signal. We also apply conditions
on the energy release $E_2$ in the 20 cm wide cluster of highest energy
of the second plane of the EC. We require ${E_2}/{E_1}$ to lie in
the 0.2-1.1 interval, which from MC simulation we know to contain 95\% of the
electron events. In the tracking volume of the SciBar we search for a
reconstructed track pointing to the selected high energy clusters in
the two projections of the EC. The efficiency for reconstructing the
electron as a track in the SciBar is high, given the low density of the
detector (X$_{0}$ = 40 cm). Finally, we impose a fiducial volume cut
on the interaction vertex, defined as the starting point of the
electron track, and we also require that outside the selected
cluster the energy in each EC plane does not exceed 30 MeV.

\begin{figure}
  \begin{center}
    \includegraphics[width=8cm]{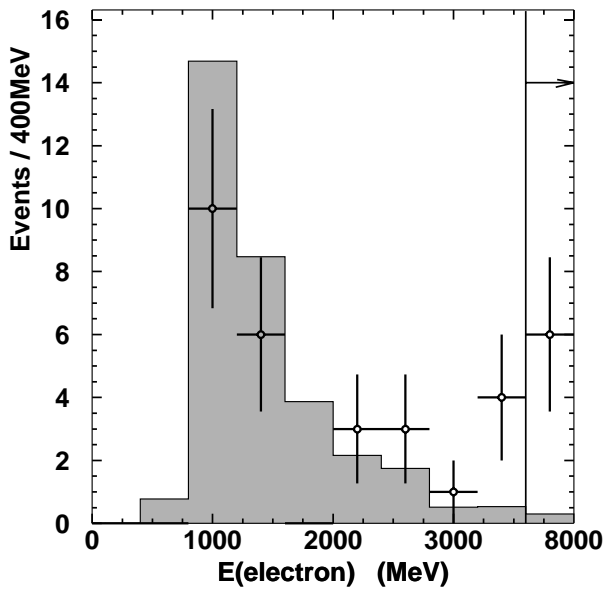}
    \caption{Electron energy spectrum for the candidates of $\nu_{e}$ interactions
                      selected in SciBar. The MC spectrum (histogram) is normalized to the
                      data separately for signal (30 events) and background (3 events).
                      The contents of the highest energy bin is integrated between 3.6 and 8 GeV.}
    \label{fig:ene}
  \end{center}
\end{figure}

The
selected sample consists of 42 electron candidate events.  The
visual examination of the display of these events allows us to discard
9 events, easily identified as background from neutrino interactions
originating outside the fiducial volume of the SciBar, or as interactions
with $\pi^{0}$ production. From the number of $\pi^{0}$ events
identified at the scanning level we can estimate, with a correction
obtained from a MC simulation, the number of $\pi^{0}$ events which
cannot be distinguished from electron events, and constitute an
irreducible background. Our final sample, with 90\% electron purity,
contains 33 events, with a background estimate of 3 $\pm$ 2 events.
The characteristics of the selected events are compared to those of
MC events resulting from a full simulation of $\nu_{e}$ interactions
in the SciBar and including also a 10\% background from $\nu_{\mu}$
interactions. The data-MC comparison is fair, except for an excess
of high energy electrons in the data. The energy spectrum of the
electrons is shown
in Fig. $\ref{fig:ene}$. The electron energy is obtained by correcting
the energy measured in the EC for the energy lost in the tracking
section of the SciBar and for the longitudinal leakage, the average
correction being of the order of 20\%. The excess may indicate an
underestimate of kaon production in the beam simulation, but the
statistics is too small to draw firm conclusions. Finally, we use
the MC simulation to extrapolate our measurement, which is only
sensitive to electron energies larger than 500 MeV,  to the full
energy range. Our result for the interaction ratio
${\nu_{e}CC}/{\nu_{\mu}CC}$ is 1.6 $\pm$ 0.3(stat.) $\pm$
0.2(syst.)\%, consistent with the MC prediction of 1.3\%.

The consistency of the measurements in the LG and SciBar between
themselves and with the MC predictions confirms the quality of the
measurements and of the MC simulation. However, since the
measurements are limited to a restricted energy region, for all the
analyses described in the paper we use the $\nu_e$ component as
given by the MC simulation.

\section{Measurement of neutrino spectrum at the near detector}
 The neutrino energy spectrum before oscillation
is measured with near detector 1KT, SciFi and SciBar
CC event samples. 
 The neutrino energy is reconstructed from the muon kinematics
parameters $p_{\mu}$ and $\theta_{\mu}$ assuming a QE interaction
as given in Eq.~\ref{eq:Enurec}.
The two-dimensional distributions of $p_{\mu}$
versus $\theta_{\mu}$ are used to measure the neutrino energy spectrum.
The spectrum is fitted by using a $\chi^{2}$ method to
compare observed ($p_{\mu}$, $\theta_{\mu}$) distributions 
to MC expectations.

\subsection{The fitting method}

\begin{figure}
    \includegraphics[width=0.45\textwidth]{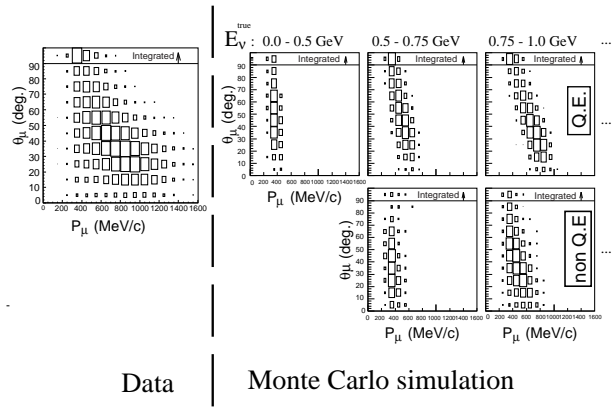}
  \caption{Schematic view of the binning of the data and Monte Carlo events
for the spectrum fit.  
The left plot shows a correlation between $p_{\mu}$ and $\theta_{\mu}$
for FC1R$\mu$ events in the 1KT data used for
the spectrum fit. The right plots show those for the MC sample
separately prepared for each neutrino energy bin and 
for QE and non-QE interactions.}

\label{fig:2dimtemplate}
\end{figure}

 In order to obtain the neutrino energy spectrum,
the ($p_{\mu}$, $\theta_{\mu}$) distribution is fit with
the MC expectation as shown in Fig.~\ref{fig:2dimtemplate}.
 The neutrino energies are divided into eight bins as shown in
Table~\ref{tbl:enubin}.
 For the MC expectation, the $(p_{\mu}, \theta_{\mu})$ distribution
is prepared for each $E_{\nu}$ bin and separately
 for QE and nonQE interactions;
  $8\times2$ distributions are prepared in total for each
event sample.
 
 The free parameters in the fit are the neutrino energy spectrum 
parameters for eight energy bins~$(f^{\phi}_{1},...,f^{\phi}_{8}) $
and a parameter, $\Rnqe$, 
which represents the relative weighting of CC-nonQE 
events to CC-QE events.
 The systematic uncertainties, such as nuclear effects, the energy
scale, the track finding efficiency, 
and other detector related systematics, are also incorporated as
the fitting parameters~($\mbox{\boldmath $f$}$).
 The contents in 
$(m,n)$-th bin of the $(p_{\mu}, \theta_{\mu})$ distribution,
$N^{\MC}_{m,n}$, is expressed with the 16 templates and the fitting 
parameters as,

\begin{multline}
\label{eq:mc-exp}
   N^{\MC}_{m,n} \equiv  
   P\cdot \sum_{i=1}^{8} \fphi_i \cdot
   \left[ N^{\MC(\QE)}_{m,n,i} +
    \Rnqe \cdot N^{\MC(\nQE)}_{m,n,i} \right],
\end{multline}

\noindent
where $P$, $N^{\MC(\QE)}_{m,n,i}$ and
$N^{\MC(\nQE)}_{m,n,i}$ are a normalization parameter, the number of
expected contents in the $(m,n)$-th bin for QE interaction and that for non-QE
interaction for the $i$-th neutrino energy bin.
 We take the $\chi^{2}$ between the observed distributions,
$N^{\mathrm{obs}}_{m,n}$, and $N^{\MC}_{m,n}$.

 During the fit, the flux in each energy bin and $\Rnqe$ are
re-weighted relative to the nominal values in the MC simulation.
 The flux for $E_{\nu} = 1.0-1.5$~GeV bin is fixed to unity for the
normalization, and another set of parameters is prepared
for relative normalization of each detector.

 The $\chi^{2}$ functions are separately defined for each
detector and then summed to build a combined $\chi^{2}$ function as
\begin{eqnarray}
   \cND = \cKT + \cSF + \cSB
\end{eqnarray}
 Finally, a set of the fitting
parameters~($f^{\phi}_{i},\Rnqe:\mbox{\boldmath $f$} $)
is found by minimizing the $\chi^{2}$ function.
 The best fit values, their error sizes and
the correlations between them are used as inputs to the oscillation
analysis, as described in Sec~\ref{sec:oscana}.  The following subsections
will describe 
the definition of $\chi^2$ for each subdetector and the results of fit.

\begin{table*}
   \begin{center}
    \caption{The $E_{\nu}$ interval of each bin.}
    \begin{tabular}{|c|cccccccc|}
     \hline
        & $f^{\phi}_{1}$ & $f^{\phi}_{2}$ & $f^{\phi}_{3}$ &
        $f^{\phi}_{4}$ & $f^{\phi}_{5}$ & $f^{\phi}_{6}$ & 
        $f^{\phi}_{7}$ & $f^{\phi}_{8}$ \\
     \hline
     $E_{\nu}$~[GeV] & 0.0$-$0.5 & 0.5$-$0.75 & 0.75$-$1.0 & 1.0$-$1.5 
     & 1.5$-$2.0 & 2.0$-$2.5 & 2.5$-$3.0 & 3.0$-$ \\
     \hline
    \end{tabular} 
    \label{tbl:enubin}
    \end{center}
\end{table*}
 
\subsection{Definition of $\chi^2$ for 1KT}

\begin{figure}
    \includegraphics[height=7.1cm]{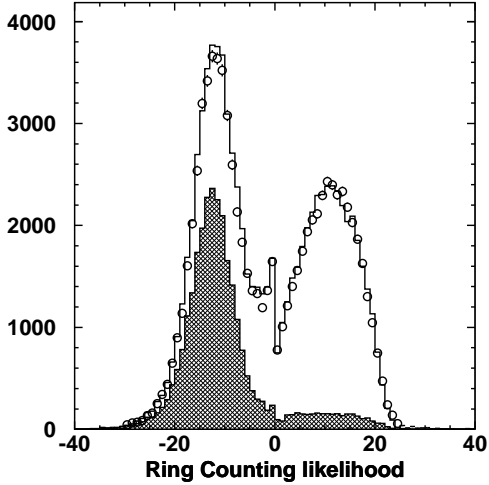}
  \caption{The distribution of ring counting likelihood for the 1KT.  Those events that have a likelihood less
    than or equal to 0.0 are considered to have one ring; those above 0.0 are
    considered to be multi-ring.  
In this plot, data are the circles and the MC simulation is the
    histogram. The hatched histogram shows the CCQE component.
Only statistical errors are shown for data.}
\label{fig:dlfct}
\end{figure}

\begin{figure}
    \includegraphics[height=7.1cm]{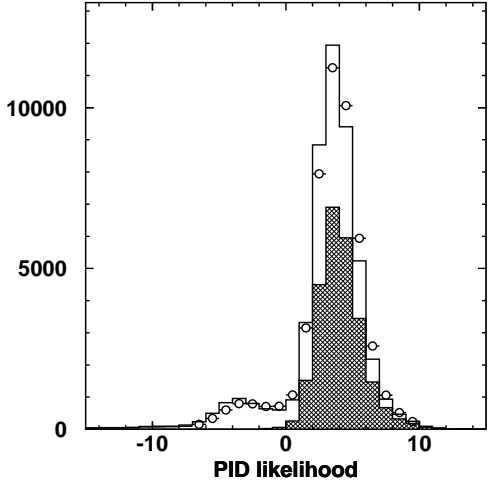}
  \caption{The distribution of particle identification likelihood for the 1KT.
    The events with a likelihood greater than 0.0 are $\mu$-like while
  those less than or equal to 0.0 are e-like.  Data are the circles
  and the MC simulation is the histogram with the CCQE component shown as the
  hatched area.}
\label{fig:pid}
\end{figure}

\begin{figure}
    \includegraphics[height=7.1cm]{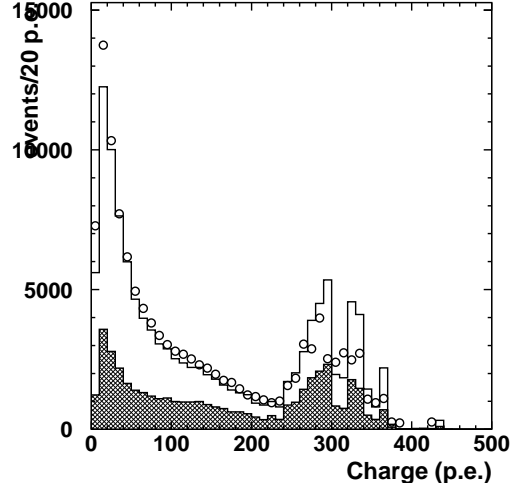}
  \caption{The largest charge in a PMT for a 1KT event.
    Events that have a charge less than 200 p.e. are considered FC events.  The
  rest are PC events.    Data are the circles
  and the MC simulation is the histogram, with the CCQE component shown as the
  hatched area.}
\label{fig:pomax}
\end{figure}

To measure the neutrino energy spectrum, we select a QE enriched 
data sample called the fully contained one-ring 
$\mu$-like (FC1R$\mu$) sample.  In addition to the 1KT event rate
selection (see Section VI.A), we require four additional conditions.  These are
that all visible particles are inside the 1KT detector (fully contained), 
one Cherenkov ring is found (one-ring), the particle identification is a muon 
($\mu$-like), and the reconstructed muon momentum is greater than 200 MeV/$c$. 
This last condition is to ensure the quality of the event reconstruction. 
After these cuts, the fraction of CCQE events is about 60\%.  See Table
\ref{tbl:ktsummary} for the data summary and CCQE fraction. 
The requirement of full containment in the 1KT suppresses
events at high momentum compared to the other detectors.
Figs.~\ref{fig:dlfct},
\ref{fig:pid}, and \ref{fig:pomax} show the ring number likelihood,
particle identification likelihood, and the fully contained versus
partially contained event (FC/PC) separation, respectively, used in
the FC1R$\mu$ event selection.  Any discrepancies between data and the MC
simulation observed are used to estimate some of the systematic errors in the 1KT
mentioned below.

\begin{table}
   \begin{center}
    \caption{The summary table for the number of observed
             events in the FC1R$\mu$ sample, the efficiency and
             the purity
             of the CCQE events estimated with the MC simulation.}
    \begin{tabular}{cccc}
     \hline
     \hline
      &~ \# of events &~ CCQE efficiency~(\%) &~ CCQE purity~(\%) \\
     \hline
     FC1R$\mu$    &~ 52110  &  53.7 &  57.9 \\
     \hline
     \hline
    \end{tabular}
    \label{tbl:ktsummary}
    \end{center}
 \end{table}

The selected data and the MC simulation are then binned into 2-dimensional distributions of 
muon momentum versus the scattering angle ($\theta_{\mu}$).  The
momentum is divided into 16 100 MeV/c bins from 0 - 1600 MeV/c.  The
scattering angle is divided into 10 bins where the first nine are in
increments of 10 degrees from 0$^{\circ}$-90$^{\circ}$ and the final
bin contains all events with the angle greater than 90$^\circ$.  
The bins for which the expected number of events is greater than 10
are used for the analysis (80 bins in total).

The neutrino spectrum is derived by comparing the observed data and
the weighted sum of Monte Carlo expectations using a $\chi^2$ test.
The $\chi^2$ is defined as:
\begin{eqnarray}
\chi^2_{KT} &=& \sum_{m,n} \frac{\left( N^{\rm obs}_{m,n} - N^{\rm MC}_{m,n} \right )^2}{\sigma^2_{m,n}} + \frac{\left( 1- \epsilon \right)^2}{\sigma^2_{energy}}
\end{eqnarray}
where $N^{\rm obs}_{m,n}$ is the number of observed events for data for $(m,n)$-th bin,
$N^{\rm MC}_{m,n}$ is the number of MC events as given by equation
\ref{eq:mc-exp}, $\epsilon$ is the fitting parameter for energy scale
($\epsilon$ = 1 is a nominal value) which scales the muon momentum,
$\sigma_{m,n}$ is the error including statistical and systematic
errors, and $\sigma_{energy}$ is the estimated uncertainty of the
energy scale, +3$/-$4\%.


The parameters $f^{\phi}_{i}$, $R_{\rm nQE}$, P, and $\epsilon$ are
the fitting parameters as discussed above.  

The systematic errors that are introduced in the fit come mainly from
the event selection.  Those errors are the ring counting likelihood,
the particle identification likelihood, the FC/PC cut, the event vertex and direction
reconstruction, and the energy scale.  The other systematic errors
are from the detector calibration and the axial vector mass ($M_{A}$), which
is used for our neutrino interaction model.

\subsection{Definition of $\chi^2$ for SciFi}
The SciFi tracking detector can observe charged particle tracks produced 
in neutrino-water interactions. Since SciFi uses the same target
material (water) as 1KT and SK, systematic uncertainty due to different
target nuclei is reduced.
The data taken during K2K-Ib and K2K-IIa periods
have been analyzed.

We choose charged current events in which a muon
track starts from an interaction in the SciFi fiducial volume 
and stops in the MRD detector. The fiducial region is defined as 
a rectangle 1.1 $m$ to each side of the detector's center
in both $x$ and $y$,  
covering the first  to 17$^{th}$ water containers. 
The fiducial mass is 5.59 $\pm$ 0.07 metric tons. 
The primary (muon) track should match 
a hit in the downstream scintillator hodoscope, with no matching hit in 
the upstream one.  It should also match a reconstructed track which has
at least penetrated one piece of steel and produced hits in two layers
in the MRD. 
The muon momentum threshold is 675 MeV/c for period-Ib and 550 MeV/c for
period IIa.

For the K2K-Ib data, to enhance sensitivity to low energy neutrino 
interactions, we also include events in which the muon
stops in the Lead Glass calorimeter (LG) and events which penetrate 
even single active layer in the MRD.  
Here, the response of the matched LG cluster should be greater than 
100 MeV, which prevents proton tracks from being identified as muon tracks.
By including these samples the $p_\mu$ threshold for period-Ib data
is reduced to 400 MeV/c.  

\begin{table}
\begin{tabular}{l|c|c}
\hline
\hline
      & Ib  & IIa \\
\hline
Detector Configuration & SciFi+LG+MRD & SciFi+MRD \\ 
$\rm{POT_{SciFi}}$ ($\times$10$^{18}$) & 39.70 & 22.04 \\
\hline
SciFi-MRD(track match)     & 6,935 & 5,188 \\
SciFi-MRD(cell match only) & 1,403 & 1,743$^*$\\
SciFi-LG    & 2,666 & --    \\
\hline
\hline
\end{tabular}
\caption{Number of events in each event category observed in SciFi. 
$^*$The events are not used for the spectrum reconstruction.}
\label{Tab.scifistat}  
\end{table}        
The number of events for each event category, together with the total POT 
of the beam spills used for the analyses, 
are summarized in Table~\ref{Tab.scifistat}.  
In total 17,935 charged current candidate events have been collected
for the data corresponding to $\rm{POT_{SciFi}}$=6.17$\times 10^{19}$ 
POT. After correcting for the changing detector configuration and efficiency, 
the event rate per POT was stable from month to month.

\begin{figure} [htpb]
 \begin{center}
  \includegraphics[width=8cm,clip]{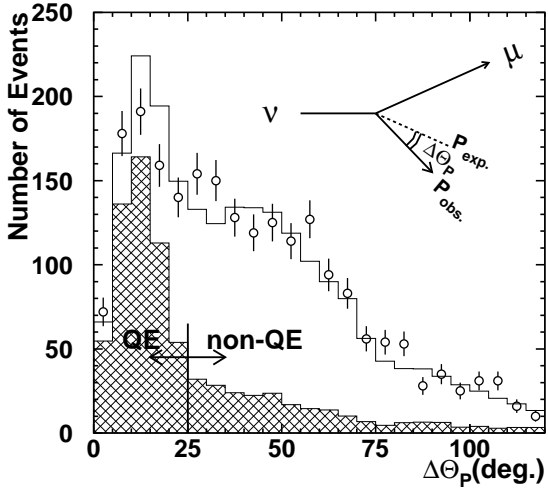}
 \end{center}
   \caption{Example of the distribution of $\Delta\theta_P$, the difference
between the observed and predicted (assuming QE interaction) angle of the
second track.  This distribution is for SciFi 2-track events for the K2K-I 
period.  The data (circles) and MC (histogram) are compared, and the shaded region shows the 
QE fraction in the MC.}
   \label{Fig.sfcosdth}
\end{figure}

In the spectrum reconstruction, 
there must be one or two tracks in the event, including the muon; events
with three or more tracks are discarded, which amounts to
about 3\% of the total.  The
two-track sample is further subdivided into a QE and a non
quasi-elastic sample.  For QE events, the muon momentum and muon angle are
sufficient to predict the angle of the recoil proton because of the 
simple two-body kinematics.
If the observed second track matches this prediction, within 25 degrees, then
it is likely the event is QE.  If the second track is more than 25 degrees
from the predicted angle, then it is very likely not a QE event, as shown
in Fig.~\ref{Fig.sfcosdth}.

\begin{table}
\begin{center}
\caption{QE efficiency and QE purity of all events for K2K-I and K2K-IIa for each SciFi subsample, estimated from the MC before fitting.}
\begin{tabular}{cccccc}
\hline
\hline
  &\multicolumn{2}{c}~ QE efficiency~(\%) &\multicolumn{2}{c}~ QE purity~(\%) \\
\hline 
 & K2K-I & K2K-IIa & K2K-I & K2K-IIa \\
1-track\hspace{0.2in} & 39 & 36 & 50 & 57 \\
QE & 5 & 5 & 53 & 58 \\
nonQE & 2 & 2 & 11 & 12\\
\hline total & 46 &  42 & &  \\
\hline
\hline
\end{tabular}
\label{tbl:scifisummary}
\end{center}
\end{table}

The above data sample gives three subsamples each for K2K-I and K2K-IIa.
Tab.~\ref{tbl:scifisummary} shows the QE efficiency (fraction of QE MC
events ending up in each subsample) and QE purity (fraction of
events in the subsample which are QE) for K2K-I and K2K-IIa.
The LG stopping events are separated from the rest of the K2K-I
data, giving another three subsamples -- a total of nine.  These samples
are then divided into seven angle bins from 0 to 60 degrees, in 10 degree
increments, and also into eight muon momentum bins, which have the same
intervals as the neutrino energy bins in
Tab.~\ref{tbl:enubin}.
The kinematics of these interactions are such
that the bins at both high angle and high momentum contain neither
data nor MC events and are not used in the analysis.  
The high muon thresholds also mean there are no data in the lowest 
energy region, and it is also not used in the fit.

In addition to the energy spectrum and R$_{nQE}$ parameters, the fit
includes several systematic parameters specific to the SciFi detector.
Three refer to uncertainties in the reconstructed energy of each
event.  Since the muon momentum is reconstructed from the range of the
muon, uncertainty in the material assay, and in the dE/dx used in the
GEANT Monte Carlo simulation, as well as the uncertainty in the
outgoing muon energy from a neutrino interaction, are important. A muon
energy scale parameter $P^{\mathrm{SF}}_{\mathrm{Escale}}$ 
with an uncertainty of $\pm$2.7\% is applied to
the measured muon momenta.
The second parameter $P^{\mathrm{SF}}_{\mathrm{LG-density}}$
accounts for an uncertainty of $\pm$ 5\% in the
muon energy loss specifically when it passes through the LG, and we consider this to be an uncertainty in the LG
density.  A final energy scale applies only to the energy
reconstructed from the visible energy clusters in the LG for LG stopping 
events.  
This is expressed as the parameter 
$P^{\mathrm{SF}}_{\mathrm{LG-cluster}}$
and is given as
an energy shift in GeV.  The uncertainty used in the fit for the LG
cluster energy calibration is $\pm$ 30 MeV.  The muon momentum for LG
stopping events is reconstructed only from the range in SciFi and the
cluster energy observed in the LG.

There are two other systematic parameters:  a migration between one-track
and two-track events,  $P^{\mathrm{SF}}_{\mathrm{2nd-track-eff}}$,
which accounts for the tracking efficiency
for short, second tracks; and a migration between the two-track QE and
non-QE samples, $P^{\mathrm{SF}}_{\mathrm{rescattering}}$,
to account for nuclear effects such as the uncertainty in
the re-scattering of the proton as it leaves the nucleus from a QE
interaction.  Both parameters are simple linear migrations, and we take a
5\% uncertainty in the former and a 2.5\% uncertainty in the latter.  
They move a fraction of events from a particular $p_\mu$ and $\theta_\mu$
bin from one of the three subsamples to the same bin in another of the
three subsamples.  

There are 286 bins of SciFi data included in the fit, and six
SciFi specific systematic parameters  including normalization $P^{\mathrm{SF}}_{\mathrm{Norm}}$.
Our $\chi^2$ is defined as the negative of the logarithm
of the Poisson likelihood for the binned data, plus $\chi^2$
terms arising from the pull of five systematic errors (but not
normalization):
\begin{eqnarray}
\nonumber \chi^2 & = & -2 \ln \lambda(\theta) \\ 
\nonumber & = & 2 \sum_{m,n} [ N^{\mathrm{MC}}_{m,n}(\theta) - N^{\mathrm{obs}}_{m,n}
+ N^{\mathrm{obs}}_{m,n} \ln (N^{\mathrm{obs}}_{m,n} / N^{\mathrm{MC}}_{m,n}(\theta))] \\
\nonumber & + & \chi^2_{\mathrm{Escale}} + \chi^2_{\mathrm{LG\ density}}
 + \chi^2_{\mathrm{LG\ cluster\ energy}} \\
& + &  \chi^2_{\mathrm{2nd\ track\ eff}} + \chi^2_{\mathrm{Rescattering}}.
\end{eqnarray}
in which $N^{\mathrm{MC}}_{m,n}(\theta)$ and $N^{\mathrm{obs}}_{m,n}$ are the predicted and observed values
in the ({\em m,n})-th bin for some values of the parameters $\theta$. 
This is the simplified version given in Ref.~\cite{Eidelman:2004wy}\ 
and its minimum follows a $\chi^2$ distribution.
To this, we add $\chi^2$ terms arising from the systematic errors.

\subsection{Definition of $\chi^2$ for SciBar}
 The CC event selection in the SciBar detector is similar to that of the SciFi detector.
The fiducial volume of SciBar is defined as a rectangle that extends $2.6 \times 2.6$~m$^2$ around the beam axis, from the second to 53$^{rd}$ layer of scintillator.
The fiducial mass is 9.38~tons.
 Events with any track starting in the SciBar fiducial 
volume and matched with a track or hits in the MRD are selected 
as CC candidates~(SciBar-MRD sample).
 This requirement imposes a threshold of 450~MeV/c on 
$p_{\mu}$, reconstructed from its range through SciBar and MRD. 
 According to the MC simulation, 98\% of the events selected by this
requirement are CC induced events, and the rest are neutral current~(NC)
interactions accompanied by a charged pion or proton which penetrates 
into the MRD.
 The $p_{\mu}$ resolution, $p_{\mu}$ scale uncertainty and
$\theta_{\mu}$ resolution are 80~MeV/c,~2.7\% and 1.6~degrees,
respectively.

Events with one or two reconstructed 
tracks are selected from the SciBar-MRD sample.
 The two-track events are subdivided into two 
categories --- QE samples and nonQE samples --- by using kinematic 
information, $\Delta \theta_{p}$, like SciFi.
 The number of observed events and the fraction and efficiency of QE
interactions estimated with the MC simulation for each event category
are summarized in Table~\ref{tbl:evesummary}.

\begin{table}
   \begin{center}
    \caption{Number of observed 
             events in each event category and the efficiency and purity 
             of the QE events estimated with the MC simulation for the SciBar detector.}
    \begin{tabular}{cccc}
     \hline
     \hline        
      &~ \# events &~ QE efficiency~(\%) &~ QE purity~(\%) \\
     \hline
     1-track    &~ 7256  &  50.0 &  57.8 \\  
       QE       &~ 1760  &  15.4 &  71.3 \\
      nQE       &~ 2014  &   3.7 &  15.9 \\
     \hline
     Total      &~ 11030 &  69.1 &  ---  \\
     \hline
     \hline
    \end{tabular} 
    \label{tbl:evesummary}
    \end{center}
 \end{table}

 The SciBar term of the $\chi^{2}$ consists of two components,
\begin{eqnarray}
 \chi^{2}_{\mathrm{SB}} = \chi^{2}_{\mathrm{dist}} + \chi^{2}_{\mathrm{syst}}.
\end{eqnarray}

 The $\chi^{2}_{\mathrm{dist}}$ is calculated by the binned
likelihood method using the $(p_{\mu}, \theta_{\nu})$ distribution.
 The bin widths are 0.1~GeV/c for $p_{\mu}$ and 10~degrees 
 for $\theta_{\mu}$.
 Bins with the expected number of events greater than five are used for the fit.
 In total, 239 bins are used for the analysis.
The number of observed events in each bin, $N^{\mathrm{obs}}_{m,n}$, is assumed to 
follow a Poisson distribution with the mean $N^{\MC}_{m,n}$.
A bin-by-bin systematic uncertainty on hadron contamination is 
implemented by the convolution of a Poisson with a Gaussian distribution.
The $\chi^{2}_{\mathrm{dist}}$ is thus defined as:
\begin{eqnarray}
&&\chi^{2}_{\mathrm{dist}} =
 -2\sum_{m,n}\ln\frac{\mathcal{L}\left(N^{\mathrm{obs}}_{m,n}; 
                                    N^{\MC}_{m,n};\mathbf{\sigma}\right)}
                                 {\mathcal{L}\left(N^{\mathrm{obs}}_{m,n}; 
                                    N^{\mathrm{obs}}_{m,n};\mathbf{\sigma}\right)}, \\
&& \nonumber \\
&&      \mathcal{L}\left(N^{\mathrm{obs}}_{m,n};
      N^{\MC}_{m,n}; \mathbf{\sigma}\right) 
       \equiv \nonumber \\
&&  ~~~\prod_{m,n}\int^{\infty}_{0}\frac{1}{\sqrt{2\pi}\sigma_{m,n}}
      \exp\left[-\frac{\left(x -
			N^{\MC}_{m,n}\right)^{2}}{2\sigma^{2}_{m,n}}\right] 
      \cdot 
      \frac{x^{N^{\mathrm{obs}}_{m,n}}e^{-x}}{N^{\mathrm{obs}}_{m,n}!}dx. \nonumber \\
\end{eqnarray}

 The normalization parameter of each sample 
is given by
\begin{eqnarray}
 P^{\mathrm{Norm}}_{\mathrm{1track}} &=&
 P^{\mathrm{SB}}_{\mathrm{Norm}} \\
 P^{\mathrm{Norm}}_{\mathrm{QE}} &=&
 P^{\mathrm{SB}}_{\mathrm{Norm}} \cdot
 \Ptrack  \\ 
 P^{\mathrm{Norm}}_{\mathrm{nonQE}} &=&
 P^{\mathrm{SB}}_{\mathrm{Norm}} \cdot
 \Ptrack  \cdot
 \Pdtheta
\end{eqnarray}
 where $P^{\mathrm{SB}}_{\mathrm{Norm}}$ is the overall
normalization factor, and $\Ptrack$ and
$\Pdtheta$ are the parameters to vary
the ratio of the number of 2-track events to that of 1-track events 
 and the ratio of the number of CC-nonQE events to that of CC-QE events, respectively, 
within their systematic uncertainties.
We consider the uncertainties from nuclear effects and detector systematics.
The possible difference between carbon and oxygen is included in the nuclear effect uncertainty.
Because the nuclear effects are common source of the uncertainties 
for $\Ptrack$ and $\Pdtheta$, their correlation is also estimated.
In addition, $ \Ppscale$ is introduced to 
account for the uncertainty of the energy scale of the muon reconstruction.
The momentum scale uncertainty is 2.7\% as described above.

  The $\chi^{2}_{\mathrm{syst}}$ is calculated with constraint
parameters, including their correlation:
\begin{eqnarray} 
 \chi^{2}_{\mathrm{syst}} &=& (\mbox{\boldmath $P_{syst}$} -
 \mbox{\boldmath $P_{0}$})^{t}
\mbox{\boldmath $V^{-1}$}(\mbox{\boldmath $P_{syst}$} -
 \mbox{\boldmath $P_{0}$})
\end{eqnarray}

\noindent
where $\mbox{\boldmath $P_{\mathrm{syst}}$} $ represents a set of 
systematic parameters, $\mbox{\boldmath
 $P_{0}$}$ are their nominal values, and $\mbox{\boldmath $V$}$ is a 
covariance matrix.     
Three systematic parameters ,
$\Ppscale$, $\Ptrack$, and $\Pdtheta$ are included in 
$\mbox{\boldmath $P_{\mathrm{syst}}$} $;
they are defined as relative weighting factors to the 
nominal MC expectation and all components of 
$\mbox{\boldmath $P_{0}$}$ are set to unity.
The uncertainties and correlation among the parameters are evaluated
to be 
\begin{equation}
  \mathbf{V} \equiv \bordermatrix{            & \Ppscale  & \Ptrack & \Pdtheta \cr
                           \Ppscale  & +(0.027)^2 & 0 & 0 \cr
                           \Ptrack   & 0 & +(0.059)^2 & +(0.017)^2 \cr
                           \Pdtheta  & 0 & +(0.017)^2 & +(0.058)^2}  .
\end{equation}
The dominant error sources are track
finding efficiency for $\Ptrack$ and the nuclear effects uncertainty for
$\Pdtheta$.

\subsection{Fit results}
\begin{figure} []
 \begin{center}
  \resizebox{6.0cm}{!}{
   \includegraphics[trim=0cm 0.0cm 0cm
   0cm,clip]{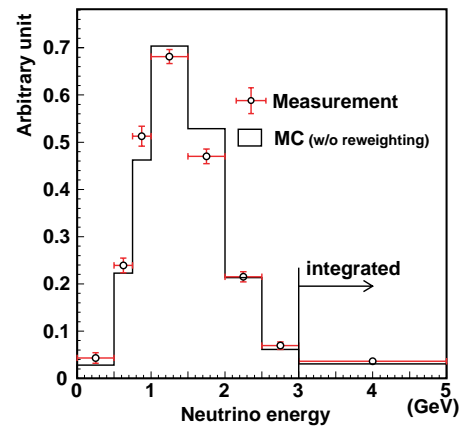}}
   \caption{The neutrino energy spectrum measured at the near site, assuming CC-QE.
   The expectation with the MC simulation  without reweighting is also shown.}
   \label{fig:merged-resultwBeamMC}
 \end{center}
\end{figure}

 The minimum $\chi^{2}$ point in the multi-parameter space
is found by changing the spectrum shape, $R_{\mathrm{nQE}}$ and
the systematic parameters, where the MINUIT program
library~\cite{James:1975dr} is employed. 
 The central values and the errors of the fitting parameters
are summarized in Table~\ref{tbl:merged-result}.
 All the systematic parameters stay within their estimated errors.
 The result of the spectrum measurement is shown in
Fig.~\ref{fig:merged-resultwBeamMC} with the prediction of the
beam MC simulation.

 The results of the measurements with individual detector data are also shown
in Table~\ref{tbl:merged-result}.
In the fit with only 1KT data, the energy spectrum 
parameters are fixed to their default values for the high energy region 
where there is little or no acceptance.
For the same reason, the low energy region is fixed for SciFi and SciBar.
 All the fitting parameters are in good agreement within their errors each
other except for $R_{\mathrm{nQE}}$.

\begin{figure*} []
 \begin{center}

\includegraphics[height=0.28\textheight]{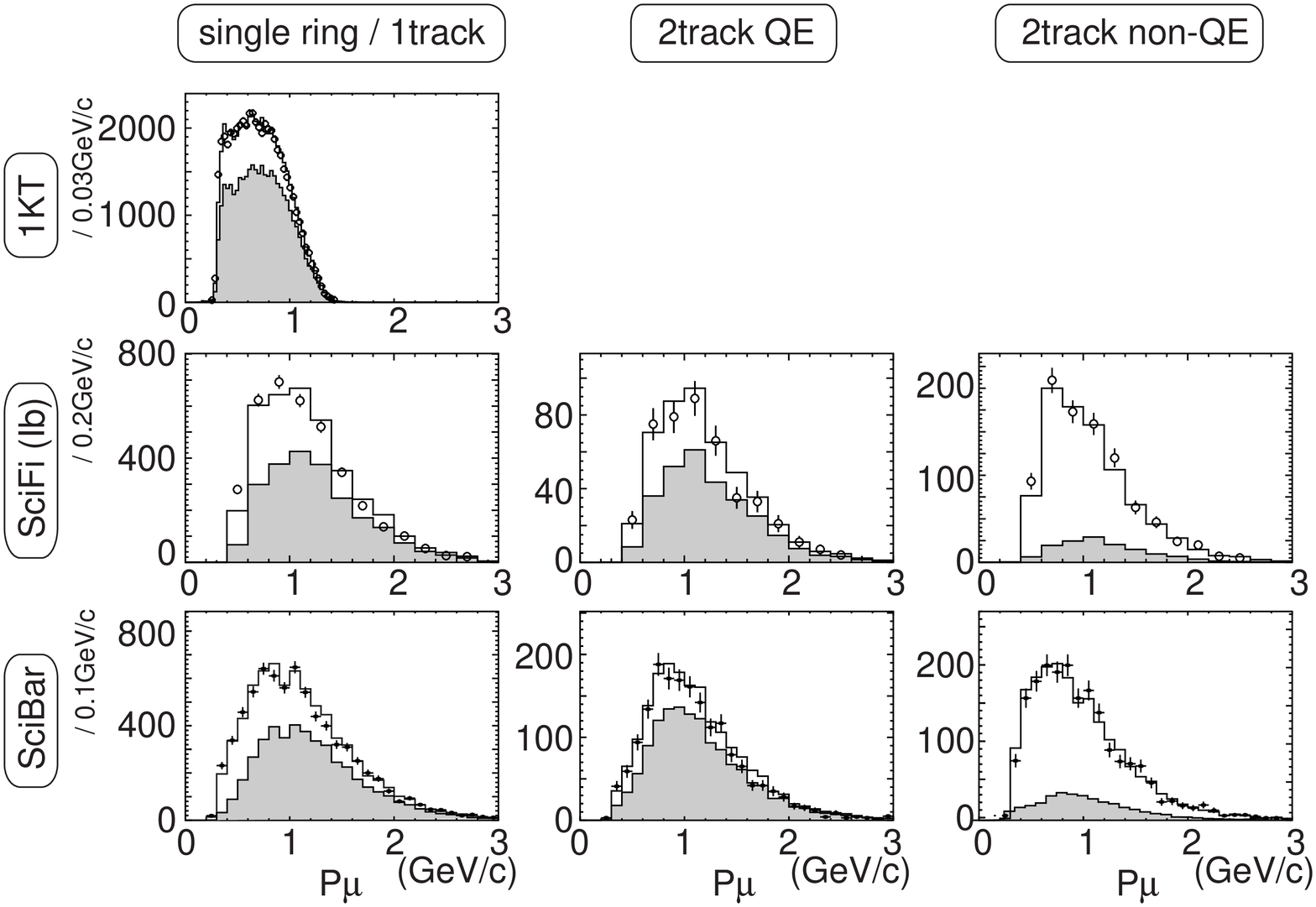}
  \caption{The $p_{\mu}$ distributions for each event sample of  all
  near detectors with the MC simulation after fitting, given by open
  histograms.  
   The hatched areas are the CCQE components in the MC distributions.
  }
  \label{fig:pmu-after}


\includegraphics[height=0.28\textheight]{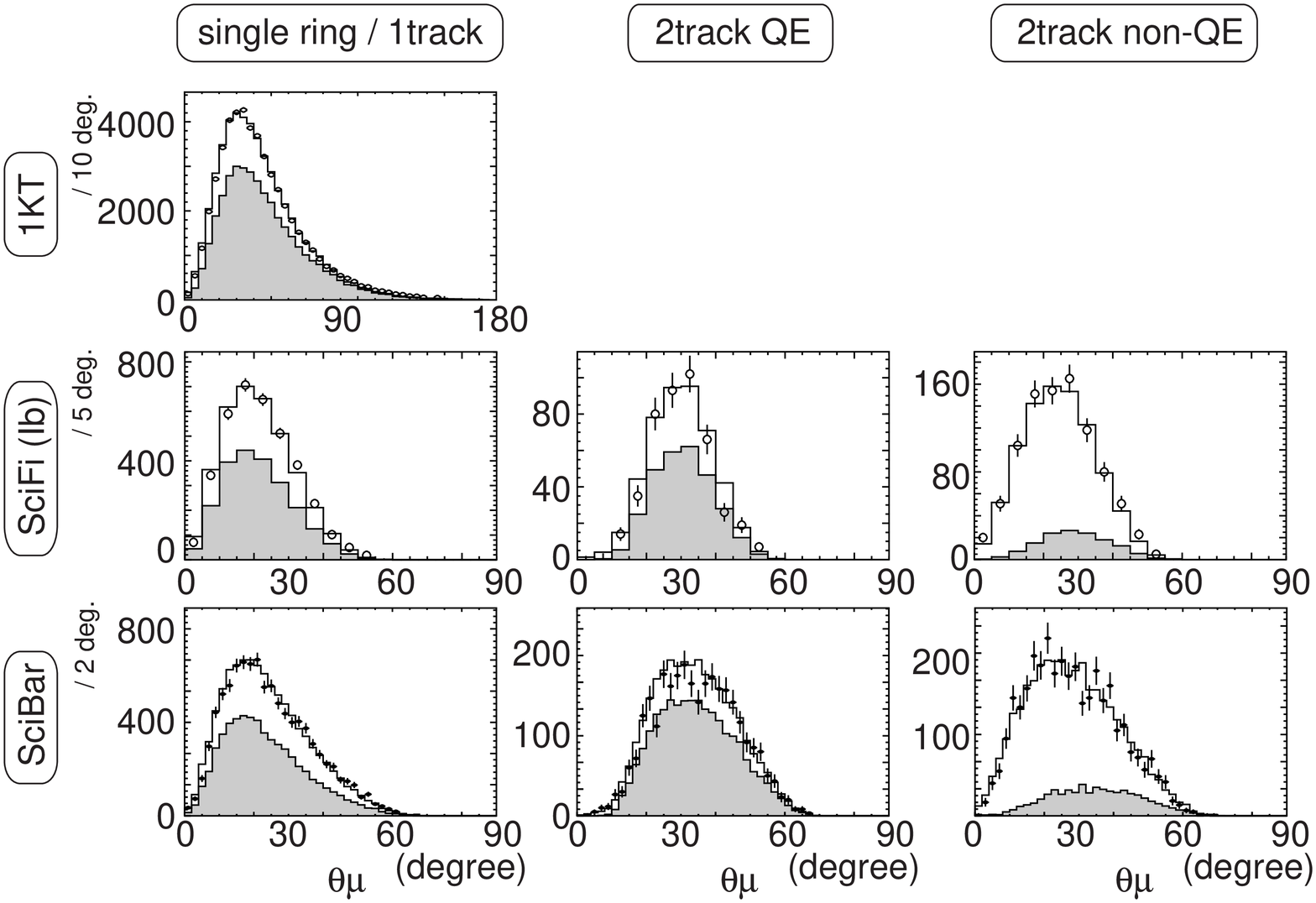}
  \caption{The $\theta_{\mu}$ distributions for each event sample 
  of all near detectors with the MC simulation after fitting, 
  given by open histograms.  
   The hatched areas are the CCQE components in the MC distributions.
  }
  \label{fig:qmu-after}

\includegraphics[height=0.28\textheight]{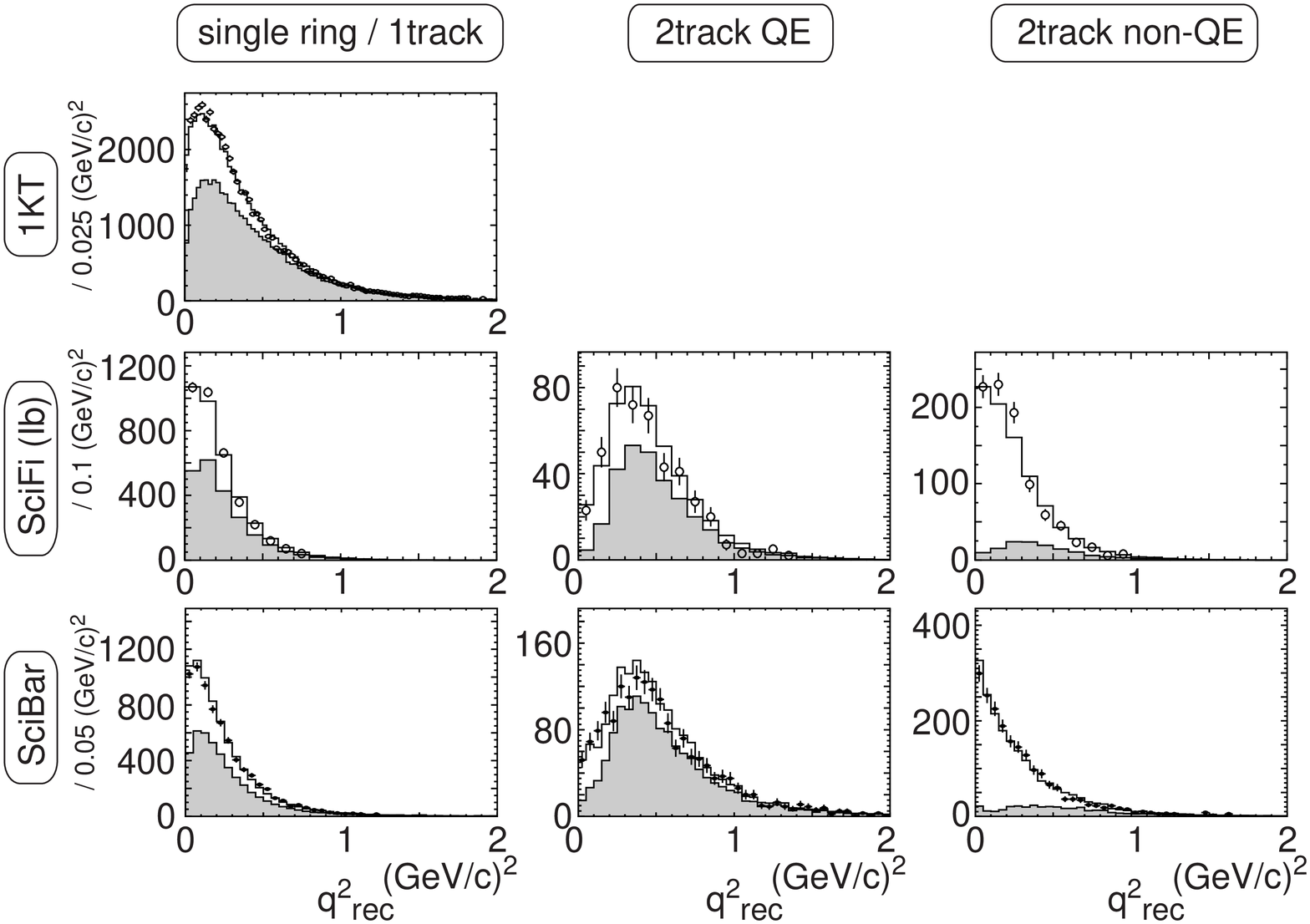}
  \caption{The $q^{2}_{\mathrm{rec}}$ distributions for each event
  sample of all near detectors with the MC simulation after fitting, 
  given by open histograms.  
   The hatched areas are the CCQE components in the MC distributions.
  }
  \label{fig:q2-after}
 \end{center}
\end{figure*}

 The $p_{\mu}$, $\theta_{\mu}$ and $q^{2}_{\mathrm{rec}}$ distributions
for the 1KT, SciFi and SciBar samples 
are shown in Fig.~\ref{fig:pmu-after}--\ref{fig:q2-after}.  In these
figures, the reconstructed $Q^2$ distributions($q^{2}_{\mathrm{rec}}$) are
constructed by assuming that the interaction was CC-QE and
using the reconstructed energy under this assumption.  The expected
distributions of the MC simulation with the best-fit parameters are
also shown.  The MC simulation reproduces all the distributions well.

 The discrepancy in $R_{\mathrm{nQE}}$ is treated as a systematic error.
 However, the value of $R_{\mathrm{nQE}}$ is strongly correlated with
the $E_{\nu}$ spectrum as well as the other systematic parameters such as
$\Pdtheta$.
In order to evaluate $R_{\mathrm{nQE}}$ with each detector data set
under identical fitting condition, a second fit is performed.
 In the second fit, the $E_{\nu}$ spectrum and the systematic
parameters, except for the overall normalization, are
fixed at the best fit values obtained with all the three detectors.
 The best fit value of $R_{\mathrm{nQE}}$ for each detector 
 in the second fit is (1KT, SciFi, SciBar) = (0.76, 0.99, 1.06), respectively,
while the fit result with three detectors is 0.96.
 Therefore, an additional error of 0.20 is assigned 
to $R_{\mathrm{nQE}}$ in order to account for the discrepancy.

The errors of the measurement are provided in the form of an
error matrix.
 Correlations between the parameters
 are taken into account in the oscillation analysis with this matrix.
 The full elements in the error matrix are shown in
Table~\ref{tbl:merged-error-matrix}.

\begin{table*}
   \begin{center}

    \caption{Results of the spectrum measurement. 
    The best fit value of each parameter is listed for the fits 
    with all the detectors' data,
    with the 1KT data, with the SciFi data and with the SciBar data,
    respectively.
    The reduced $\chi^{2}$~($\chi^{2}_{\mathrm{total}}$/DOF)
    and the averaged $\chi^{2}$ of each
    detector~($\chi^{2}$/N$_{\mathrm{bin}}$) are also shown.}

    \begin{tabular}{|l|c|ccc|}
     \hline
     parameter  & {\bf Combined} & 1KT only & SciFi only & SciBar only \\
     \hline        
     \hline 
     $f_{1}$ (0.00-0.50 GeV) & $\mathbf{1.657 \pm 0.437}$ 
        & $ 2.372 \pm 0.383 $  & $ \equiv 1$  & $ \equiv 1$   \\
     $f_{2}$ (0.50-0.75 GeV) & $\mathbf{1.107 \pm 0.075}$ 
        & $ 1.169 \pm 0.072 $  & $ 0.882 \pm 0.317$ & $ 1.166 \pm 0.251
     $ \\
     $f_{3}$ (0.75-1.00 GeV) & $\mathbf{1.154 \pm 0.061}$ 
        & $ 1.061 \pm 0.065 $  & $ 1.157 \pm 0.201$ & $ 1.145 \pm 0.134
     $   \\
     $f_{4}$ (1.00-1.50 GeV) & $\mathbf{\equiv 1}$ 
        & $ \equiv 1 $ & $ \equiv 1 $ & $ \equiv 1 $  \\
     $f_{5}$ (1.50-2.00 GeV) & $\mathbf{0.911 \pm 0.044}$ 
        & $ 0.709 \pm 0.151$   & $ 0.980 \pm 0.107$ & $ 0.963 \pm 0.070
     $   \\
     $f_{6}$ (2.00-2.50 GeV) & $\mathbf{1.069 \pm 0.059}$ 
        & $\mathbf{\equiv 1}$   & $ 1.188 \pm 0.096$ & $ 0.985 \pm 0.086$
     \\
     $f_{7}$ (2.50-3.00 GeV) & $\mathbf{1.152 \pm 0.142}$ 
        & $\mathbf{\equiv 1}$   & $ 1.062 \pm 0.230$ & $ 1.291 \pm 0.283$
     \\
     $f_{8}$ (3.00-     GeV) & $\mathbf{1.260 \pm 0.184}$ 
        & $ \equiv 1 $    & $ 1.323 \pm 0.203 $ & $ 1.606 \pm 0.749 $\\
     $R_{\mathrm{nQE}}$ & $\mathbf{0.964 \pm 0.035}$ 
        & $ 0.589 \pm 0.071 $  & $ 1.069 \pm 0.060 $ & $ 1.194 \pm
     0.092$ \\
     \hline
     \hline
     $\mathrm{P}^{\mathrm{1kt}}_{\mathrm{Norm}}$    
         & $\mathbf{0.948 \pm 0.024}$ & $ 1.172 \pm 0.046 $  & --- & ---
     \\
     $\mathrm{P}^{\mathrm{1kt}}_{\mathrm{energy}}$ 
         & $\mathbf{0.984 \pm 0.004}$ & $ 0.993 \pm 0.007 $  & --- & ---
     \\
     \hline
     $\mathrm{P}^{\mathrm{SF}}_{\mathrm{Norm}}$       
         & $\mathbf{1.009 \pm 0.029}$ & --- & $0.925 \pm 0.058 $ & ---    \\
     $\mathrm{P}^{\mathrm{SF}}_{\mathrm{Escale}}$     
         & $\mathbf{0.980 \pm 0.006}$ & --- & $0.980 \pm 0.007 $ & ---    \\
     $\mathrm{P}^{\mathrm{SF}}_{\mathrm{LG-density}}$ 
         & $\mathbf{0.929 \pm 0.012}$  & --- & $ 0.928 \pm 0.012$ & ---
     \\
     $\mathrm{P}^{\mathrm{SF}}_{\mathrm{LG-cluster}}~[\mathrm{GeV}]$ 
        & $\mathbf{-0.001 \pm 0.002}$ & --- & $ -0.002 \pm 0.003$ & ---
     \\
     $\mathrm{P}^{\mathrm{SF}}_{\mathrm{2nd-track-eff}}$    
         & $\mathbf{0.959 \pm 0.014}$  & --- & $ 0.932 \pm 0.017$ & ---
     \\
     $\mathrm{P}^{\mathrm{SF}}_{\mathrm{rescattering}}$  
         & $\mathbf{1.048 \pm 0.055}$  & --- & $ 0.993 \pm 0.062$& ---    \\
     \hline

     $\mathrm{P}^{\mathrm{SB}}_{\mathrm{Norm}}$  
         & $\mathbf{0.998 \pm 0.010}$  & --- & --- & $ 1.003 \pm 0.011
         $\\
     $\mathrm{P}^{\mathrm{SB}}_{\mathrm{p-scale}}$    
         & $\mathbf{0.976 \pm 0.004}$  & --- & --- & $ 0.972 \pm 0.004
         $\\
     $\mathrm{P}^{\mathrm{SB}}_{\mathrm{2trk/1trk}}$    
         & $\mathbf{0.953 \pm 0.021}$  & --- & --- & $ 0.961 \pm 0.023
         $\\
     $\mathrm{P}^{\mathrm{SB}}_{\mathrm{nonQE/QE}}$  
         & $\mathbf{1.066 \pm 0.032}$  & --- & --- & $ 0.978 \pm 0.040 $\\
     \hline
     \hline
     $\chi^{2}_{\mathrm{total}}$/DOF  & {\bf 687.2 / 585} & 46.8 / 73 
        & 328.7 / 273 &  253.3 / 228 \\
     $\chi^{2}_{\mathrm{1kt}}$/N$_{\mathrm{bin}}$    
         & {\bf  85.4 / 80} & 47.7 / 80& --- & --- \\        
     $\chi^{2}_{\mathrm{SciFi}}$/N$_{\mathrm{bin}}$  
         & {\bf 335.6 / 286} & --- & 328.7 / 286 & ---  \\        
     $\chi^{2}_{\mathrm{SciBar}}$/N$_{\mathrm{bin}}$ 
         & {\bf 266.1 / 239}& ---  & --- & 253.3 / 239 \\        
     \hline
    \end{tabular} 
    \label{tbl:merged-result}
    \end{center}
\end{table*}

\begin{table}
   \begin{center}
    \caption{ The  
             error matrix for
             $f_{\mathrm{i}}$ and $R_{\mathrm{nQE}}$. 
              The square root of error
             matrix (sign $\left[{M_{ij}}\right] \cdot \sqrt{|M_{ij}|}$) 
             is shown here in the unit of \%.}
    \begin{tabular}{lccccccccc}
     \hline
        & $f_{1}$ & $f_{2}$ & $f_{3}$ & $f_{5}$ & $f_{6}$ &
          $f_{7}$ & $f_{8}$ & $R_{\mathrm{nQE}}$  \\
     \hline
      $f_{1}$ & 43.86 & -3.16 &  7.28 & -2.21 & -0.76 & 
        -3.48 &  0.81 & -8.62 \\
      $f_{2}$ & -3.16 &  7.51 &  1.97 &  1.90 &  0.62 &  
         1.29 &  2.43 & -5.68 \\
      $f_{3}$ &  7.28 &  1.97 &  6.00 &  3.38 &  1.63 &  
         3.44 &  1.71 & -2.99 \\
      $f_{5}$ & -2.21 &  1.90 &  3.38 &  4.04 & -1.86 & 
          4.53 & 2.20 &  1.65 \\
      $f_{6}$ & -0.76 &  0.62 &  1.63 & -1.86 & 5.28  &
         -5.85 & 5.11 &  0.94 \\
      $f_{7}$ & -3.48 &  1.29 &  3.44 &  4.53 &-5.85  &
         13.67 &-10.14&  4.09 \\
      $f_{8}$ &  0.81 &  2.43 &  1.71 & 2.20 &  5.11 &-10.14 & 
         18.35 &-11.77 \\
      $\Rnqe$ & -8.62 & -5.68 & -2.99 & 1.65 &  0.94 & 4.09  &  
        -11.77 & 20.30 \\
     \hline
    \end{tabular} 
    \label{tbl:merged-error-matrix}
    \end{center}
 \end{table}

\section{SK data}
\subsection{SK data reduction}
\label{sec:skreduc}

Beam-induced neutrino events in SK are selected according
to criteria described in this section.
Following selection, events which are fully contained in the SK
fiducial volume are reconstructed using similar methods to the
SK atmospheric neutrino analysis and then used in the K2K
oscillation analysis.

In order to select those neutrino interactions which come from the
accelerator at KEK, two Universal Time Coordinated time stamps from the
worldwide GPS system are compared. Both $T_{KEK}$ for the KEK-PS beam
spill start time, and $T_{SK}$ for the SK trigger time are recorded.
The time difference $\Delta T = T_{SK} - T_{KEK} -TOF$, where $TOF$ is
a time of flight, is distributed from 0 and 1.1 $\mu sec$ matching the
timing width of the beam spill of the KEK-PS.  The maximum difference of the
synchronization between two sites is measured to be less than 200~ns
by using an external atomic clock. For this reason we require the
$\Delta T$ for selected events to be between -0.2 to 1.3 $\mu$ sec.

\begin{table}
   \begin{center}
    \caption{SK event reduction summary.}
    \begin{tabular}{lrr}
     \hline        
     Reduction step & K2K-I & K2K-II \\
     \hline
     ${\rm |\Delta T|<500\mu sec}$,& 107892 & 470469 \\
     no pre-activity               &        & \\
     total number of p.e. within   &  36560 &  16623 \\
     300 n sec timing window       &~ &~ \\
     $>$200(K2K-I),94(K2K-II)       &~ &~ \\
     Fully contained event         &    153 &     99 \\
     flasher cuts                  &      97 &     88 \\
     visible Energy $>$30MeV         &    95 &     85 \\
     fiducial volume cut           &     56 &     59 \\
     ${\rm |\Delta T|=-0.2 - 1.3 \mu sec}$ & 55 & 57 \\
     \hline
    \end{tabular} 
    \label{tbl:SKredsummary}
    \end{center}
 \end{table}
 
\begin{table}
   \begin{center}
    \caption{SK event summary. For oscillated expectations,
${\rm sin^22\theta=1}$ and ${\rm \Delta m^2=2.8\times10^{-3}eV^2}$ are assumed.}
    \begin{tabular}{l|r|r|r|r|r|r}
     \hline        
     ~ & \multicolumn {3} {c|} {K2K-I} & \multicolumn {3} {c} {K2K-II} \\
     \hline
     ~ & data & \multicolumn {2} {c|} {expected} & data & \multicolumn {2} {c} {expected} \\
     ~ & ~ & w/o osc. & w/ osc. & ~ & w/o osc. & w/ osc. \\
     \hline
     Fully contained          & 55 & 80.8 & 54.8 & 57 & 77.3 & 52.4 \\
     ~~1-ring                 & 33 & 51.0 & 31.1 & 34 & 49.7 & 30.5 \\
     ~~~~$\mu$-like           & 30 & 47.1 & 27.7 & 28 & 45.2 & 26.7 \\
     ~~~~e-like               &  3 &  3.9 &  3.4 &  6 &  4.5 &  3.8 \\
     ~~multi-ring             & 22 & 29.8 & 23.7 & 23 & 27.6 & 21.9 \\
     \hline
    \end{tabular} 
    \label{tbl:SKeventsummary}
    \end{center}
 \end{table}
 
\begin{figure} [htbp]
 \begin{center}
  \includegraphics[width=8cm]{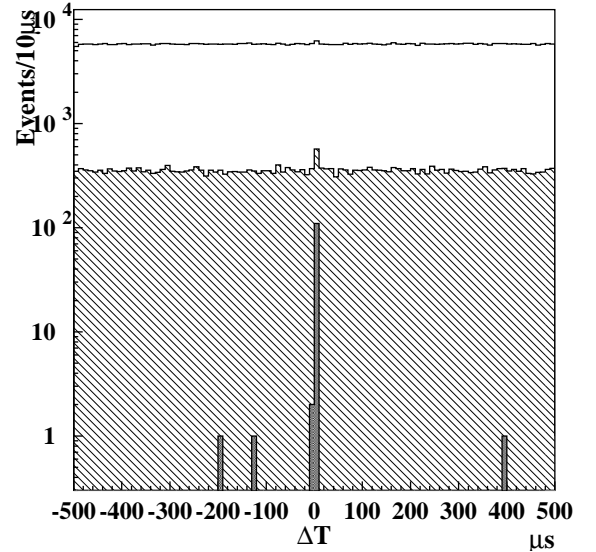}
  \caption{The $\Delta T$ distribution at each reduction step.  Clear,
    hatched and shaded histograms are after pre-activity cut, total
    p.e. cut, and fiducial volume cut, respectively.}
   \label{fig:SKtdiff}
 \end{center}
\end{figure}

\begin{figure} [htbp]
 \begin{center}
  \includegraphics[width=8cm]{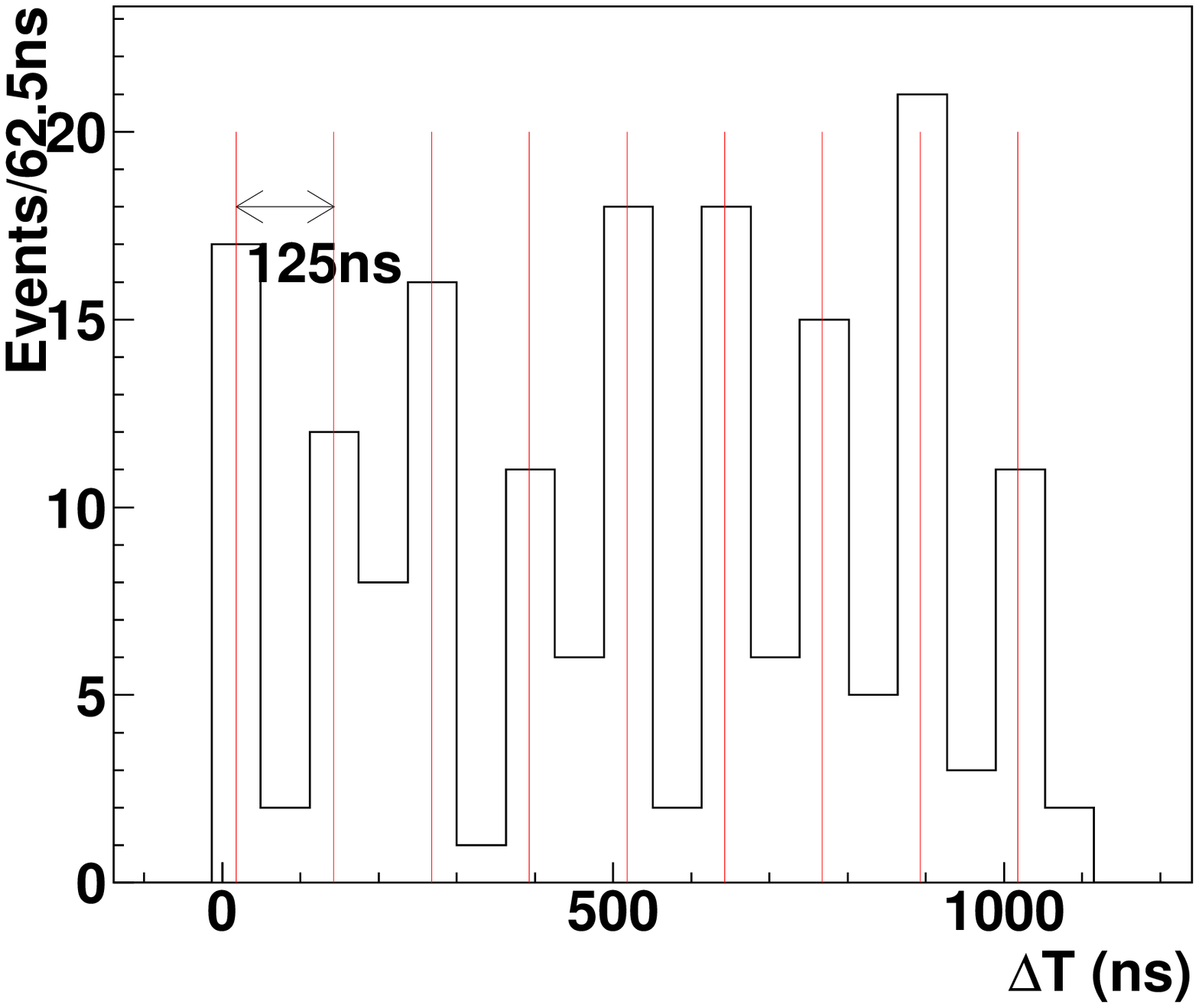}
  \caption{The $\Delta T$ distribution for fully contained events.
    The nine micro-bunch structure present in the beam is clearly
    seen.}
   \label{fig:SK9bunch}
 \end{center}
\end{figure}

In addition to the timing criteria, the following cuts are required:

\begin{enumerate}
\item In order to remove decay-electrons from the sample, events must
  have no activity in the  30~$\mu$s before the event.
\item There must be a minimum number of photo-electrons seen within a
  300~ns timing window.  The required number of photo-electrons are
  200 for K2K-I and 94 for K2K-II.
\item Fully contained events are selected by requiring 
  no activity in the outer detector. 
\item A selection is made to remove events with PMTs which sometimes
  begin to produce light because of a discharge around the dynode.
  These events have easily identified characteristics such as a timing
  distribution which is much broader than neutrino events, and a
  repeating pattern of light in the detector.
\item At least 30~MeV energy must be deposited in the inner detector.
\item The events are selected to come from the 22.5~kiloton fiducial
  volume by requiring the reconstructed vertex position be at least 2m
  away from the inner detector wall.
\end{enumerate}

Tab.~\ref{tbl:SKredsummary} shows the reduction summary for K2K-I and
K2K-II.  
The efficiency for these cuts are 77.2\% for
K2K-I and 77.9\% for K2K-II. The majority of the  inefficiency is due
to NC interactions which are not selected by these criteria.
In total, 112 accelerator produced, fully contained events, are observed
in the SK fiducial volume, with 58 events reconstructed as 1-ring
$\mu$-like.  
Tab.~\ref{tbl:SKeventsummary} summarizes the characterization
of these events and the MC expectations with and without
neutrino oscillation.

\begin{figure}[!htbp]
 \begin{center}
  \includegraphics[width=8cm]{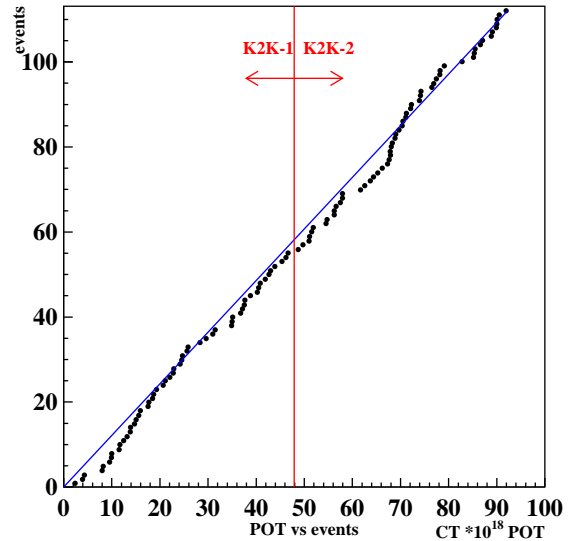}
  \caption{Event rate as a function of POT. The KS-test probability to
    observe our event rate under the assumption that the event rate is
    proportional to POT is 79\%.}
   \label{fig:SKct}
 \end{center}
\end{figure}

Figure \ref{fig:SKtdiff} shows the $\Delta T$ distribution at each
reduction step.  A clear peak at $\Delta T=0$ is seen after the
fiducial volume cut.  Three off-timing fully contained events are
observed within ${\rm \pm 500\mu sec}$ timing window which is
consistent with the 2 expected background events from
atmospheric neutrinos.  In addition, the nine-bunch timing structure
of the beam can be clearly seen in the $\Delta T$ distribution if finer
bins are used as in Fig.~\ref{fig:SK9bunch}.  Fig.~\ref{fig:SKct}
shows the event rate as a function of POT.  A KS-test performed
against the assumption that the rate is proportional to POT gives a
probability of 79\%.

\begin{figure}[!htbp]
 \begin{center}
  \includegraphics[width=8cm]{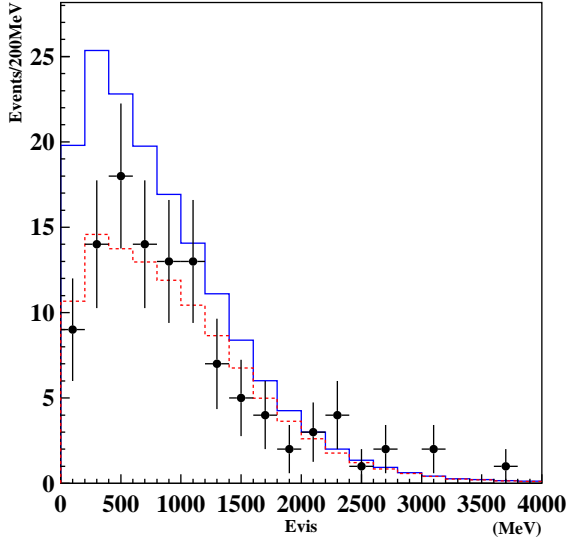}
  \caption{The visible energy distribution for fully contained
    fiducial-volume events in SK. The closed circles are the 
    observed data. The solid histogram is the MC expectation based on 
    the ND measurement without neutrino oscillation, and the dashed
    one is the MC expectation with neutrino oscillation of
    ${\rm sin^22\theta=1}$ and ${\rm \Delta m^2=2.8\times10^{-3}eV^2}$.}
   \label{fig:SKevis}
 \end{center}
\end{figure}

The energy distribution of the events is compared against expectation
in several ways.  Fig.~\ref{fig:SKevis} shows the visible energy
distribution, which is estimated from the energy deposit in the inner
detector for all of the fully contained fiducial volume events.  In
this figure, the observed data is compared with the 
MC expectation based on the ND measurement without neutrino oscillation.

\begin{figure} [htpb]
 \begin{center}
  \resizebox{8cm}{!}{
   {\includegraphics[trim=0cm 0.0cm 0cm
    0.cm,clip]{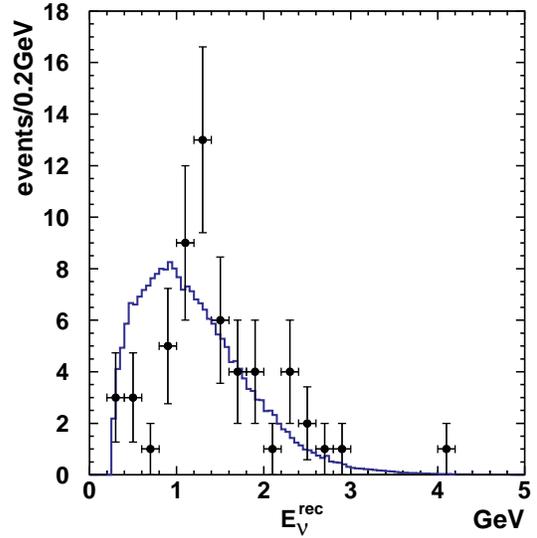}}}
  \caption{ The reconstructed $E_{\nu}$ distribution for the SK 1-ring
           $\mu$-like sample. Points with error bars are data. 
            The solid line is the expectation without oscillation.
            The histogram is normalized by the number of events 
           observed~(58).}
   \label{fig:Enunull}
 \end{center}
\end{figure}

Figure~\ref{fig:Enunull} shows the expected energy spectrum together
with the observation for the one-ring $\mu$-like events. 
The expectation is normalized by the number
of observed events~(58).  The neutrino energy is reconstructed using
the reconstructed muon momentum and the known beam-direction while assuming
there was a QE interaction and ignoring the Fermi momentum.  As can be
seen, compared to the MC expectation there is a deficit of 1R$\mu$
events in the low energy region, as is expected from the oscillation
hypothesis.

\subsection{Systematic error from reconstruction at SK}
\label{sec:skreduc-error}

The systematic uncertainties for estimating ${\rm N_{SK}}$ and the
reconstructed neutrino energy in SK are evaluated using atmospheric
neutrinos as a control sample.  Tab.~\ref{tbl:SKnsksyserr} shows the
systematic errors for ${\rm N_{SK}}$.  The dominant uncertainty for
estimating ${\rm N_{SK}}$ comes from the vertex reconstruction.  Since
a cut is made on fiducial volume a systematic shift in or out of this
volume will either over or underestimate the number of events expected.  It
is evaluated comparing the number of events for atmospheric neutrino
data with the MC expectation in the fiducial volume using two
different vertex reconstruction programs.  

\begin{figure} [htbp]
 \begin{center}
  \includegraphics[width=8cm]{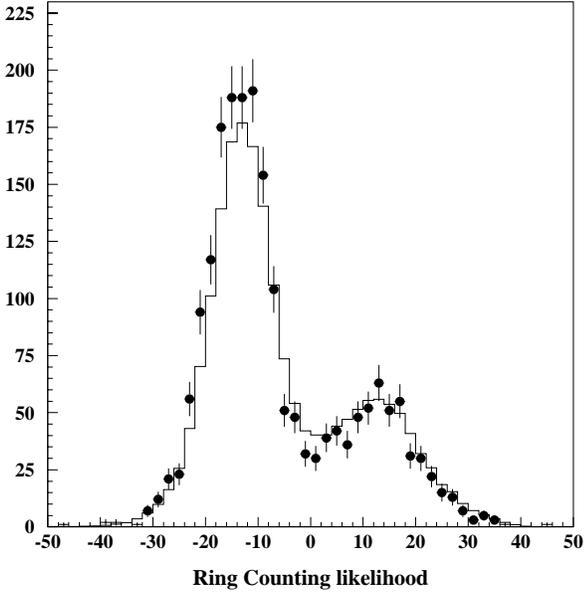}
  \caption{Ring counting likelihood distribution for SK-II atmospheric
    fully contained neutrino events. Closed circles are data and the
    histogram is MC expectation normalized by live time assuming neutrino
    oscillation at atmospheric best-fit parameters.  
    Events with Likelihood $<$ 0 are assigned
    to be one-ring. }
   \label{fig:SKring}
 \end{center}
\end{figure}

\begin{figure} [htbp]
 \begin{center}
  \includegraphics[width=8cm]{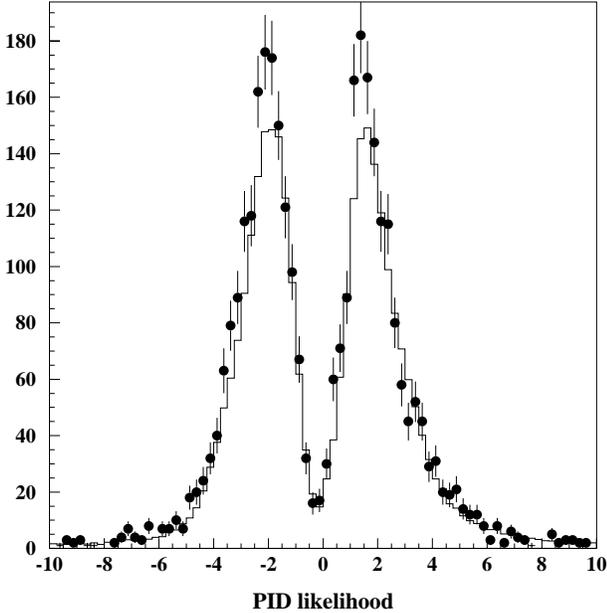}
  \caption{Particle identification likelihood distribution for SK-II
    atmospheric fully contained one-ring events.  Closed circles are
    data and the histogram is MC expectation normalized by live time
    assuming neutrino
    oscillation at atmospheric best-fit parameters.  
    Events with Likelihood $<$ 0
    are assigned to be e-like. }
   \label{fig:SKpid}
 \end{center}
\end{figure}

Systematic errors due to the reconstruction algorithms themselves are
also taken into account in the oscillation analysis.  Systematic
errors due to reconstruction are shown in table
\ref{tbl:SKspecsyserr}.  Uncertainties coming from the ring counting
and particle identification are evaluated by comparing the likelihood
distributions for data and MC, and varying the selection criteria.  Figure
\ref{fig:SKring} and \ref{fig:SKpid} show the ring counting and
particle identification likelihood distributions of atmospheric
neutrino data compared with the MC expectations in SK-II.  The MC
expectations reproduce the data well.  The uncertainty for the energy
scale are also estimated by using cosmic ray muons, the $\pi^0$
invariant mass and decay electrons. The energy scale uncertainty at SK
is estimated to be 2.0\% for K2K-I and 2.1\% for K2K-II.

\begin{table}
   \begin{center}
    \caption{Systematic errors for ${\rm N_{SK}}$.}
    \begin{tabular}{lrr}
     \hline        
      & K2K-I & K2K-II \\
     \hline
     reduction                 & $<$1\% & $<$1\% \\
     fiducial volume cut       & 2\% & 2\% \\
     decay electron background & 0.1\% & 0.1\% \\
     MC statistics             & 0.6\% & 0.6\% \\
     \hline
     Total                     & 3\% & 3\% \\
     \hline
    \end{tabular} 
    \label{tbl:SKnsksyserr}
    \end{center}
 \end{table}

\begin{table}
   \begin{center}
    \caption{Systematic errors for reconstructed neutrino energy spectrum.
    The errors are shown in \%, and the five columns refer to different bins
    of neutrino energy, as shown in the table in units of GeV.}
    \begin{tabular}{lrrrrr}
     \hline        
     K2K-I (GeV)   & 0-0.5 & 0.5-1.0 & 1.0-1.5 & 1.5-2.0 & 2.0- \\
     \hline
     ring counting & 3.4\% & 2.7\% & 3.0\% & 4.5\% & 4.5\% \\
     particle ID   & 0.9\% & 0.3\% & 0.5\% & 0.4\% & 0.4\% \\
     vertex        & 2.0\% & 2.0\% & 2.0\% & 2.0\% & 2.0\% \\
     \hline
     total         & 4.1\% & 3.4\% & 3.6\% & 4.9\% & 4.9\% \\
     \hline        
     \\
     \hline        
     K2K-II (GeV)  & 0-0.5 & 0.5-1.0 & 1.0-1.5 & 1.5-2.0 & 2.0- \\
     \hline
     ring counting & 5.3\% & 4.1\% & 3.7\% & 3.8\% & 3.8\% \\
     particle ID   & 2.6\% & 0.4\% & 0.3\% & 0.6\% & 0.6\% \\
     vertex        & 2.0\% & 2.0\% & 2.0\% & 2.0\% & 2.0\% \\
     \hline
     total         & 6.2\% & 4.6\% & 4.2\% & 4.3\% & 4.3\% \\
     \hline        

    \end{tabular} 
    \label{tbl:SKspecsyserr}
    \end{center}
 \end{table}

\section{Neutrino Oscillation Analysis}
\subsection{Oscillation analysis method}\label{sec:oscana}
 
A two-flavor neutrino oscillation analysis is performed based on a
comparison between the observation and the expectation by use of a
maximum-likelihood method.  The signatures of neutrinos oscillating
from $\nu_\mu$ to $\nu_{\tau}$ are both a reduction in the total
number of observed neutrino events and a distortion in the neutrino
energy spectrum.  Thus, the likelihood function is defined as the
product of the likelihoods for the observed number of events in the SK
fiducial volume~($\Lnorm$) and the shape of the $\enurec$
spectrum~($\Lshape$).
In addition, the systematic uncertainties are also treated as fitting
parameters in the likelihood.  They are included in a constraint
likelihood term~($\Lsyst$) where they are varied within their
uncertainties, thus modifying the expectation.  The total likelihood
function is then defined as:

\begin{eqnarray}
 \Lcal &=& \Lnorm \times \Lshape \times \Lsyst.
\end{eqnarray}

The oscillation parameters, $\dms$ and $\sstt$, are obtained by
maximizing the likelihood function.  One-hundred twelve FC events are
used in $\Lnorm$ and fifty eight FC 1R$\mu$ events are used for $\Lshape$, respectively.  
The systematic parameters in the likelihood consist of 
the neutrino energy spectrum at the near detector site,
the \fn flux ratio,
the neutrino-nucleus cross-section,
the efficiency and the energy scale of SK,
and the overall normalization.

\subsection{Prediction of the number of events and the energy spectrum at Super-Kamiokande}

\subsubsection{Number of neutrino events}

The expected number of neutrino events in SK (\nsk) is derived by
extrapolating the measured number of interactions in the 1KT (\ktint,
calculated in Eq.~\ref{eq:1ktint}) with the ratio of the expected
neutrino event rate per unit mass,
$\rho^\mathrm{SK}/\rho^\mathrm{1KT}$.  Taking into account the
difference of fiducial mass ($M$) and the number of protons on target
(POT) used in the analysis for 1KT and SK, \nsk\ is written as:

\begin{multline} \label{eq:Nexp}
 \nsk(\dms, \sstt) \\
 \equiv  \ktint  \cdot
 \frac{\rho^\mathrm{SK}}{\rho^\mathrm{1KT}}
 \cdot \frac{\MSK}{\MKT} \cdot \frac{\POTSK}{\POTKT} \cdot C_{\nu_e},  
\end{multline}

\noindent
where superscripts ``1KT'' and ``SK'' denote the variable for SK and
1KT, respectively,
and $C_{\nu_e}$ is the correction factor for the difference of the
electron neutrino contamination in the neutrino beam at 1KT and SK.
The value of $C_{\nu_e}$ is estimated to be 0.996 with the MC
simulation.

The expected event rate at each detector, $\rho$, is calculated from
the neutrino flux $\Phi$, the neutrino-water interaction cross-section
$\sigma$, and the detector efficiency $\epsilon$ estimated with the MC
simulation:

\begin{equation}
\rho = \int dE_{\nu} \Phi(\enu) \cdot \sigma(\enu) \cdot \epsilon(\enu). \nonumber
\end{equation} 

\noindent
In order to account for the 
systematic uncertainty, 
we classify the neutrino interaction into three categories:
CC-QE, CC-nonQE and NC.
The event rate is calculated separately for each of the three interaction types and then summed.
The neutrino flux at SK, $\LSK$, is estimated from the F/N flux ratio \rfn and 
the measured ND spectrum $\LND$:

\begin{equation}
\LSK = \rfn(\enu) \cdot \LND(\enu) \cdot (1-P(\enu; \dms, \sstt)),  \nonumber
\end{equation}

\noindent
where $P(\enu; \dms, \sstt)$ is the neutrino oscillation probability given by 
Eq.~(\ref{eqn:oscillation}).

The uncertainties of 
$\LND$ and their correlation 
are obtained from the ND analysis as shown in Tab.~\ref{tbl:merged-error-matrix}.
Those for 
$\rfn$ are derived from the HARP $\pi^+$ measurement and the beam MC
simulation, and are summarized in Tab.~\ref{tab:harpfluxratio_errormatrix}.
In order to be insensitive to the absolute cross-section uncertainty,
we incorporate the uncertainties in neutrino-nucleus cross-section as
the cross-section ratio relative to CC-QE interactions.
The uncertainty of the CC-nonQE/CC-QE cross-section ratio is taken
from the ND measurements.  For the NC/CC-QE
cross-section ratio, we assign 11\% uncertainty to NC single $\pi^0$
production based on the measurement with the
1KT~\cite{Nakayama:2004dp}.  The other NC interaction modes are
assigned 30\% uncertainty based on past
experiments~\cite{Monsay:1978}.  Taking into account the detection
efficiency in SK, 15\% is assigned as the net uncertainty on NC/CC-QE
ratio.
The uncertainties from event reconstruction at SK are summarized 
in Tab.~\ref{tbl:SKnsksyserr}. 
The uncertainty of the overall normalization of the number of events
in each period is estimated from the fiducial mass error of 1KT and SK
and the uncertainty in the difference of the number of protons on
target used for the analysis.
In total, the normalization error is estimated to be 
$\pm 5.1$\% for both Ib and II periods.

For the period Ia, the beam configuration is different from other
periods and we do not have a ND measurement of the energy spectrum.  We
employ different treatment of the systematic errors for this period.
All the uncertainties, including those from the \fn flux ratio, energy
spectrum and cross-section, are incorporated into the error of the
single normalization parameter.  The total normalization uncertainty
for the Ia period is estimated to be +9.0/-9.8\%.  The largest
contribution to this uncertainty is from the energy spectrum, which is
estimated with the HARP $\pi^+$ measurement and the beam MC simulation
and gives $+5.8$\%/$-7.0$\%.  Other contributions come from the \fn
flux ratio ($\pm4.3$\%) and the fiducial volume uncertainties of 1KT
($\pm4.3$\%) and SK ($\pm3.0$\%).

We estimate the expected number of events without neutrino
oscillation while incorporating all of the known systematic uncertainties
by use of a MC technique.
Many sets of the systematic parameters 
are randomly generated with proper consideration of their
correlation.
For each systematic parameter set, 
$\nsk$ is calculated using Equation~(\ref{eq:Nexp}), setting oscillation parameters to zero.
The number of FC events without neutrino
oscillation is estimated to be $158.1^{+9.2}_{-8.6}$.  
This technique
allows us to determine the contributions from the individual
systematics to the the total error by selectively including only some
errors during the generation.
We find that the dominant error sources are the fiducial volume uncertainties
in 1KT and SK~$\left(^{+4.9\%}_{-4.8\%}\right)$ and the \fn
ratio~$\left(^{+2.9\%}_{-2.9\%}\right)$.

\subsubsection{Reconstructed neutrino energy spectrum} \label{sec:skspec}
The expected spectrum shape of the reconstructed neutrino energy at SK, 
$\phi_{\mathrm{exp}}^{\mathrm{SK}} (\enurec )$,
is calculated as:

\begin{equation} \label{eq:skspec}
\phi_{\mathrm{exp}}^{\mathrm{SK}} = 
\int d\enu \cdot \LSK(\enu) \cdot
\sigma(\enu) \cdot \Effsrm(\enu)
  \cdot r(\enu ; \enurec),
\end{equation}

\noindent
where $\Effsrm$ is 
the detection efficiency for 1R$\mu$ events in SK
 and $r(\enu; \enurec )$ is 
 the probability of reconstructing an event with true energy $\enu$ as \enurec.  
 Both of them are estimated with the MC simulation.
In the actual procedure, the $\enu$ and $\enurec$ are binned with an interval
of 50 MeV, and hence the integral over the true neutrino energy is 
replaced by a summation over true energy bins.

The uncertainties from the neutrino energy spectrum at the ND, the \fn
flux ratio, and the cross-section ratios are incorporated as described
above.  The uncertainties from 1R$\mu$ event reconstruction at SK are
shown in Tab.~\ref{tbl:SKspecsyserr}.  The energy scale uncertainty in
SK is 2.1\% for SK-I and 2.0\% for SK-II, respectively, as described
in section~\ref{sec:skreduc-error}.

\begin{figure} []
 \begin{center}
  \resizebox{6.0cm}{!}{
   \includegraphics[trim=0.0cm 0.0cm 0.0cm
   0.0cm,clip]{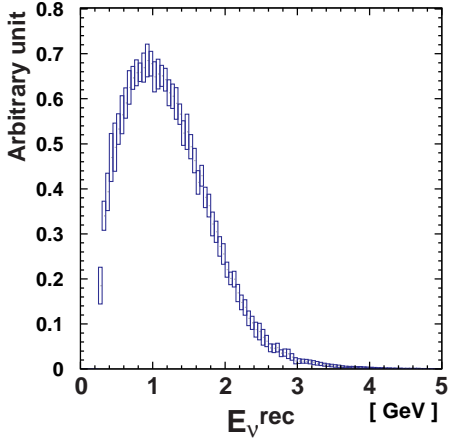}}
   \caption{Expected reconstructed neutrino energy spectrum shape in the case of
   null oscillation. 
   Height of boxes indicate the size of error.}
   \label{fig:shapetoymc}
 \end{center}
\end{figure}

\begin{figure} []
 \begin{center}

 \resizebox{\columnwidth}{!}{
   \includegraphics[trim=0.0cm 0.0cm 0.0cm
   0.0cm,clip]{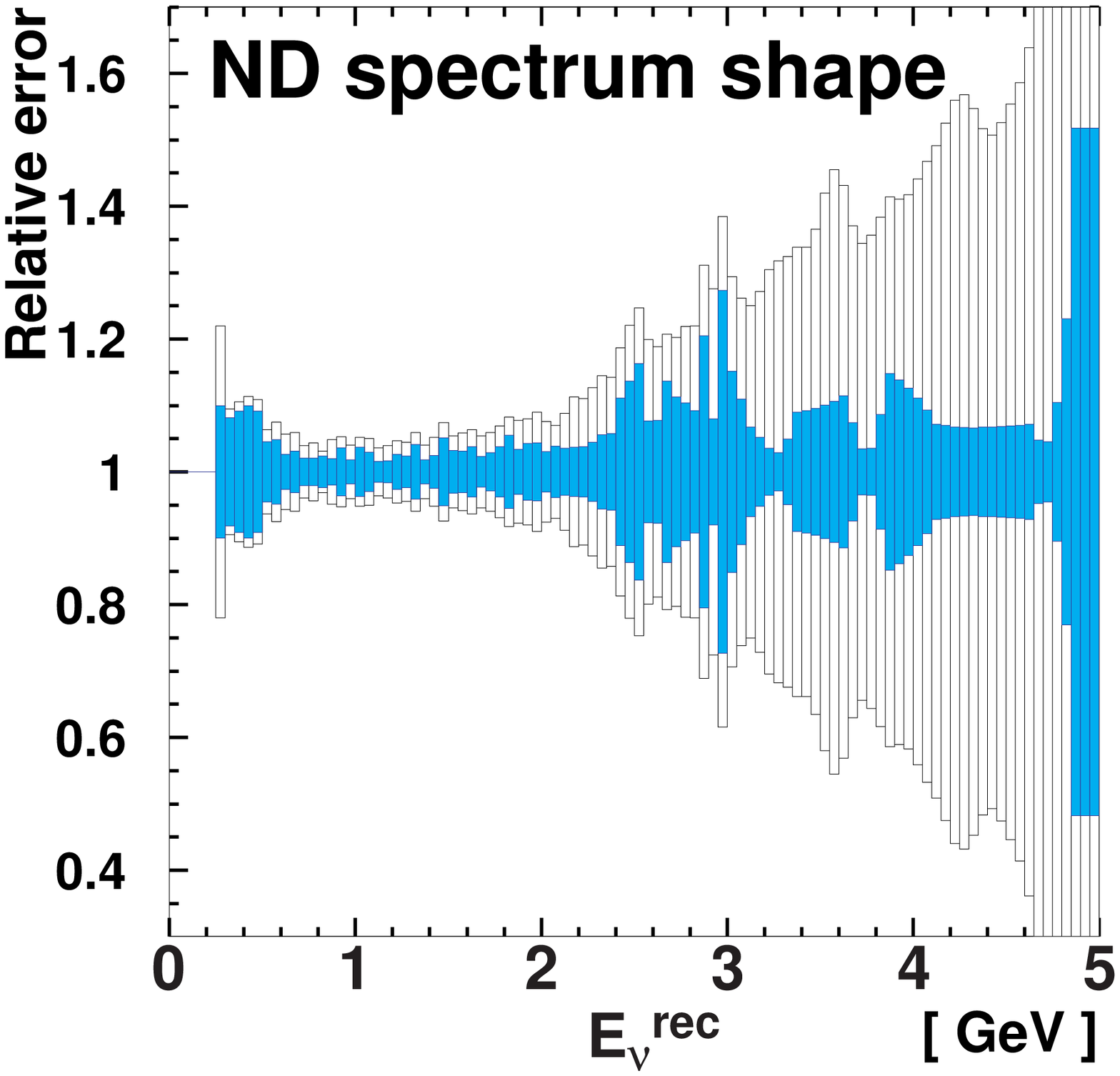}
   \includegraphics[trim=0.0cm 0.0cm 0.0cm
   0.0cm,clip]{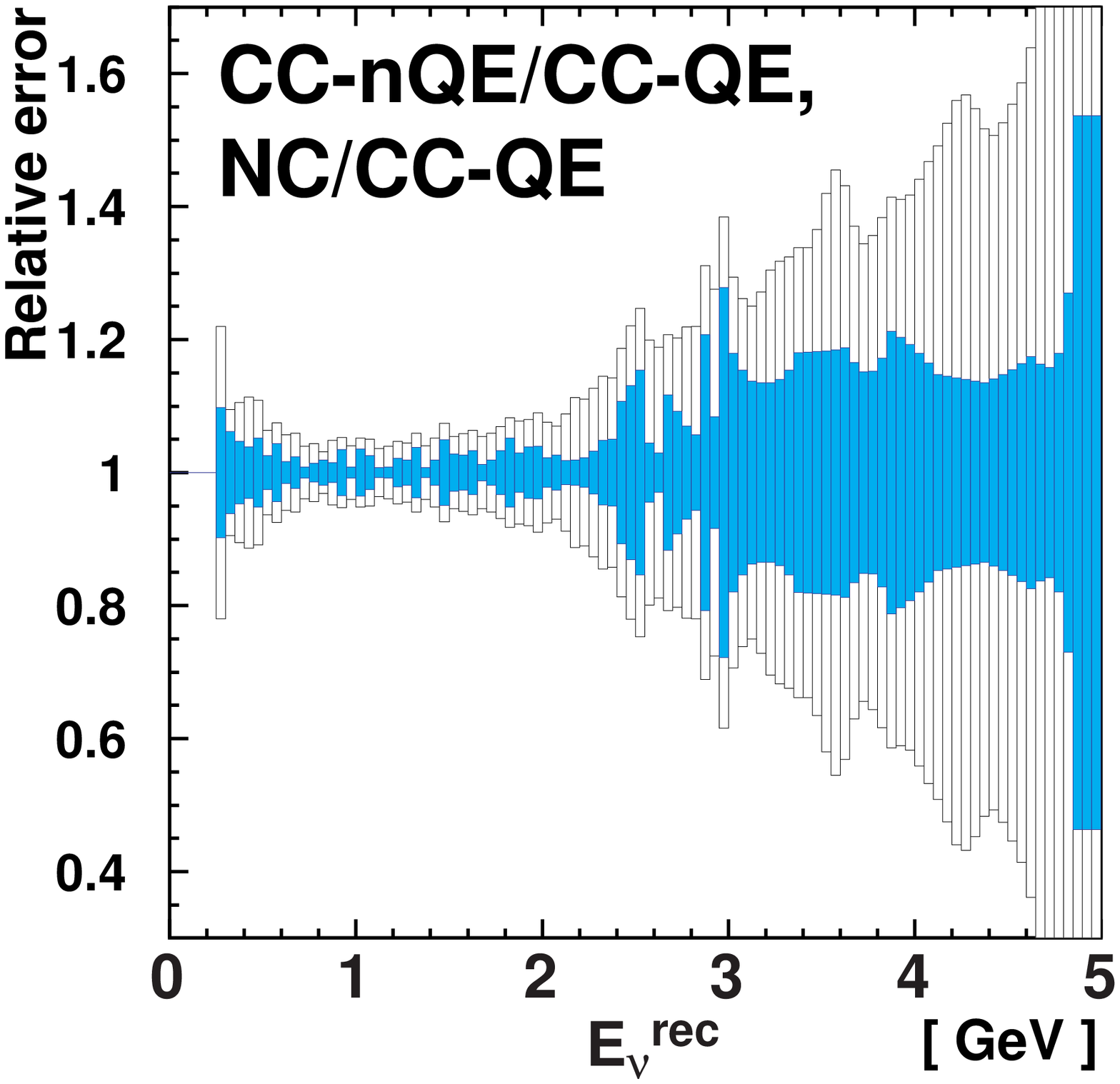}
}
 \resizebox{\columnwidth}{!}{
   \includegraphics[trim=0.0cm 0.0cm 0.0cm
   0.0cm,clip]{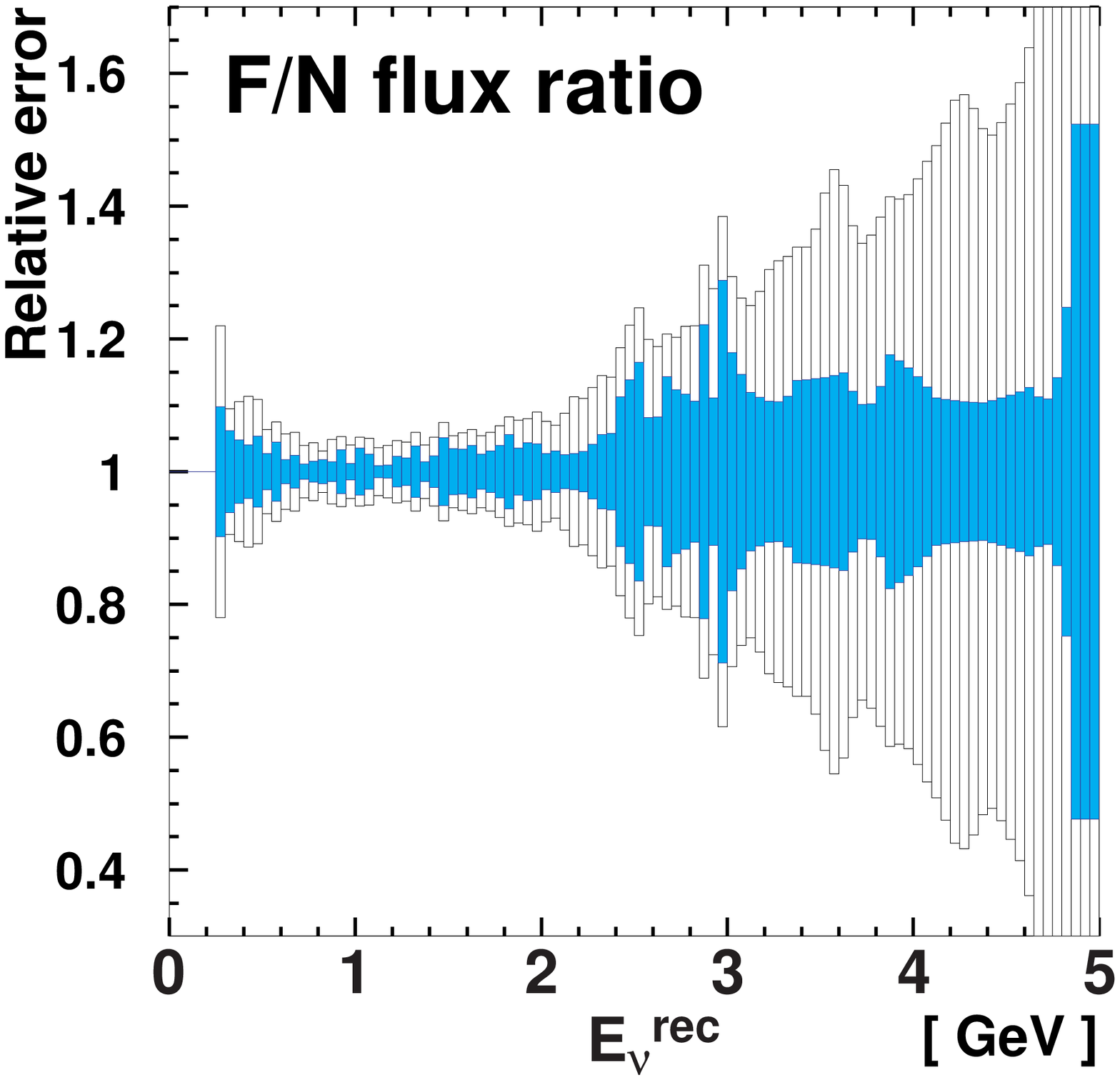}
   \includegraphics[trim=0.0cm 0.0cm 0.0cm
   0.0cm,clip]{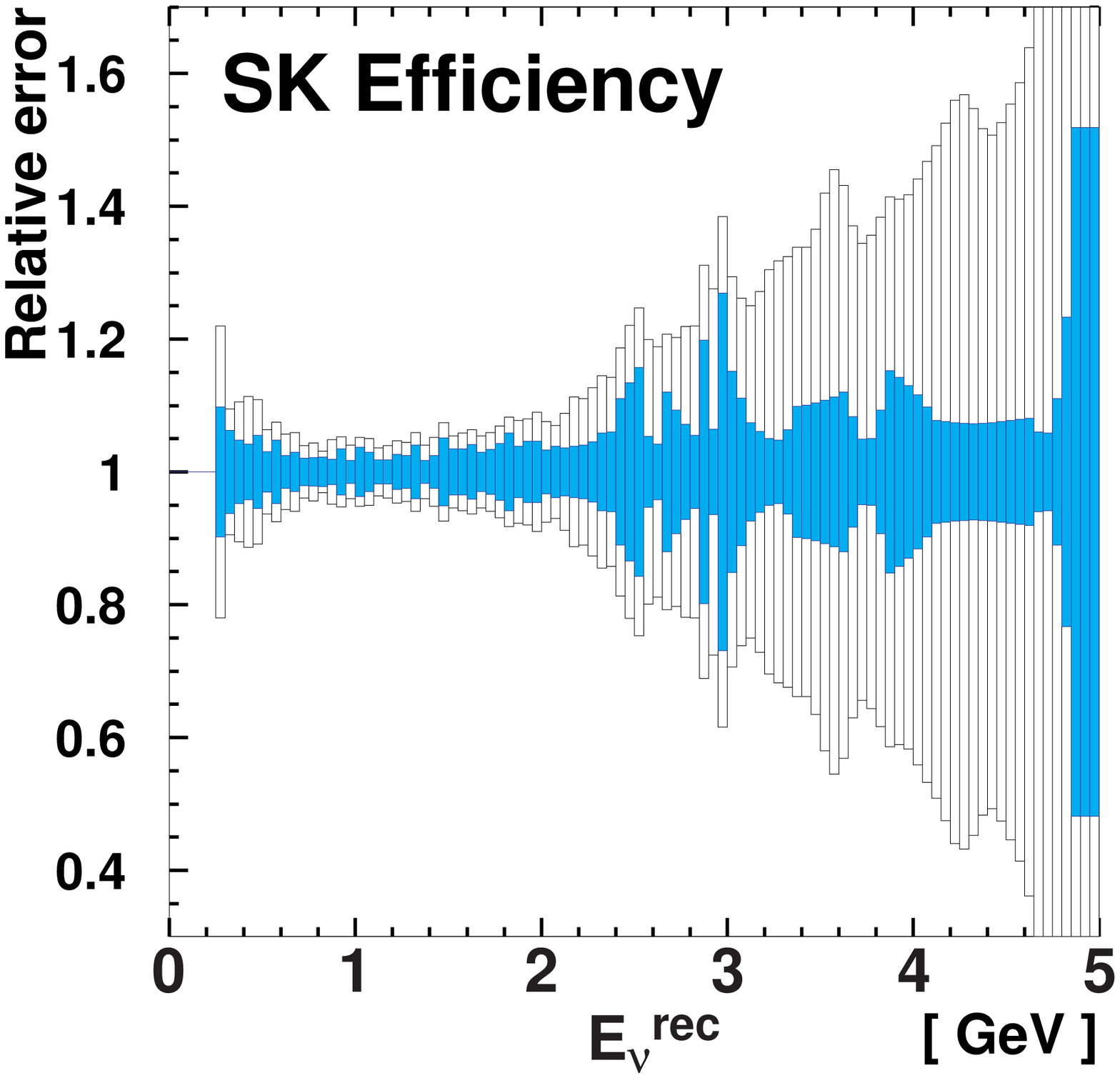}
}
 \resizebox{\columnwidth}{!}{
   \includegraphics[trim=0.0cm 0.0cm 0.0cm
   0.0cm,clip]{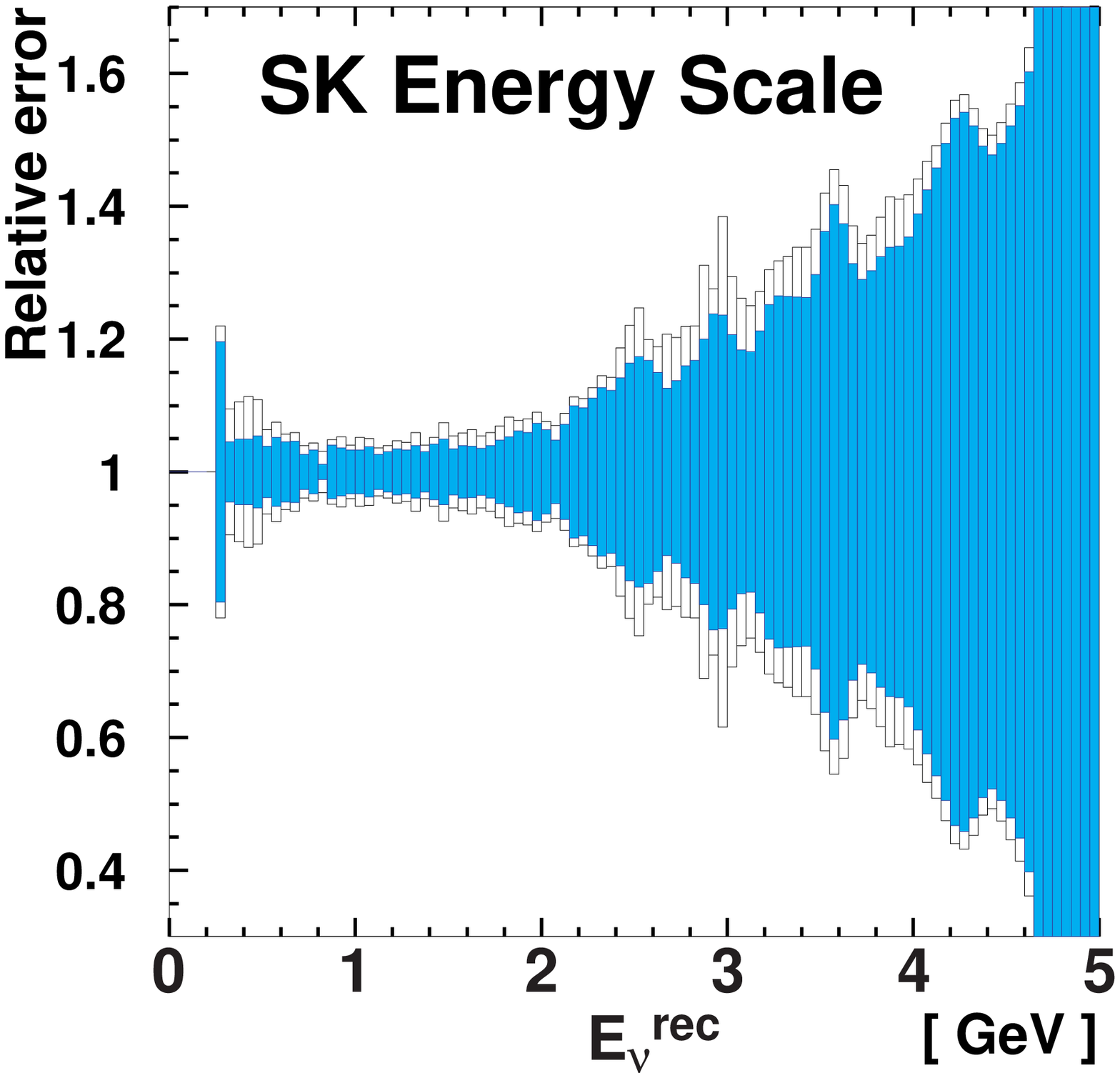}
   \includegraphics[trim=1.0cm 1.0cm 1.0cm
   1.0cm,clip]{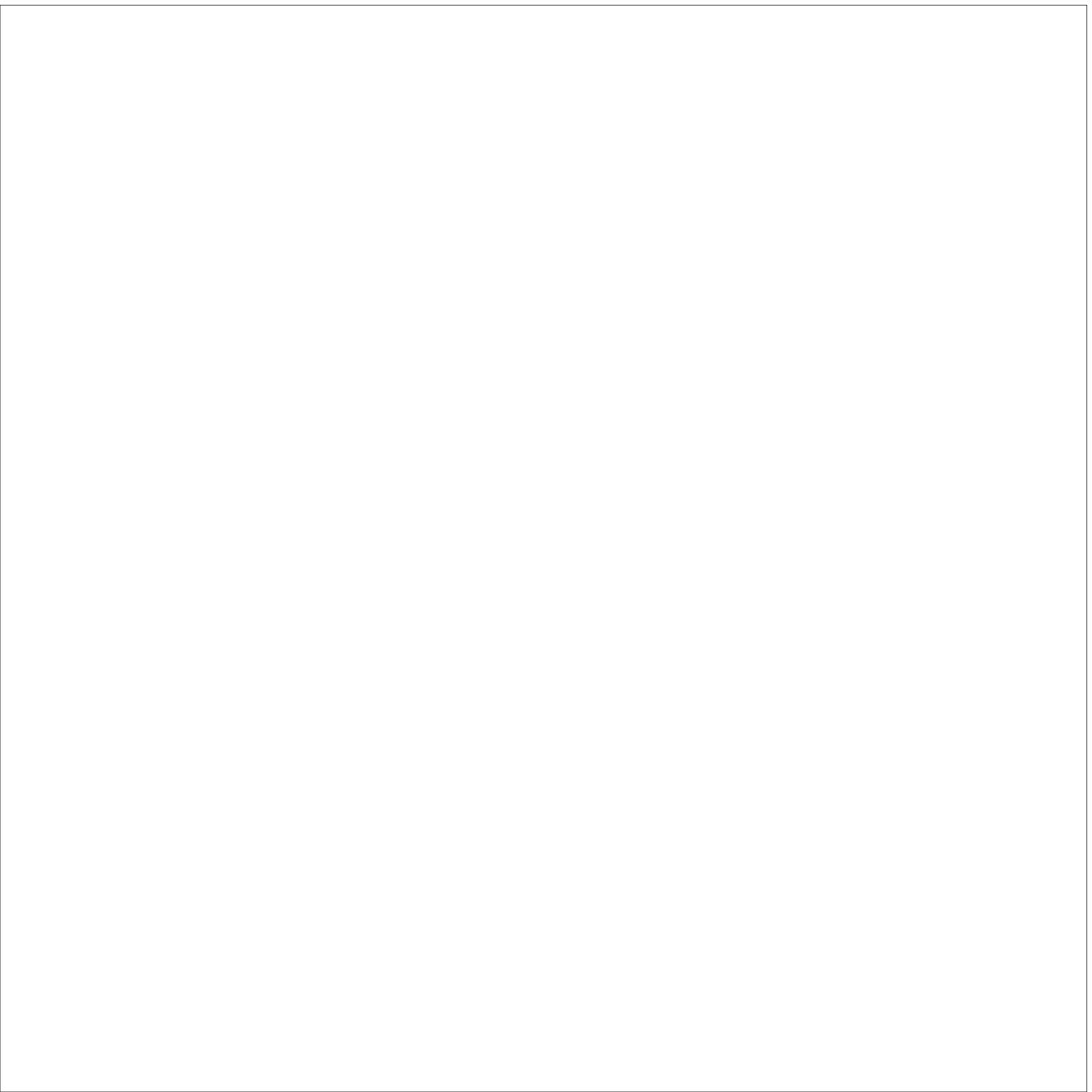}
}

   \caption{
         Contribution of each systematic error to the reconstructed
        neutrino energy spectrum.
         Vertical axis is relative error of the spectrum. 
         Source of uncertainty is indicated in each plot.
         Blank and filled bars represent the sizes of the total error
         and the contribution from the source being considered,
         respectively.
        }
   \label{fig:shapesystoymc}
 \end{center}
\end{figure}

The expected $\enurec$ spectrum shape for null oscillation case and its
error are estimated using the same technique as the number of events
and shown in Fig.~\ref{fig:shapetoymc}.  
The height of the box represents the size of estimated error in each bin.
The contribution of each systematic uncertainty is estimated by turning
each uncertainty on exclusively one by one, as shown in Fig.~\ref{fig:shapesystoymc}.  
We find that the error on the spectrum shape is dominated by the SK energy scale.

\subsection{Definition of likelihood}

\subsubsection{Normalization term}
The normalization term, $\Lnorm$, is defined as the Poisson probability to
observe $\Nobs$ events when the expected number of events is
$\Nexp$:

\begin{eqnarray}
 \Lnorm = \frac{(\Nexp)^{\Nobs}}{\Nobs!}e^{-\Nexp}.
\end{eqnarray}

\noindent
In order to account for the difference of the experimental configuration,
the expectation for each experimental period is separately calculated
using Eq.(\ref{eq:Nexp}) and then
summed as:

\begin{eqnarray}
\label{eq:NexpEqn}
  \Nexp = \Nexp^{\mathrm{Ia}} + \Nexp^{\mathrm{Ib}} +
  \Nexp^{\mathrm{II}}.
\end{eqnarray} 

\subsubsection{Energy spectrum shape term}
The energy spectrum shape term is defined as the product of the probability for each 1R${\mu}$ event to be
observed at reconstructed neutrino energy $\enurec$.
We use the expected neutrino energy spectrum, given in Eq.~(\ref{eq:skspec}), as the probability density function.
The probability density function is separately defined for each experimental period:
\begin{eqnarray} 
\Lshape &=& \prod_{i=1}^{N^{\mathrm{Ib}}_{1R{\mu}}} 
          \phi^{\mathrm{SK}}_{\mathrm{exp}, \mathrm{Ib}}(E^{\mathrm{rec}}_{\nu, i}; \dms, \sstt)
          \nonumber \\
        &\times&
          \prod_{i=1}^{N^{\mathrm{II}}_{1R{\mu}}} 
          \phi^{\mathrm{SK}}_{\mathrm{exp}, \mathrm{II}}(E^{\mathrm{rec}}_{\nu, i}; \dms, \sstt)
\end{eqnarray}
where $N^{\mathrm{Ib}}_{1R{\mu}}=30$ and
$N^{\mathrm{II}}_{1R{\mu}}=28$ are the number of observed FC 1R${\mu}$
events for period Ib and II, respectively.  There is no 1R$\mu$ event
in the Ia run period.

\subsubsection{Systematic term}
The systematic parameters are treated as fitting parameters, and are
assumed to follow a Gaussian distribution.  They are constrained
within their uncertainties by constraint terms expressed as:

\begin{equation}
\label{eq:sysosc}
\Lsyst \equiv \prod_{j=1}^{N_\mathrm{syst}} 
\exp(- {\Delta \Sysf_j}^{\mbox{t}} (M_j)^{-1} \Delta \Sysf_j),
\end{equation}

\noindent
where $N_\mathrm{syst}$ is the number of parameter sets, 
$\Delta \Sysf_j $ represents the deviations of the
parameters from their nominal values and $M_j$ is the error matrix
for $j$-th set of parameters.

\subsection{Results}

\begin{figure} [htpb]
 \begin{center}
  \resizebox{\columnwidth}{!}{
   {\includegraphics[trim=0cm 0.0cm 0cm
    0.cm,clip]{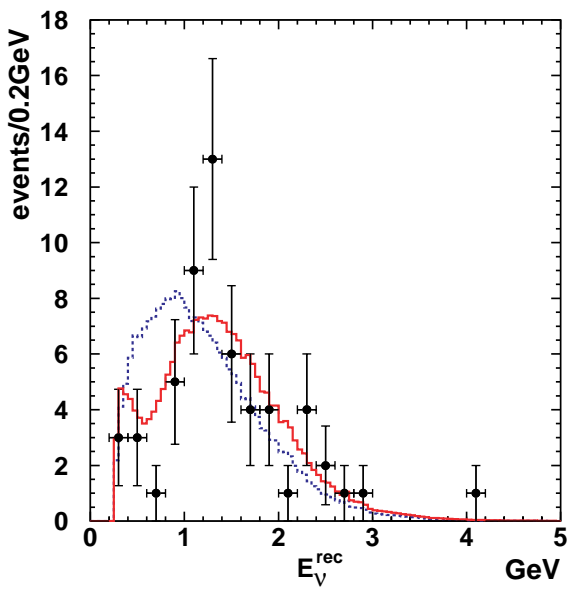}}}
  \caption{ The reconstructed $E_{\nu}$ distribution for the 1-ring
           $\mu$-like sample. Points with error bars are data. 
            The solid line is the best fit spectrum with neutrino oscillation 
             and the dashed line is the expectation without oscillation.
            These histograms are normalized by the number of events 
           observed~(58).}
   \label{fig:EnuFitResult}
 \end{center}
\end{figure}

The likelihood is maximized in the $\Delta m^{2} $ --
$\sin^{2}2\theta$ space and the best fit point within the physical
region is found to be at $(\Delta m^{2}, \sin^{2}2\theta) = (
2.8 \times 10^{-3}\mathrm{eV}^{2}, 1.0)$.
The values of all systematic parameters at the best fit point are
within 1$\sigma$ of their estimated errors.
 At this point, the expected number of events
is 107.2, which agrees well with the 112 observed 
within the statistical uncertainty.
 The observed $\enurec$ distribution is shown in
Fig.~\ref{fig:EnuFitResult} together with both the 
expected distributions for the best-fit parameters, 
and the expectation without oscillations.
The consistency between the observed and the best-fit $\enurec$
distributions is checked using a Kolmogorov-Smirnov~(KS) test.  For
the best fit parameters, the KS probability is 37~\%, while for the
null oscillation hypothesis is 0.07~\%.  The observation agrees with
the expectation of neutrino oscillation.
The highest likelihood is found at $(\Delta m^{2}, \sin^{2}2\theta) =
(2.6 \times 10^{-3}\mathrm{eV}^{2}, 1.2)$, which is outside of the
physical region.  The probability that we would get $\sstt \ge 1.2$ 
if the true parameters are at our best fit point is 26.2\%, based on
the virtual MC experiments.

The probability that the observations can be explained equally well by
the no oscillation and by the oscillation hypotheses
is estimated by computing
the difference of log-likelihood between the null oscillation case and
the best fit point with oscillation.  
The null oscillation probability
is calculated to be 0.0015~\%~(4.3$\sigma$).  When only
normalization~(shape) information is used, the probability is
0.06\%~(0.42\%).

The null oscillation probability calculated separately for each
sub-sample or each likelihood term is shown in
Tab.~\ref{tbl:nullsummary}.  In addition, Tab.~\ref{tbl:nullsys} shows
the effect of each systematic uncertainty on the null oscillation
probability.  The effect is tested by turning on the error source
written in the first column in the table. As shown in the table, the
dominant contributions to the probabilities for the normalization
information are from the \fn flux ratio and the normalization error,
while the energy scale is the dominant error source for the
probability with the \enurec shape information
consistent with the results found using the MC test described in
Sec.~\ref{sec:skspec}.

\begin{table}
 \begin{center}
  \caption[Null oscillation probability.]
  {Summary of the null oscillation probability.
   Each row is classified by the likelihood term used,
   and each column represents the data set.}
  \begin{tabular}{lccc} \hline \hline
    & ~K2K-I+II \quad & K2K-I only & K2K-II only \\ \hline
    Shape + Norm. & \bf 0.0015\%~(4.3$\sigma$) &
    0.18\%~($3.1\sigma$) & 0.56\%~(2.8$\sigma$) \\
    \hline
    Shape only    & 0.42\%~(2.9$\sigma$)   &  7.7\%  & 5.2\% \\
    Norm. only    & 0.06\%~(3.4$\sigma$)   &  0.6\%  & 2.8\% \\ \hline \hline
  \end{tabular}
  \label{tbl:nullsummary}
 \end{center}
\end{table}

\begin{table}[t]
 \begin{center}
  \caption{Effect of each systematic uncertainty on the null oscillation
  probability. The numbers in the table are null oscillation
  probabilities when only the error written in the first column is turned
  on.  
  }
  \begin{tabular}{lccc} 
  \hline  \hline
       & Norm-only & Shape-only & Combined  \\  
  \hline
  \hline
       Stat. only         & 0.01\%  & 0.22\%  & 0.0001\%\\ 
       FD spectrum         & 0.01\%  & 0.24\%  & 0.0002\%\\ 
       nQE/QE, NC/CC       & 0.01\%  & 0.23\%  & 0.0002\%\\ 
       Far/Near            & 0.02\%  & 0.23\%  & 0.0003\%\\ 
       $\epsilon^{1R\mu}$  &  ---    & 0.23\%  & 0.0002\%\\ 
       Energy scale        &  ---    & 0.38\%  & 0.0002\%\\ 
       Normalization       & 0.03\%  &  ---    & 0.0005\%\\ 
  \hline
  \hline
       All errors          & 0.06\%  & 0.42\%  & 0.0015\%\\ 
  \hline   \hline
  \end{tabular}
  \label{tbl:nullsys}
 \end{center}
\end{table}

 The allowed region of oscillation parameters are evaluated based on the 
difference of log-likelihood between each point and the best fit point:

\begin{eqnarray}
 \Delta\lnL(\dms,\sstt)
 &\equiv& \ln \left(
 \frac{\mathcal{L}_{\mathrm{max}}^{\mathrm{phys}}}{\Lcal(\dms,\sstt)}
 \right) \nonumber \\
 &=& \ln\mathcal{L}_{\mathrm{max}}^{\mathrm{phys}} - \lnL(\dms,\sstt),
 \nonumber \\
\end{eqnarray}
where $\mathcal{L}_{\mathrm{max}}^{\mathrm{phys}}$ is the likelihood at the best-fit point and
$\Lcal(\dms,\sstt)$ is the likelihood at $(\dms,\sstt)$ 
with systematic parameters that maximize the likelihood at that point.

\begin{figure} []
 \begin{center}
  \resizebox{\columnwidth}{!}{
   \includegraphics[trim=0.0cm 0.0cm 0.0cm
   0.0cm,clip]{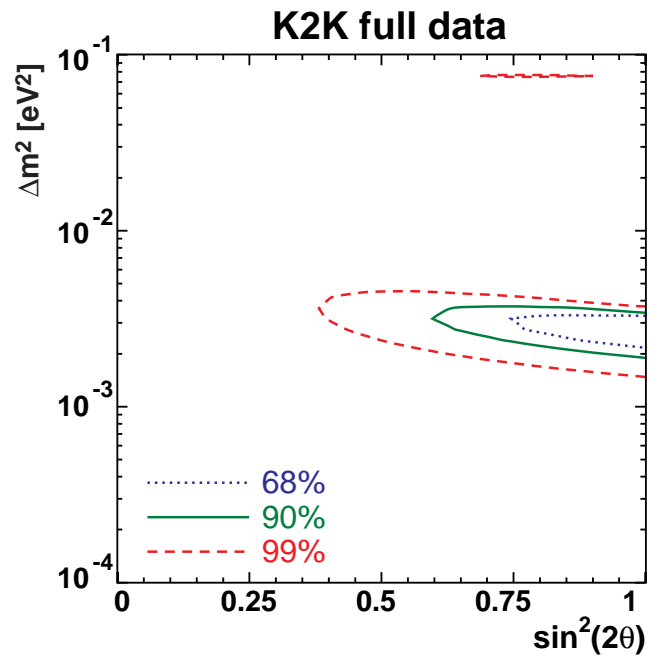}}
   \caption{Allowed regions of oscillation parameters. 
   Three contours correspond to the 68\%~(dotted line),
   90\%~(solid line) and 99\%~(dashed line) CL. allowed regions,
   respectively.}
   \label{fig:contour-all}
 \end{center}
\end{figure}

\begin{figure} []
 \begin{center}
  \resizebox{\columnwidth}{!}{
   \includegraphics[trim=0.0cm 0.0cm 0.0cm
   0.0cm,clip]{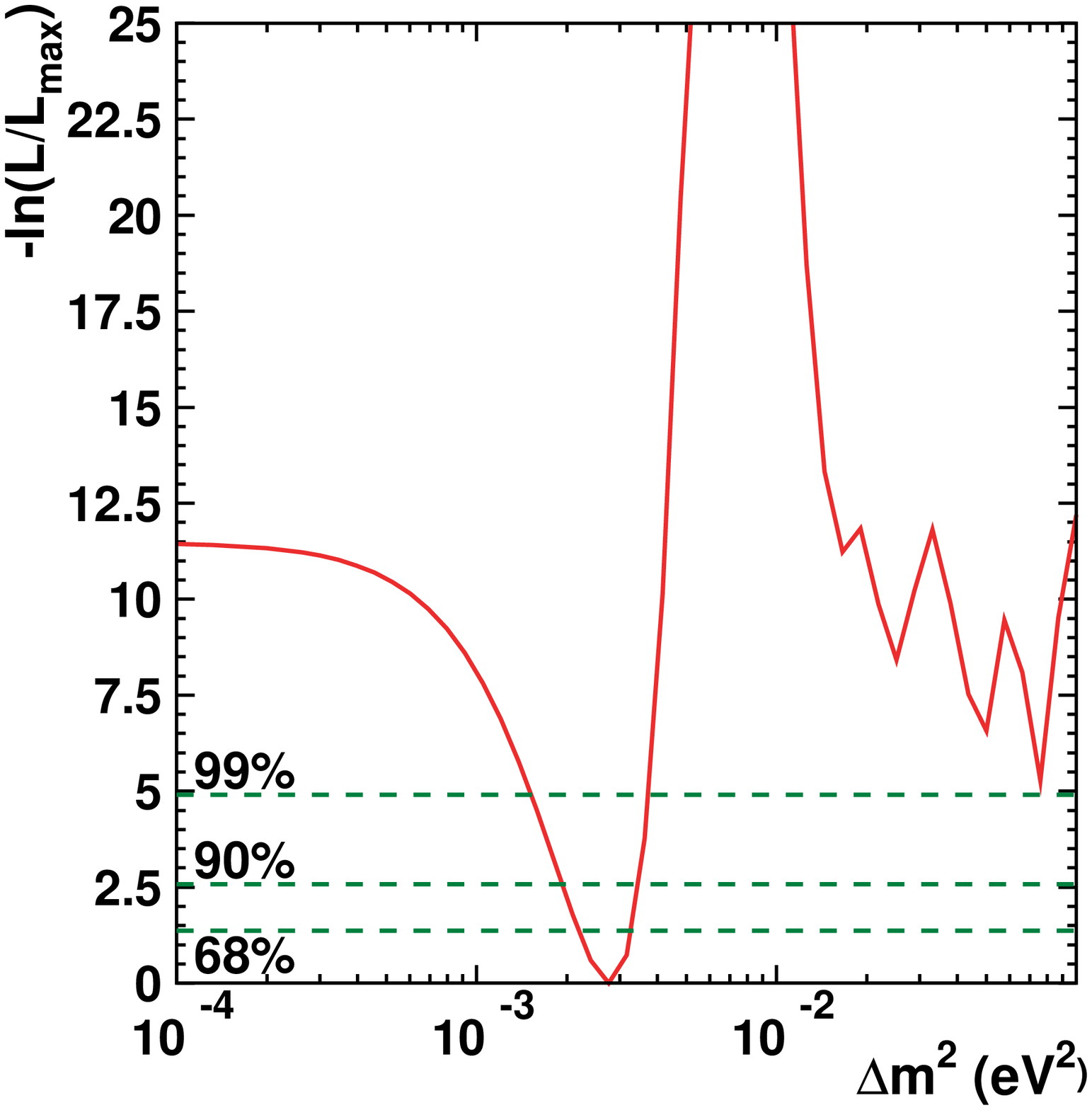}
   \includegraphics[trim=0.0cm 0.0cm 0.0cm
   0.0cm,clip]{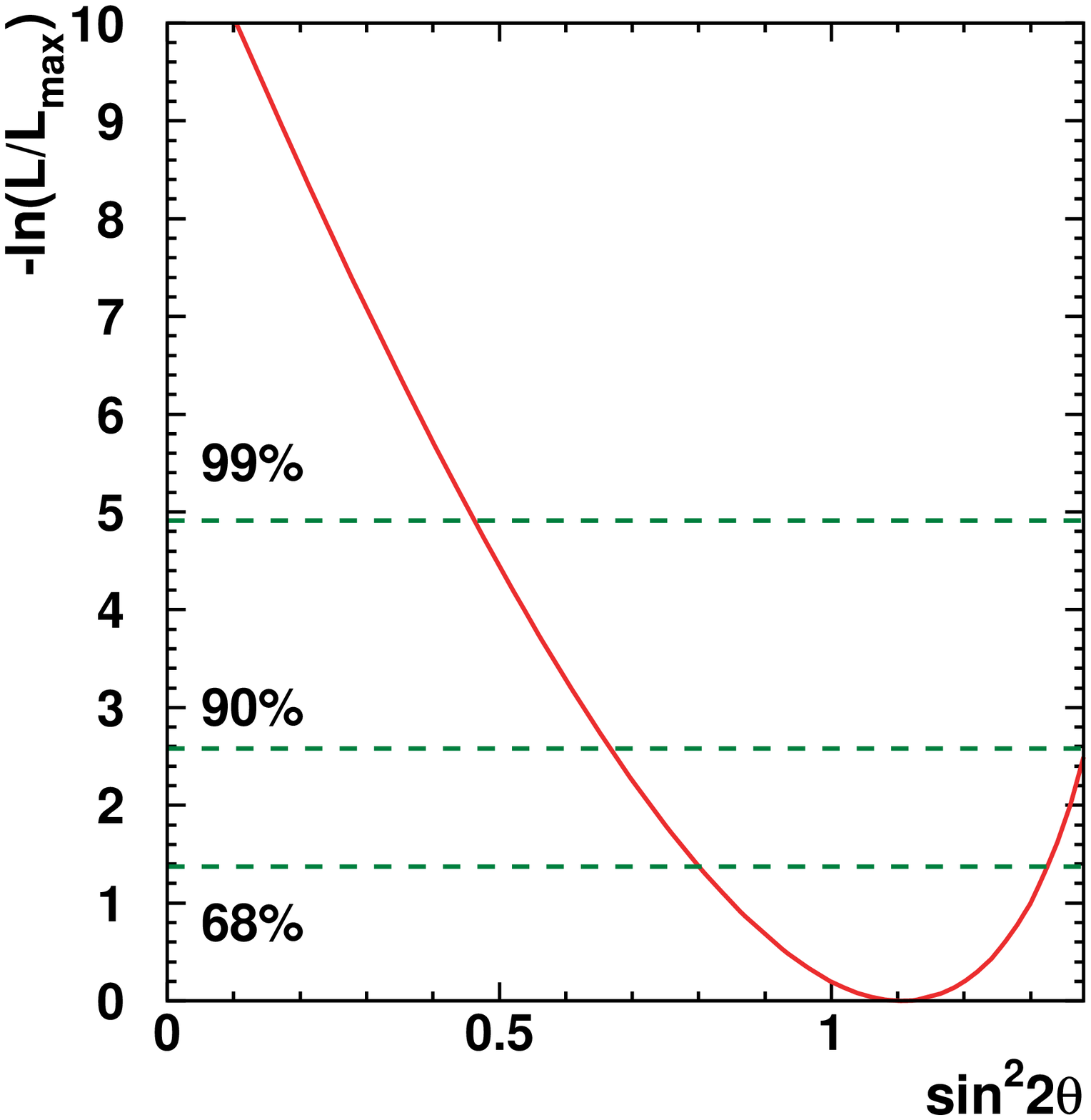}}
   \caption{$\ln\mathrm{L}_{\mathrm{max}}^{\mathrm{phys}} - \lnL(\dms,\sstt)$ distribution as a
   function of $\Delta m^{2}$~(left) and $\sin^{2}2\theta$~(right).
   $\sin^{2}2\theta$ is set to be 1.0 in the
   left-hand figure and $\Delta m^{2}$ is set to be $2.8 \times 10^{-3}
   \mathrm{eV}^{2}$ in the right-hand figure. Three horizontal lines
   correspond to the 68\%, 90\% and 99\% CL interval from the bottom
   one, respectively.}
   \label{fig:axis-scan}
 \end{center}
\end{figure}

The allowed regions in the neutrino oscillation parameter space,
corresponding to the 68\%, 90\% and 99\% confidence levels~(CL)
  are shown in Fig.~\ref{fig:contour-all}. 
They are defined as the contour lines with $\ln\Lcal =
\ln\mathcal{L}_{\mathrm{max}}^{\mathrm{phys}} - 1.37$, $-2.58$
and $-4.91$, respectively.  
These regions are derived by
using the two-dimensional Gaussian approximation from the maximum in the
unphysical region~\cite{Eidelman:2004wy}.
 The 90\% C.L. contour crosses the $\sin^{2}2\theta = 1$ axis at $\Delta
m^{2} = 1.9~\mathrm{and}~3.5 \times 10^{-3}~\mathrm{eV}^{2}$.
 Figure~\ref{fig:axis-scan} shows the distributions of
$\ln\mathcal{L}_{\mathrm{max}}^\mathrm{phys} - \ln\mathcal{L}(\dms,\sstt)$
as a function of $\sin^{2}2\theta$ and $\Delta m^{2}$,
with a slice at either $\Delta m^{2} = 2.8 \times 10^{-3} \mathrm{eV}^{2}$ or $\sin^{2}2\theta = 1.0$. 

\begin{figure} []
 \begin{center}
  \resizebox{\columnwidth}{!}{
   \includegraphics[trim=0.0cm 0.0cm 0.0cm
   0.0cm,clip]{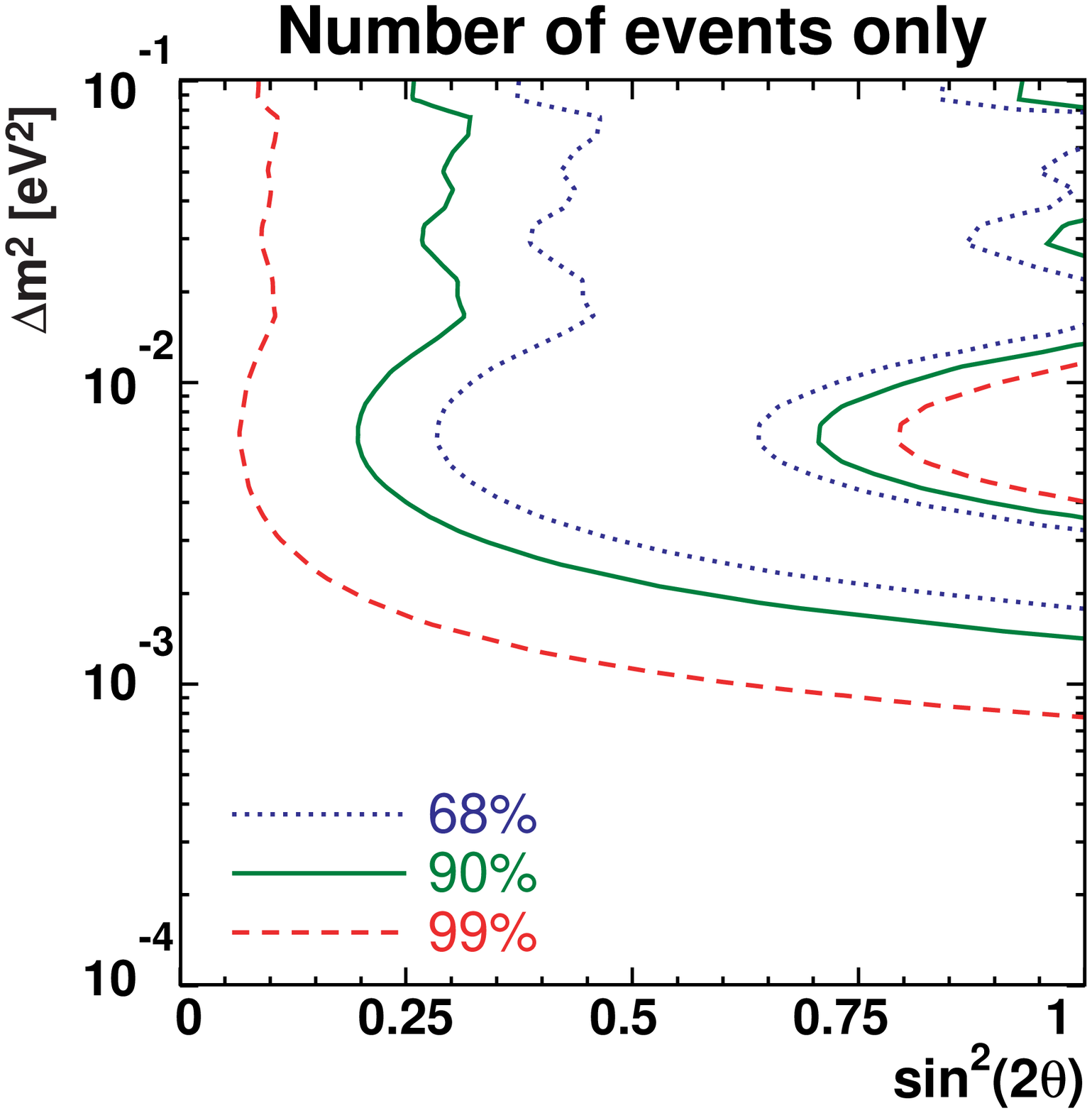}
   \includegraphics[trim=0.0cm 0.0cm 0.0cm
   0.0cm,clip]{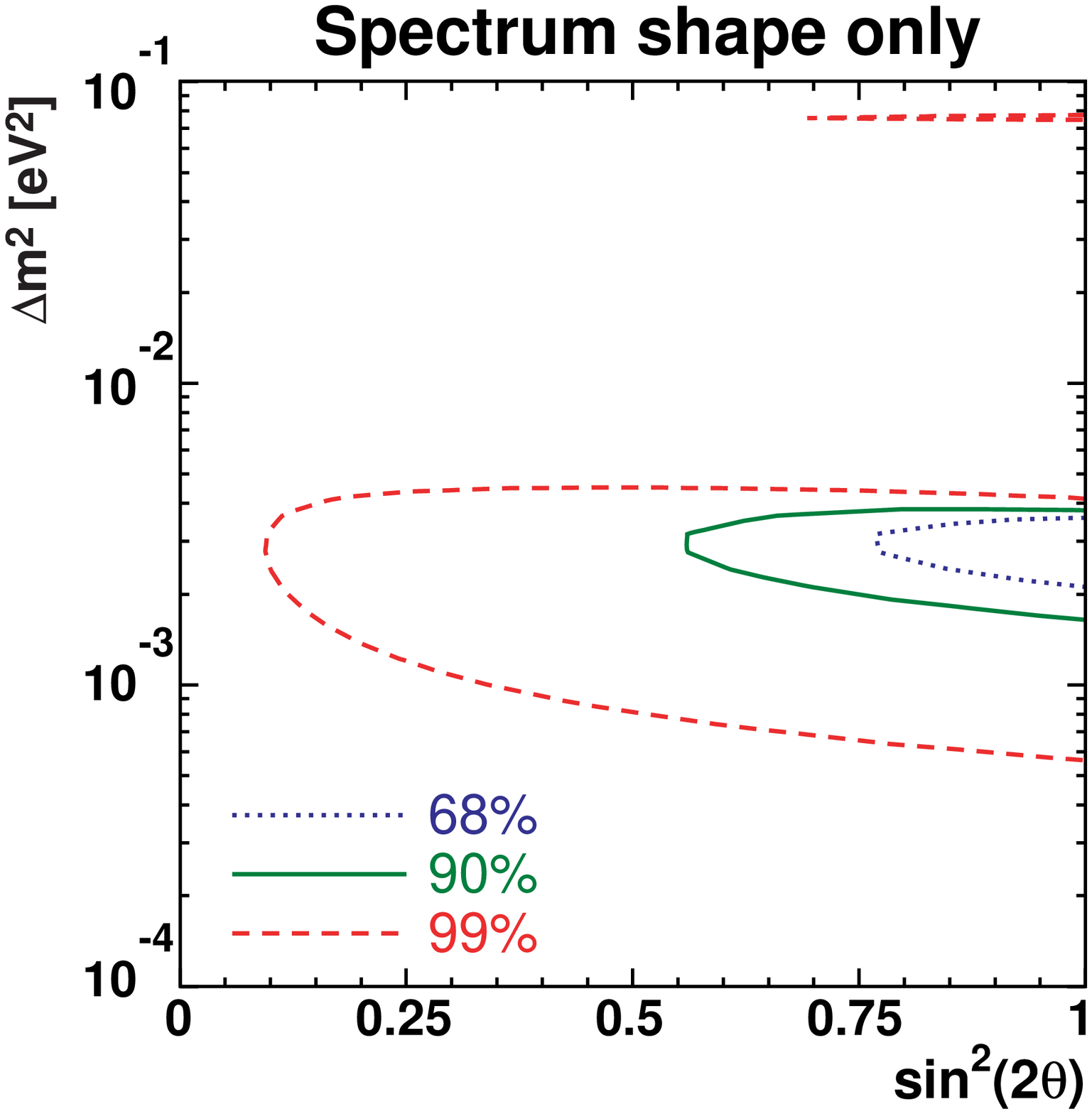}}
   \caption{Allowed region of oscillation parameters evaluated with the
   number of events only~(left) and the $E^{\mathrm{rec}}_{\nu}$
   spectrum shape only~(right). Both information allow the consistent region on
   the parameters space.}
   \label{fig:contour-partialdata}
 \end{center}
\end{figure}

\begin{figure} []
 \begin{center}
  \resizebox{\columnwidth}{!}{
   \includegraphics[trim=0.0cm 0.0cm 0.0cm
   0.0cm,clip]{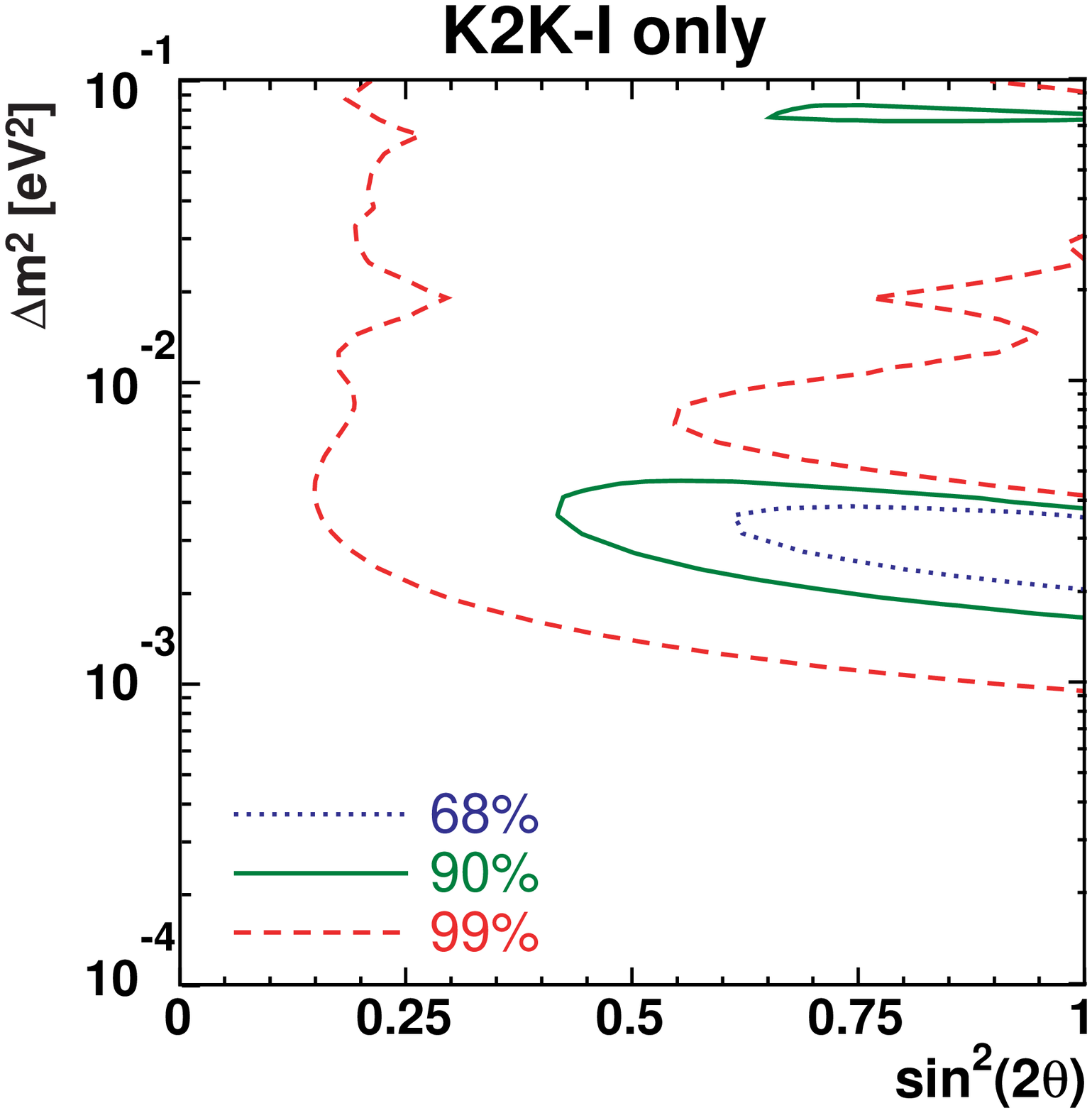}
   \includegraphics[trim=0.0cm 0.0cm 0.0cm
   0.0cm,clip]{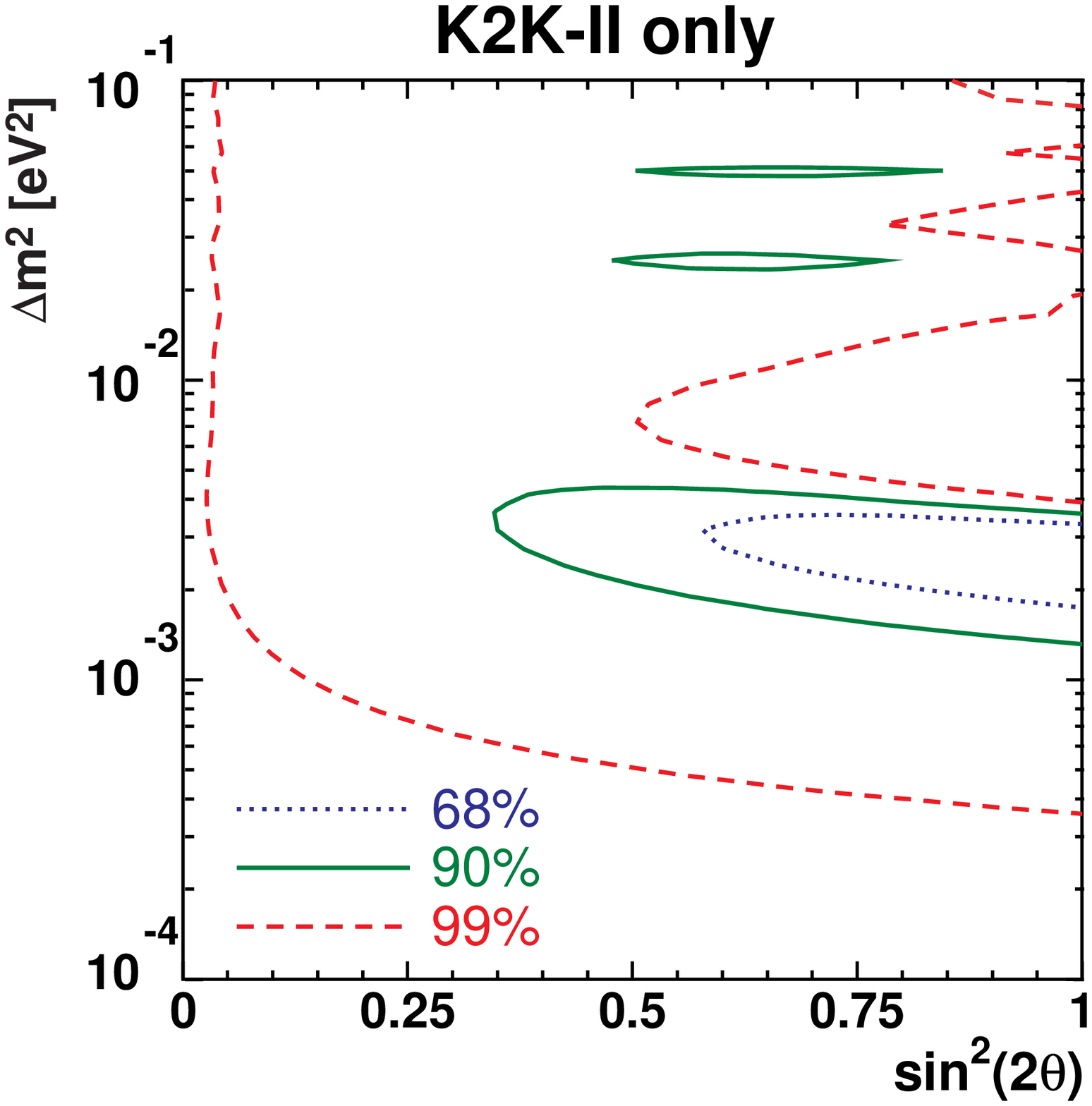}}
   \caption{Allowed region of oscillation parameters evaluated with
   partial data of K2K-I-only~(left)/K2K-II-only~(right). Both data
   allow the consistent region on the parameter space.}
   \label{fig:contour-splitdata}
 \end{center}
\end{figure}

We also check the consistency of the fit results performing the
analyses with only the normalization term or spectrum shape term, and
with the K2K-I or K2K-II sub-samples separately.  The fit results are
summarized in Tab.~\ref{tbl:FitSummary}.  There is no entry for the
normalization term only, because the two parameters cannot be
simultaneously determined from only one number.  The oscillation
parameters allowed by the normalization and the spectrum shape alone
agree with each other, as shown in both Tab.~\ref{tbl:FitSummary} and
Fig.~\ref{fig:contour-partialdata}.  The allowed regions calculated
with only K2K-I and K2K-II data are also consistent as shown in
Tab.~\ref{tbl:FitSummary} and Fig.~\ref{fig:contour-splitdata}.
 
\begin{figure} []
 \begin{center}
  \resizebox{\columnwidth}{!}{
   \includegraphics[trim=0.0cm 0.0cm 0.0cm
   0.0cm,clip]{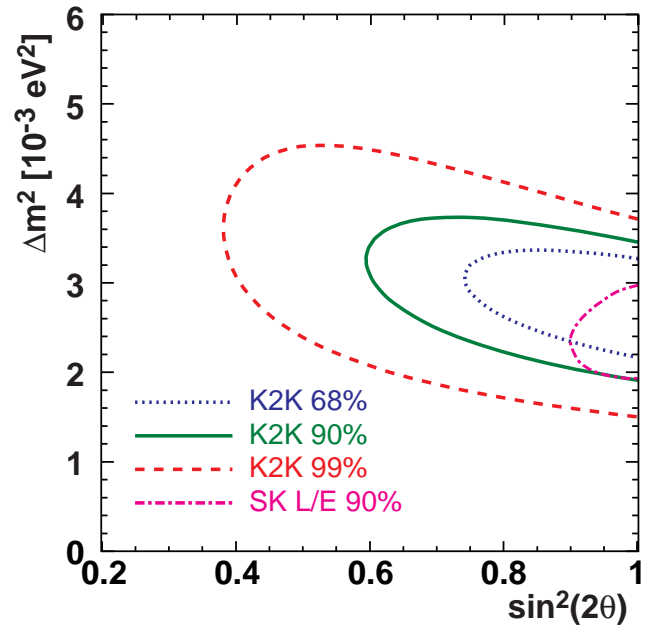}}
   \caption{
   Comparison of K2K results with the SK atmospheric neutrino measurement~\cite{Ashie:2004mr}.
Dotted, solid, dashed and dash-dotted lines represent 68\%, 90\%, 99\% C.L. allowed regions of K2K and 90\% C.L. allowed region from SK atmospheric neutrino, respectively.
}
   \label{fig:comparison-sk}
 \end{center}
\end{figure}

Finally, we compare our result with the parameters found by the
measurement of atmospheric neutrino oscillation by the
Super-Kamiokande collaboration~\cite{Ashie:2004mr}.
Figure~\ref{fig:comparison-sk} shows the allowed regions of
oscillation parameters found in this analysis
together with the SK result.  The K2K result is in good agreement with
the parameters found using atmospheric neutrinos, thereby confirming
the neutrino oscillation result reported by SK.
 
 \begin{table}
\begin{center}
  \caption{Summary of the oscillation parameters at the best fit point
  for each fitting condition.}
  \begin{tabular}{llcccc} \hline \hline
   \multicolumn{2}{c}{} & \multicolumn{2}{c}{All region} &
   \multicolumn{2}{c}{Physical region} \\ \cline{3-6}
   \multicolumn{2}{c}{} & $\dms~[\mathrm{eV}^2]$  & $\sstt$& 
   $\dms~[\mathrm{eV}^2]$  & $\sstt$ \\ \hline \hline
   All data & shape + norm. & $\exv{2.55}{-3}$ & 1.19  
   &  $\bf {\exv{2.75}{-3}}$ & \bf 1.00 \\ \cline{2-6}
    & shape only  & $\exv{2.77}{-3}$ & 1.25 & $\exv{2.95}{-3}$ & 1.00 
   \\
 \hline
   K2K-I     & shape + norm. & $\exv{2.77}{-3}$  & 1.08  &
   $\exv{2.89}{-3}$ & 1.00 \\ \hline
   K2K-II    & shape + norm. & $\exv{2.36}{-3}$ & 1.35 &
   $\exv{2.64}{-3}$ & 1.00  \\ \hline  \hline
  \end{tabular}
  \label{tbl:FitSummary}
 \end{center}
\end{table}

\section{Summary}

Data taken by the K2K experiment between June 1999 and November 2004
is used to observe and measure the parameters of neutrino oscillation
using an accelerator-produced neutrino beam.  The K2K experiment is
the first long-baseline neutrino experiment to operate at a distance
scale of hundreds of kilometers.  The neutrinos are measured first by
near detectors located approximately 300 meters from the proton
target, and then by the Super-Kamiokande detector 250~km away.  The
near detector complex consists of a 1~kiloton water Cherenkov
detector, and a fine grained detector system.  The energy spectrum and
flux normalization measured at the near detectors are used to predict
the signal seen at Super-K.  The results found are consistent with the
neutrino oscillation parameters previously measured by the
Super-Kamiokande collaboration using atmospheric neutrinos.

One hundred and twelve beam-originated neutrino events are observed in
the fiducial volume of Super-Kamiokande with an expectation of
$158.1^{+9.2}_{-8.6}$ events without oscillation.  The spectrum
distortion expected from oscillation is also seen in fifty-eight
single-ring muon-like events which have had their energy
reconstructed.  A likelihood analysis was performed and the
probability that the observations are explained by a statistical
fluctuation with no neutrino oscillation is $0.0015\%$~($4.3\sigma$).
In a two flavor oscillation scenario, the allowed $\dms$ region at
$\sstt=1$ is between $1.9$ and $3.5 \times 10^{-3}$~$\rm eV^2$ at the
90~\% C.L.  with a best-fit value of $2.8 \times 10^{-3}$~$\rm eV^2$.

\section{Acknowledgment}
We thank the KEK and ICRR directorates for their strong support and
encouragement.  K2K is made possible by the inventiveness and the
diligent efforts of the KEK-PS machine group and beam channel group.
We gratefully acknowledge the cooperation of the Kamioka Mining and
Smelting Company.  This work has been supported by the Ministry of
Education, Culture, Sports, Science and Technology of the Government
of Japan, 
the Japan Society for Promotion of Science, the U.S. Department of
Energy, the Korea Research Foundation, the Korea Science and
Engineering Foundation, NSERC Canada and Canada Foundation for
Innovation, the Istituto Nazionale di Fisica Nucleare (Italy), 
the Ministerio de Educaci\'on y Ciencia and Generalitat Valenciana (Spain),
the Commissariat \`{a} l'Energie Atomique (France), and Polish KBN grants:
1P03B08227 and 1P03B03826.

\bibliography{references}

\end{document}